# QUASI-NORMAL MODES IN RANDOM MEDIA

by

Jing Wang

2012

Abstract

QUASI-NORMAL MODES IN RANDOM MEDIA

by

Jing Wang

Advisor:  Professor Azriel Z. Genack

This thesis is an experimental study of microwave transmission through quasi-one-dimensional random samples via quasi-normal modes.

We have analyzed spectra of localized microwave transmitted through quasi-one-dimensional random samples to obtain the central frequency, linewidth and field speckle pattern of the modes for an ensemble of samples at three lengths. We find strong correlation between modal field speckle patterns. This leads to destructive interference between modes which explain strong suppression of steady state transmission and of pulsed transmission at early times. At longer times, the rate of decay of transmission slows down because of the increasing prominence of long-lived modes.

We have also studied the statistics of mode spacings and widths in localized samples. The distribution of mode spacings between adjacent modes is close to the Wigner surmise predicted for diffusive waves, which exhibit strong level repulsion. However, a deviation from Wigner distribution can be seen in the distribution of spacings beyond the nearest ones. A weakening in the rigidity of the modal spectrum is also observed as the sample length increases because of reduced level repulsion for more strongly localized waves. In contrast to residual diffusive behavior for level spacing statistics, the distribution of level widths are log-normal as predicted for localized waves. But the residual diffusive behavior can be seen from the smaller variance of the normalized mode width as compared to predictions for strongly localized waves.

We also measured the steady state and dynamic fluctuations and correlation of localized microwave transmitted through random waveguides. We find the degree of intensity correlation first increases, and then decays with time delay, before increasing dramatically. The variation in the spatial correlation of intensity with time delay is due to the changing effective number of modes that contribute to transmission. A minimum in correlation is reached when the number of modes contributing appreciably to transmission peaks. At long times, the degree of intensity correlation and the variance of total transmission increase dramatically. This reflects the reduced role of short-lived overlapping states and the growing weight of long-lived spectrally isolated modes.

## Acknowledgements

First and foremost, I am truly and deeply indebted to my advisor, Professor Azriel Z. Genack for his invaluable guidance, encouragement, patience and caring in all the time of research and writing of this thesis. His perpetual energy and enthusiasm in research had motivated all his advisees, including me. In addition, he was always accessible and willing to help his students not only with their research but also with their life. I also would like to thank Prof. Alexander Lisyansky, Prof. Tsampikos Kottos, Professors Micha Tomkiewicz and Prof. Leon Cohen for being my supervisory committee and for their valuable advices.

I would like to thank Professors Zhao-Qing Zhang and Andrey Chabanov for their great contribution to portions of the research presented in this thesis. Many thanks also go to the former and current students and postdoctoral fellows in this group, Dr. Bing Hu, Dr. Sheng Zhang, Zhou Shi and Dr. Matthieu Davy for their help, valuable discussion and friendship. I also want to thank Professor Jerome Klosner for long discussions and for continued detailed questions throughout the work. I would like to thank Professor Amnon Moalem for reviewing this thesis and for helpful suggestions. I am also happy to thank Howard Rose, Robert Bunch and Yelfim Radomyselskiy for their great technical assistance on machinery and electronics.

My deepest gratitude goes to my family for their everlasting love and support throughout my life; this dissertation would simply be impossible without them. Although my mother is no longer with us, she is forever remembered. I am sure she shares our joy and happiness in the heaven.

Finally, I would like to thank my wife Juan Shao. She was always there cheering me up and stood by me through good times and bad.

# Table of Contents

# LIST OF FIGURES

generate time-frequency spectrogram. (b) Spectrogram of total transmission at different delay time. The arrows indicate the initial guessing positions of underlying modes. ...........................................................................................................54

# CHAPTER 1

# INTRODUCTION

## 1.1 General introduction

Multiple scattering of waves is an inescapable part of our environment. Understanding wave propagation in random media such as clouds, fog, white paint, human tissue, electronic and optical devices, and metamaterials can help us probe and image our world and to transmit energy and information.

Multiply scattered waves in disorder media follow innumerable trajectories or partial waves with a wide distribution of path lengths. After a random sequences of scattering events, some of the trajectories may return to the same coherence volumes in the sample of volume, $V_c \sim (\lambda/2)^d$, where $d$ is the dimensionality and $\lambda$ is the wavelength in the medium, and form closed loops. For each loop there are two partial waves following the same sequences of scatterers but in opposite order, as shown schematically in Fig. 1-1. The coherent sum of the complex amplitudes for one looped trajectory $A_{\circlearrowleft}$ and its reversed loop $A_{\circlearrowright}$ may be written as $A = A_{\circlearrowleft} + A_{\circlearrowright}$. When time-reversal invariance holds, the amplitudes and phase associated with the time reversed partial waves are identical and the probability of return, $P_{\text{return}}$, is proportional to the absolute square of the amplitude of the coherent sum $|A|^2 = |A_{\circlearrowleft} + A_{\circlearrowright}|^2 = |2A_{\circlearrowleft}|^2 = 4|A_{\circlearrowleft}|^2$, which is twice the return probability obtained for the incoherent sum, $P_{\text{return}}^{\text{inc}} \sim |A_{\circlearrowleft}|^2 + |A_{\circlearrowright}|^2 = 2|A_{\circlearrowleft}|^2$. The enhanced return probability due to constructive interference leads to a reduction in transport [1-8]. This can be seen directly in enhanced retroreflection from a surface by a factor of two over the incoherent background, known as coherent backscattering or weak localization [3-5].

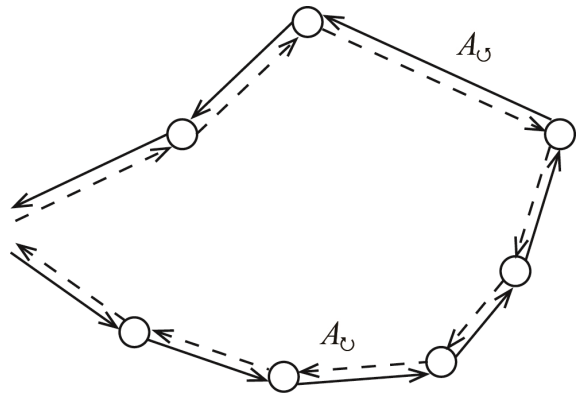

Fig. 1-1 Schematic representation of the two sequences of scattering events traversed in opposite directions. It is the source of the coherent backscattering effect.

In the weak scattering limit, the impact of weak localization is small and average transport can still be well described by the diffusion equation for average intensity, which corresponds to the random walk model of photons. When scattering within a medium is strong, the probability of waves returning back to the same coherent volume increases and wave transport is strongly suppressed by the constructive interference between returning partial waves. When scattering is sufficiently strong, the wave can be trapped inside the medium and the wave can become exponentially localized. The average of transmission then decays exponentially beyond the localization length, $\xi$, with $<T(L)> \sim \exp(-L/2\xi)$. $\xi$ corresponds to the average decay length of exponentially peaked within the sample, $L$ is the sample length, and $<...>$ represents the average over an ensemble of random sample configurations. This can be described as the vanishing of an effective diffusion coefficient, known as Anderson localization. Such strong localization can be distinguished from weak localization [1]. Both weak and strong localization are "mesoscopic" effects [9-13] that arise in multiply-scattered samples for waves that are temporally coherent within the sample.

In unbound systems, the return probability for the incoherent partial waves can be calculated in the diffusion model [14], $P_{\text{return}}^{\text{inc}} = \frac{V_c}{\tau_c} \int\limits_{2\tau}^{\infty} P(\mathbf{r} = 0, t) dt$. Here $P(\mathbf{r}, t) = \frac{1}{(4\pi D_B t)^{d/2}} \exp(-\frac{\mathbf{r}^2}{4 D_B t})$ is the Green function of the diffusion equation and it is the probability of particle diffusion with a displacement $\mathbf{r}$ at time $t$. $\tau_c$ is the time needed to travel a length in which an incident field loses coherence $\tau_c = \lambda/2v$, where $v$ is the transport velocity [15]. The lower limit of the integral is the earliest time of return, which is twice the mean free time between scattering events $\tau = \ell/v$ where $\ell$ is the transport mean free path. This corresponds to the time to return for a single scattering event at a distance of a single mean free path and, in which the direction of wave is randomized. $D_B = v\ell/d$ is the bare or Boltzmann diffusion coefficient. Then the return probability of incoherent partial waves can be written as $P_{\text{return}}^{\text{inc}} = \frac{(\lambda/2)^d}{\tau_c (4\pi D_B)^{d/2}} \int\limits_{2\tau}^{\infty} t^{-d/2} dt$ which is finite only for $d > 2$. This suggests the wave will return back to the coherent volume eventually after a certain delay time. Thus, the wave is localized independently of the strength of scattering for $d \leq 2$. For $d = 3$, expressing $D_B$ and $\tau$ in terms of the mean free path gives $P_{\text{return}} \sim 1/(k\ell)^2$, where $k = 2\pi/\lambda$ is the wave vector. For $k\ell \gg 1$, the return probability is small and the impact of wave interference on average transport is therefore small and propagation may be described by particle diffusion. However, when $k\ell \sim 1$, the probability with which waves will return to a coherence volume approaches unity and interference impedes the escape from this volume. This leads to the Ioffe-Regel criterion [16] for localization, $k\ell \sim 1$.

For electromagnetic radiation, the Ioffe-Regel criterion is hard to fulfill [17]. The electromagnetic scattering strength from a single particle vanishes in the long-wavelength Rayleigh limit, as $a^6/\lambda^4$, where $a$ is the particle radius. As a result, it is not possible to enhance scattering by packing many small scattering particles into volumes comparable to the wavelength since the density of particles only increases by $1/a^3$ as $a$ shrinks. Maximum scattering for spherical particles can be achieved at Mie resonances, when the particle radius is comparable to $\lambda$. But, even then, it is difficult to create strong scattering with $\ell$ as short as $\lambda$ since the maximum strength of scattering from individual particles is achieved only in dilute systems when the density of scatterers is low. However, in one- and two-dimensional samples or in samples with highly reflective boundaries, the average number of returns of a wave to a coherence volume can increase and exceed unity as the sample size increases. As a result, the conductance will always decrease with increasing sample size $L$ for $d \leq 2$ and waves will always become localized in sufficiently large samples.

Since the details of a configuration of a random sample are generally unknown and fluctuations in intensity are large, a statistical approach is required in studying multiply-scattered waves. Unlike electronic systems, in which only the conductance can be easily measured, a variety of less averaged propagation variable can be measured. Three principle transmission variables that can be measured are shown in Fig. 1-2. The coefficient $t_{ab}$ refers to the coefficient of field transmission from an incoming channel $a$ to an outgoing channel $b$. The indexes $a$ and $b$ may also refer to different points or polarizations on the input or output surface. The intensity transmission coefficient is $T_{ab} = |t_{ab}|^2$, and summation over all outgoing channels yields the total transmission, $T_a=\Sigma_b T_{ab}$. Further summing over all incoming channels gives the transmittance $T=\Sigma_a T_a$, which is associated with the dimensionless conductance $g$ in electron system via Landauer formula [18-20],

$$g = G/(e^2/h) = <T> = \sum_a <T_a> = \sum_{a,b} <T_{ab}>. \tag{1-1}$$

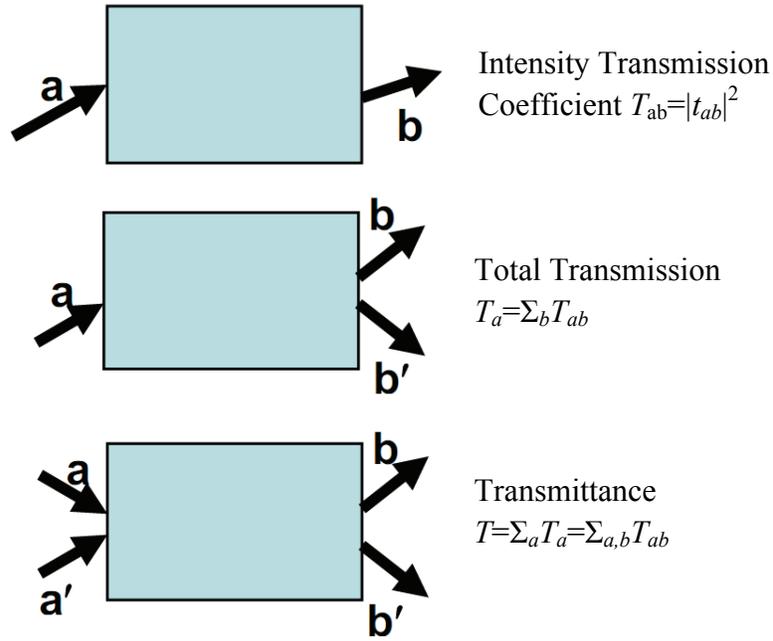

Fig. 1-2 Illustration of three principle transmission quantities.

Here $G$ is the conductance, $e$ is the electron charge, $h$ is Planck's constant and we associate $g$ with the ensemble average. In quasi-one-dimensional (quasi-1D) samples, with constant cross section and reflecting side walls with length $L$ much longer than the transverse dimensions, incident channels are completely mixed. As a result, the coupling between all pairs of incident and output transverse modes is equivalent. The conductance is then $g = <T> = N<T_a>$, where $N\sim 2\pi A/\lambda^2$ is the total number of incident transverse modes and $A$ is the cross-sectional area. In the diffusive regime, $<T_a> \sim \ell/L$, and the conductance is Ohmic with $g = N\ell/L >> 1$. Since $g$ will fall as the sample length increases in quasi-1D samples, $g$ will fall below unity and the wave will be localized in sufficiently long samples. Since $g = 1$ at the localization threshold, the localization length in quasi-1D samples is $\xi \sim N\ell$.

In the diffusive limit in the absence of inelastic processes, the variance of the relative total transmission and of transmittance are calculated using random matrix theory (RMT) to be, $\mathrm{var}(s_a) = <(\delta s_a)^2> = 2/3g$ and $\mathrm{var}(s) = <(\delta s)^2> = 2/15g^2$ [21,22] where $\delta s_a$ is the difference of the normalized total transmission relative to the ensemble average from its average value, $\delta s_a = s_a - <s_a> = s_a - 1$ and $\delta s = s - <s> = s - 1$. The distribution of total transmission normalized to its ensemble average $P(s_a = T_a/<T_a>)$ in quasi-1D samples is found to be a function of a single parameter, $g$ [21,23,24], and can be expressed as:

$$P(s_a) = \frac{1}{2\pi i} \int_{-i\infty}^{+i\infty} \exp(q s_a) F(3q\beta/2) dq \,,$$
$$F(q) = \exp\left(-\frac{2\ln^2(\sqrt{1+q}+\sqrt{q})}{3\beta}\right).$$

(1-2)

where $\beta = \mathrm{var}(s_a) = 2/3g$. Once the field in the speckle pattern is normalized by the square root of the average value of the intensity in the speckle pattern, it becomes a Gaussian random variable over the entire ensemble [21,23,25]. The speckle pattern is the fine-scale granular, random spatial pattern of intensity. In quasi-1D samples, the distribution of normalized transmitted intensity for a single polarization $s_{ab} = T_{ab}/<T_{ab}>$, is thus obtained by mixing the distribution of normalized total transmission of Eq. 1-2 with the negative exponential distribution of intensity normalized by the average intensity within the speckle pattern, $s_{ab}/s_a$, found by Rayleigh for a Gaussian field pattern [21,23,25],

$$P(s_{ab}) = \int_0^{+\infty} P(s_a) \frac{\exp(-s_{ab}/s_a)}{s_a} ds_a \,.$$

(1-3)

This is confirmed by experiments for both diffusive and localized waves [26,27]. This leads to the relationships between the moments of $s_a$ and $s_{ab}$ [21],

$$< s_{ab}^n > = n! < s_a^n > ,$$  (1-4)

and between the variance of the normalized total transmissions and transmitted intensities

$$\text{var}(s_{ab}) = 2\text{var}(s_a) + 1 .$$  (1-5)

where $\text{var}(s_{ab}) = < (\delta s_{ab})^2 >$ . The variances of total transmission and of transmittance can be related as $\text{var}(s_a) = g + \text{var}(s) + g\,\text{var}(s)$ [22].

## 1.2 Anderson localization

In 1958, Anderson explained that the anomalously slow relaxation of electron spins in doped semiconductors, which suggested that the interaction between electrons and their surroundings is less than expected, was due to the absence of diffusion of electrons in strongly scattering disordered lattices. It was not appreciated for over 25 years that localization was the result of wave interference [2,28,29] and was thus a general wave effect rather than a quantum mechanical effect that would apply to classical as well as to quantum waves. Anderson localization has been observed for electrons [30] and cold atoms [31-33], as well as for classical waves [34] such as microwave radiation [35,36], plasmons [37] and ultrasound [38,39]. Photon localization has been observed in 1D [40-42], 2D [43,44], quasi-1D [35,36], layered samples [45-48], and in the transverse dimension [49,50] for samples which are uniform in the propagation direction [51].

Many theoretical approaches and models have been developed to explain the wide variety of localization phenomena. In 1979, Abrahams, Anderson, Licciardello, and Ramakrishnan, proposed a scaling theory [52] of localization in which scaling of $g$ depends only on a single parameter, $g$, itself. In the scaling theory of localization, the variation of the scaling parameters, $g$, depends only upon the parameter itself [52,53]. In the early 1980s, Vollhardt and Wölfle developed a self-consistent localization theory (SCLT) [54] in which transport could be described by an effective diffusion coefficient as a function of the frequency of modulation of the incident intensity $\Omega$, $D(\Omega)$. Later, Van Tiggelen and coworkers argued that the diffusion coefficient should also depend on the depth into the sample $z$, $D(z; \Omega)$ [39,55-57]. But the SCLT failed to describe the stronger slowdown of the decay rate of the transmission in long time delay due to strong localization since it didn't include the effect of the long-lived localized states which is prominent in strong localized systems [58,59].

Another powerful approach to the study of waves propagating in disordered system is random matrix theory (RMT). RMT disregards the details of a system and assumes that the system can be described by a large random matrix with elements determined only by the symmetries of the system. RMT has been applied to two distinct problems in disordered systems. It was first applied to the study of the statistics of energy levels, associated with the eigenvalues of a random Hamiltonian matrix [60-64]. Later RMT was extended to the study of the statistics of conductance and of transmission through quasi-1D systems via the random scattering matrix [63,65-68].

Another powerful tool for treating disordered systems is the supersymmetric method introduced by Wegner [69] and Efetov [70,71]. In this approach, the problem of a single particle in a random potential is mapped onto the nonlinear $\sigma$-model in which the disorder-averaged correlation functions is calculated. The corresponding results for the two-level correlation function are the same as obtained by RMT [72,73]. Recently Tian and coworkers [59] included the effects of sharp resonances of the medium, which contribute significantly to the intensity within the medium in strongly localized samples and found a position-dependent diffusion coefficient $D(z)$ from a first-principles study of static wave transport in quasi-1D samples using the supersymmetric technique [74]. This work showed that SCLT is valid only when resonant transmission can be neglected, and so does not apply to deeply localized media or for pulsed transmission at long times.

In finite open medium, Thouless and co-workers argued [75,76] that the metal–insulator transition in disordered systems [77] could be described by a single parameter, the ratio of the average width and spacing of electronic energy levels, known as the Thouless number, $\delta=\delta E/\Delta E$ [75,76]. Such levels correspond to resonances of an open system. These are often referred to as quasi-normal-modes and we will refer to these simply as modes. The width of modes or levels is closely linked to the sensitivity of level energies to changes at the boundary since both are proportional to the ratio of the strengths of the mode at the boundary relative to the interior of the sample. When the wave is localized within the interior of a sample, the amplitude squared of the wave is exponentially small at the boundary [78] shown in Fig. 1-3 and is only weakly coupled to the surrounding medium. The mode lifetime is then long and its linewidth correspondingly narrow, so that, $\delta E < \Delta E$. On the other hand, when the wave is diffusive, modes extend throughout the medium; the wave then couples readily to its surroundings and the level is wide enough that its width overlaps several modes, $\delta E > \Delta E$. Thus, the localization threshold occurs at $\delta = 1$.

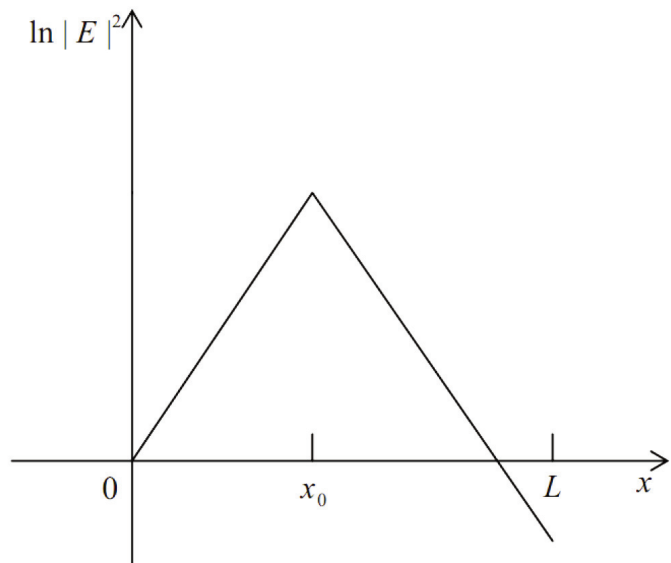

Fig. 1-3 Schematic plot of a state exponentially peaked within a random sample at position $x_0$.

However, because of the spatial and spectral overlap of modes, finding the characteristics of individual modes is difficult. But the Thouless number was related directly to a measurable quantity, the dimensionless conductance $g$. The conductance $G$ can be written as $G = \sigma A/L$, where $\sigma$ is the conductivity. The conductivity may be expressed via the Einstein relation as $\sigma=(e^2/h)Dn(v)$, where $n(v)$ is the density of states per unit volume per unit frequency and $D$ is diffusion coefficient, the dimensionless conductance

$$g = \frac{G}{e^2/h} = \frac{\sigma A}{L(e^2/h)} = \frac{DAn(v)}{L}. \tag{1-6}$$

The average level width $\delta E$ is inversely proportional to the Thouless time $\tau_{\mathrm{Th}}= L^2/D$ [75,76] which is defined as the time in which an electron diffuses through the sample. The average level spacing $\Delta E$ is the inverse of the density of states of the sample and can be written as $\Delta E \sim 1/n(v)AL$, Thus $g$ can be expressed as,

$$g = \frac{DAn(v)}{L} = \left(\frac{1}{n(v)AL}\right)^{-1} \left(\frac{L^2}{D}\right)^{-1} = \frac{\delta E}{\Delta E} = \delta, \tag{1-7}$$

and so is equal to the degree of spectral overlap of modes of the medium [52,53].

Though the details of the interactions of various waves are very different, the statistics of transport in random samples can be characterized in terms of a single localization parameter [52,53]. The many aspects of propagation and localization surely may be reflected in different parameters, but these should be simply related and provide an index of the character of transport.

The scaling theory of localization provided the variation of $g$ with the scale of the sample and the dimensionality of the sample, $d$. The numerator in the ratio $g = \delta = \delta E/\Delta E$ scales as $L^{-2}$ in the Ohmic or diffusive regime and falls exponentially in the localized limit, whereas the denominator $\Delta E$ scales inversely with the volume of the samples, as $L^{-d}$. Thus $g \sim L^{2-d}$ increases with $L$ for diffusive waves only for $d>2$, and decreases for localized waves. Thus we find again that $d=2$ is the marginal dimensionality for localization. A critical fixed point marking the localization threshold exists at $g \sim 1$ for $d > 2$ for which $g$ is invariant with sample size. For $d<2$, there is no transition since $g$ always falls with increasing sample size and multiply scattered coherent waves will always become localized in sufficiently large samples even when the scattering cross section is small. In the localized regime, for $d<2$, $\delta E$ is exponentially small and $\delta$ and hence $g$ fall essentially exponentially with sample.

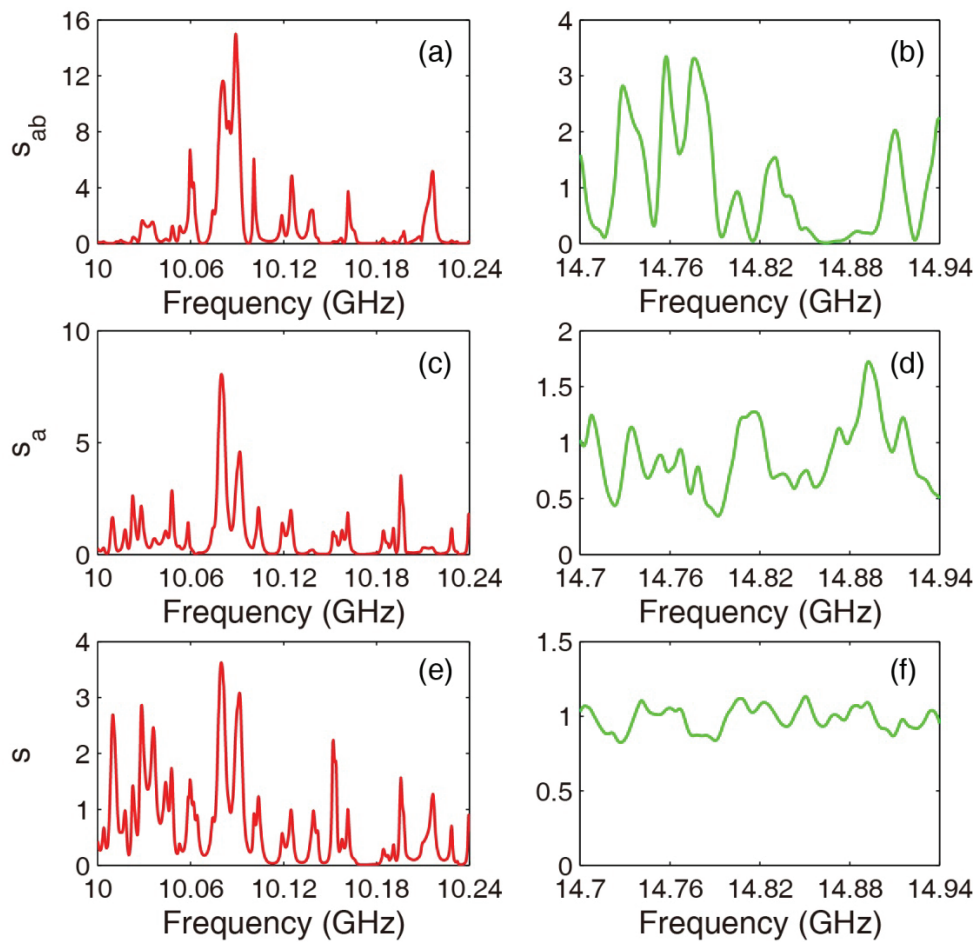

Fig. 1-4 Examples of spectra of transmitted microwave intensity $s_{ab}$, total transmission, $s_a$, and transmittance $s$ relative to the ensemble average value for (a), (c), (e) localized and (b), (d), (f) diffusive waves in quasi-1D samples.

For classical waves, it is natural to describe the width and spacing between resonances in terms of frequency rather than energy so we write $\delta = \delta v / \Delta v$ where $\delta v$ and $\Delta v$ is the mode width and spacing in frequency. In strongly localized samples in which $\delta < 1$, sharp peaks are seen in the transmission spectrum as the frequency is tuned through resonances with modes of the random medium [35]. In contrast, when resonances overlap spectrally, so that, $\delta > 1$, fluctuations of transmission relative to its average are reduced. This can be seen in the example spectra of the transmitted intensity, total transmissions and transmittance for diffusive and localized microwave radiation in quasi-1D samples shown in Fig 1-4. The variance of transmitted intensity, total transmission and transmittance normalized by its ensemble average over with statistically equivalent disorder are inversely related to the $\delta$.

## 1.3 Speckle

A high-contrast, fine-scale granular pattern of wave intensity was observed in reflection from a rough surface as soon as coherent laser sources were introduced in the early 1960s [79,80]. This random spatial pattern of field or intensity referred to as the  pattern because of its grainy appearance as shown schematically in Fig. 1-5, is produced by the superposition of randomly scattered waves in static disordered systems. The origin of these speckle patterns is the interference between multiple scattered waves. This pattern washes out if averaged over a long time in a disordered sample with moving scatterers or if the average is taken over a large number of realizations of static random scatterers. Since even small rearrangements of scatterers could modify the speckle pattern, the speckle pattern provides a fingerprint of a particular configuration of the random sample. Even though it is not in general possible to infer the internal structure of a body even from a complete set of such patterns for all incident wave vectors, essential elements of a description of wave transport can be inferred from the statistics of the speckle pattern of scattered radiation from ensembles of random samples.

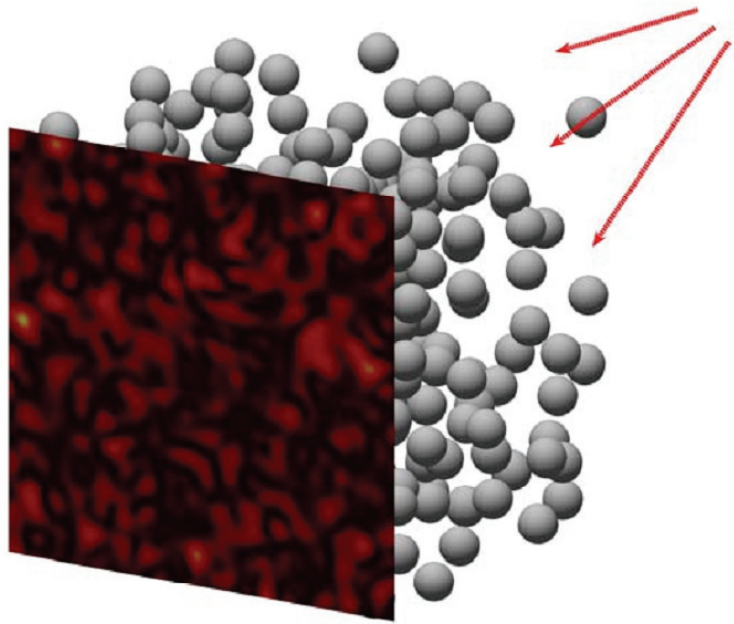

Fig. 1-5 Schematic of speckle pattern produced by multiple scattering in a random medium.

The size of speckle spots within a random medium is on the order of the wavelength and equals to the field correlation length $\delta x$, which is the displacement for the first zero of the field correlation function. Long-range correlation of intensity within the speckle pattern leads to enhanced fluctuations in total transmission and transmittance between sample configurations [81-83]. The degree of correlation $\kappa$ can be obtained from the cumulant intensity correlation function with displacement between two positions within the speckle patters for a fixed source can be written as, $C_I(\Delta r)=<\delta I(r)\delta I(r+\Delta r)>=F(\Delta r)+\kappa(1+F(\Delta r))$ [84]. Here $F$ is the square of the field correlation function $F = |F_E|^2 = |<E(\mathbf{r})E^*(\mathbf{r}+\Delta\mathbf{r})>|^2$, $\kappa$ is the value of $C_I$ when $F(\Delta r)=0$. The degree of correlation $\kappa \equiv <\delta s_{ab}\delta s_{ab'}>$ is directly related to the var($s_a$), $\kappa = \text{var}(s_a)$ [81,82,84,85].

The evolution of the speckle patterns as the frequency is tuned can be described as the changing of the weights of different modes. Near the resonance with a mode of the medium, the normalized intensity speckle evolves slowly since the speckle is dominated by the modes at resonance. Off-resonance, the intensity speckle pattern can change rapidly with frequency shift as the relative contributions of different modes changes.

## 1.4 Modes

Each mode corresponds to a volume speckle pattern of the field with an overall variation with frequency which depends only upon its central angular frequency and linewidth, $\omega_n$ and $\Gamma_n$, respectively, for the $n^{\text{th}}$ quasimode. The mode linewidth is the inverse of the lifetime of the mode, $\Gamma_n = 1/\tau_n$, and is the sum of the leakage rate from the sample and absorption rate within the sample. The spatial and spectral variation of the $j^{\text{th}}$ polarization component of the field at position $\mathbf{r}$ for the $n^{\text{th}}$ quasimode is given by,

$$A_{n,j}(\mathbf{r},\omega) = a_{n,j}(\mathbf{r}) \frac{\Gamma_n/2}{\Gamma_n/2 + i(\omega - \omega_n)}. \tag{1-8}$$

where $a_{n,j}(\mathbf{r})$ is the complex amplitude for $n^{\text{th}}$ modes. The spatial distribution of mode field corresponds to the Green's function for the Helmholtz equation. The denominator of Eq. 1-8 may be regarded as a resonance with a complex frequency with real part corresponding to the central resonant frequency the imaginary to the linewidth of the mode. In contrast to modes of a closed system, quasi-normal modes are eigenstates associated with open systems, in which energy may be absorbed within the sample or leak out through the boundaries. Although, in general, the eigenstates of a non-Hermitian operator do not form a complete basis, Leung and coworkers have demonstrated [86] that when the refractive index of materials in a leaky system varies discontinuously and approaches a constant asymptotic value sufficiently rapidly, the modes are complete and orthogonal so that an arbitrary state of the system can be expressed as a superposition of modes [41,87]. The complex frequency $\omega_n - i\Gamma_n/2$ is the pole of the scattering matrix $S$, which relates the outgoing channels to the incoming channels. So the full statistics of modes should make it possible to determine the full statistics of wave propagation in the frequency and time domains.

The Thouless number represents the average overlap of modes in a random ensemble, but the number of modes contributing substantially to transmission at any frequency in a given configuration will vary. Multiply-peaked states form in space on both sides of the mobility edge whenever a small number of modes overlap spectrally with a number of peaks approximately equal to the number of overlapping modes [41,88-91]. This was explained by Mott [88] as the hybridization of excitations due to the overlap of excitations peaked at neighboring localization centers [88-90,92]. Such overlapping states have been termed necklace states by Pendry [90,93], who showed that they dominate transmission for localized waves. Necklace states are relatively short lived and contribute strongly to transmission since the weight of intensity is closer to the sample boundaries than for localized states. Moreover, these short-lived modes impact transmission over a wide frequency range since the lines are broad. The impact on transmission of such short-lived states become even more pronounced in the presence of absorption since they are attenuated less by absorption than are long-lived. In diffusive samples, frequencies at which the degree of overlap of resonant states is not as high as the average value within the random ensemble are unusually long-lived since the spatial spread of these modes is not as great as for typical modes. Such prelocalized states [94] are of particular importance near the lasing threshold of random lasers [95].

## 1.5 Random matrix theory

Wigner first proposed that random matrices could describe the statistical properties of excited states of atomic nuclei. Later RMT was applied to other domains including atomic physics [96], mesoscopic physics [62,63,66,97], wireless communication [98] and even finance [99]. RMT has been applied to two kinds of transport problems using two types of approaches. In the first approach, the statistics of energy levels of disordered systems, such as nuclei, metal grains or chaotic structures are studied. The Hamiltonian $H$ is represented as a random matrix. In the second approach, the statistics of transport properties are described for disordered systems, such as metal wires or quantum dots with point contacts. In this case, the random matrix is the scattering matrix $S$ or its submatrix, the transmission matrix $t$.

### 1.5.1 Random Hamiltonian matrix

In 1951 Wigner [60] conjectured that the distribution of energy spacings between neighboring levels in heavy nuclei in which states overlap spatially would be the same as for the eigenvalues in an ensemble of matrices with random Gaussian elements. Statistical properties of the energy levels and wave functions for such an ensemble or many other disordered systems are universal. They are independent of the physical properties of the complex systems, such as the size or shape of the sample and scatters, or the impurity concentration, and depend only on the symmetries of the system labeled by symmetry index $\beta$: time-reversal symmetry and spin-rotational symmetry. The elements of $H$ are real, complex, or real quaternion numbers for systems with time-reversal symmetry ($\beta$=1), systems in which time-reversal symmetry is broken ($\beta$=2), and systems in which spin-rotation symmetry is broken but time-reversal symmetry ($\beta$=4) is valid, respectively [63,100]. The random ensemble with $\beta$=1 is known as the Gaussian orthogonal ensemble (GOE) since the Gaussian distribution of the random matrix is invariant under orthogonal transformation. The distribution of level spacings found in slow nuclear scattering experiments normalized by the average spacing, $s = (E_{i+1}-E_i)/<E_{i+1}-E_i>$ is well fit by the Wigner surmise for GOE, $P_W(s)=(\pi s/2)\exp(-\pi s^2/4)$ [60,64,101]. The Wigner surmise is in good agreement with measurements of nuclear and atomic level spacings and also describes the level statistics of chaotic systems such as classical waves in billiards [97,102] and electrons in quantum dots [103]. It has also been predicted to apply to disordered metals [73,104]. The vanishing of the probability density of the level spacing as $s \rightarrow 0$ is a consequence of the repulsion between levels. Level repulsion leads to rigidity in the spacing between levels which gives rise to universal conductance fluctuations since the variance of the number of states excited within the level width by a potential difference across a conductor or by an incident monochromatic classical wave is of order unity [104]. The deviations of the level statistics from the universal result of RMT become more and more prominent with increasing disorder. In the strong localization limit of the Anderson model, in which the off-diagonal elements of the random matrix vanish, eigenvalues are uncorrelated and level spacings are expected to follow the Poisson distribution, $P_\text{P}(s){=}\exp(-s)$ [64]. To study the crossover from the diffusive limit to localized limit, a power-law random banded matrix ensemble [105], whose elements are independent random variables with the amplitude decreasing from the matrix diagonal in a power law, is considered.

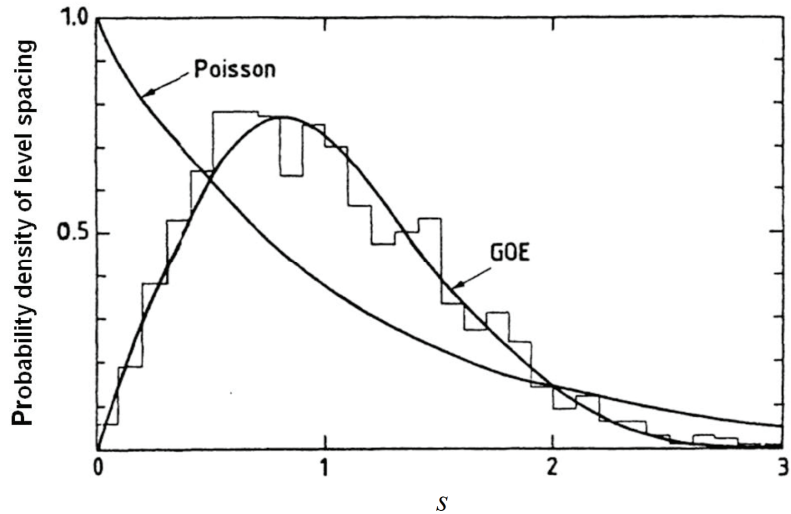

Fig. 1-6 Level spacing distribution for a large set of nuclear levels.

### 1.5.2 Random transmission matrix

RMT has been applied to transport properties through open system with wire geometry, such as metal wire for electrons or quasi-1D random media for classical waves. Here the random matrix is the scattering matrix which relates all outgoing and incoming channels or its submatrix, the transmission matrix $t$. The scattering matrix $S$ is defined as,

$$\begin{pmatrix} E_a^- \\ E_b^+ \end{pmatrix} = S \begin{pmatrix} E_a^+ \\ E_b^- \end{pmatrix} \tag{1-9}$$

where $E_a^+$ and $E_b^-$ are the incident waveguide channels, $E_a^-$ and $E_b^+$ are the outgoing waveguide channels . Channels labeled $a$ are channels on the left side of the sample and channels labeled $b$ are channels on the right side of the sample as shown schematically in Fig. 1-7. The + sign indicates the flux is towards the sample while the direction, the − sign indicates the wave is moving away from the sample. In nondissipative open systems, the matrix $S$ is unitary due to current conservation. In addition, if the system has time-reversal symmetry, $S$ is symmetric as well as unitary. The scattering matrix can be split into four submatrices, the transmission matrix $t$ and its transpose $t'$, and the reflection matrix $r$ and its transpose $r'$,

$$S = \begin{pmatrix} r & t' \\ t & r' \end{pmatrix}. \tag{1-10}$$

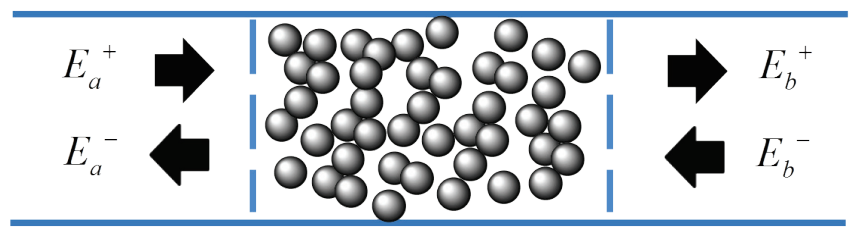

Fig. 1-7 Schematic of scattering matrix in a disordered waveguide. The scattering matrix $S$ couples incident waveguide modes $E_a^+$ and $E_b^-$ to outgoing waveguide modes $E_a^-$ and $E_b^+$ .

The elements of the field transmission matrix $t$ relate the transmitted field in propagating channel, $b$, and the channels of the incident field, $a$, $E_b = \sum_{a=1}^{N} t_{ba} E_a$. Here $N$ is the number of freely propagating incoming, or outgoing waveguide modes in a region beyond the sample but with the same cross section. The field transmission matrix is not hermitian and so cannot be diagonalized. However, $t$ can be expressed as, $t = U \Lambda V^{\dagger}$, where $U$ and $V$ are unitary matrices and $\Lambda$ is a diagonal matrix with singular values $\lambda_n$ along the diagonal. One way to study the properties of propagation is to consider the eigenvalues $\tau_n$ of the transmission matrix $t t^{\dagger}$, which is hermitian in the absence of inelastic processes. The eigenvalues $\tau_n$ are the squares of the singular values $\lambda_n$ of $t$, $\tau_n = \lambda_n^2$. The scattering matrix can be expressed in terms of the $\tau_n$ via the polar decomposition [66],

$$S = \begin{pmatrix} U & 0 \\ 0 & V \end{pmatrix} \begin{pmatrix} -\sqrt{1-T} & \sqrt{T} \\ \sqrt{T} & \sqrt{1-T} \end{pmatrix} \begin{pmatrix} U' & 0 \\ 0 & V' \end{pmatrix}, \tag{1-11}$$

where $U$, $V$, $U'$, $V'$ are $N{\times}N$ unitary matrices and $T$ is an $N{\times}N$ diagonal matrix with transmission eigenvalues on the diagonal, $T = \text{diag}(\tau_1, \tau_2, \dots, \tau_N)$. The conductance is the ensemble average of the sum of all transmission eigenvalues, $g = <\sum_n \tau_n>$. The field at the output may be represented as a sum of orthogonal speckle patterns each generated by one of a set of orthogonal incident field patterns corresponding to the singular values $\lambda_n$. One may visualize the transmission process as the coupling of the incident field to the channels of the disordered medium through the unitary matrix $V$ and the subsequently coupling to the transmitted signal by unitary matrix $U$.

The constant of proportionality 2/3 between $g$ and $1/\mathrm{var}(s_a)$, $g=2/3\mathrm{var}(s_a)$, is calculated using diagrammatic calculations [23,24] as well as RMT based under the assumption that the eigenvalues have a bimodal distribution $P(\tau) = \dfrac{g}{2} \dfrac{1}{\tau\sqrt{1-\tau}}$ [21]. While Dorokhov [65,106] predicted a constant spacing between adjacent values of the logarithm of eigenvalues of transmission matrix [107], which leads to $g=1/2\mathrm{var}(s_a)$. The departure from the bimodal distribution is observed in our experiments. This may be a consequence of internal reflection [107].

## 1.6 Outline

In this thesis, we deal with waves transmitted through multiply scattered quasi-1D random systems and provide a modal description of different facets of wave propagation and localization in random media. In the first chapter, we reviewed important aspects of wave propagation in disordered system. In Chapter 2, we describe measurements of steady state and dynamics fluctuations and correlation of localized waves with polarization rotation of both the source and detector or with displacement of the detector. The non-monotonic behavior of the dynamics of mesoscopic fluctuations of localized waves provides a window on the evolving contributions of short- and long-lived electromagnetic modes of the random medium. In Chapter 3, we present the algorithm used to decompose the transmitted field speckle patterns into a superposition of mode patterns and retrieve the central frequency and linewidth of each mode. We also show that the modal description of transmission explains both the suppression of steady state and pulsed transmission. In Chapter 4 we will present the statistics of mode parameters. Our findings and conclusions are summarized in Chapter 5.

# CHAPTER 2

# FLUCTUATIONS AND CORRELATION IN LOCALIZED QUASI-ONE DIMENSIONAL RANDOM SAMPLES

## 2.1 Introduction

Even after averaging over disorder, the correlation between transmitted waves does not disappear. Enhanced mesoscopic correlation of intensity at two points on the sample output due to two sources at the input arises is a result of interference between the corresponding four partial waves with trajectories which cross within the sample, as shown in Fig. 2-1. The crossing of trajectories $ab'$ and $a'b$ pair the four trajectories $ab$, $ab'$, $a'b$ and $a'b'$ differently. Before the crossing, the trajectories $ab$, $ab'$ and $a'b$, $a'b'$ are paired together while after the crossing the trajectories $ab$, $a'b$ and $ab'$, $a'b'$ are paired together. At the crossing, the complex amplitudes of trajectories $ab'$ and $a'b$ are permuted, so that the trajectories of each pair after the crossing have a small phase difference. Because the phase shift is small, the contributions of different possible pairings will survive averaging over an ensemble of configurations. This gives rise to spatially extended correlation, which leads to enhanced fluctuations of integrated quantities such as $T_a$ and $T$.

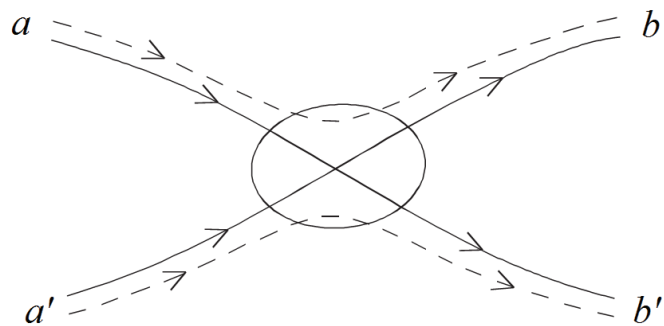

Fig. 2-1 Schematic representation of the crossing of four partial waves.

From diagrammatic calculations [85,108] for correlation, as well as RMT calculations [81], and experiments [82,109-111], the cumulant correlation function with displacement and polarization rotation of the source and detector of intensity normalized by its ensemble,

$$C_I(\Delta r_S, \Delta r_D; \Delta \theta_S, \Delta \theta_D) = \frac{< \delta I(r_S, r_D; \theta_S, \theta_D) \delta I(r'_S, r'_D; \theta'_S, \theta'_D) >}{< I(r_S, r_D; \theta_S, \theta_D) >< I(r'_S, r'_D; \theta'_S, \theta'_D) >}, \qquad (2\text{-}1)$$

can be expressed as the sum of three terms for both diffusive and localized waves. Here $\delta I$ is the deviation of the intensity from its ensemble average value $\delta I = I - <I>$, $\Delta r_S$ and $\Delta r_D$ are the displacements of the source and detector, and $\Delta \theta_S$ and $\Delta \theta_D$ are the polarization rotations of the source and detector, respectively. Each term involves only the product or sum of the complex squares of the correlation functions of the displacement and polarization on the input and output surfaces of the field normalized to the square root of the average intensity,

$$F_{in} = \left| F_E(\Delta r_S, \Delta \theta_S) \right|^2 = \left| \frac{< E(r_S, \theta_S) E^*(r'_S, \theta'_S) >}{\sqrt{< I(r_S, \theta_S) >}\sqrt{< I(r'_S, \theta'_S) >}} \right|$$

and

$$F_{out} = \left| F_E(\Delta r_D, \Delta \theta_D) \right|^2 = \left| \frac{< E(r_D, \theta_D) E^*(r'_D, \theta'_D) >}{\sqrt{< I(r_D, \theta_D) >}\sqrt{< I(r'_D, \theta'_D) >}} \right|.$$

This gives

$$C_I = C_1 + C_2 + C_3 = F_{in} F_{out} + A'_2 (F_{in} + F_{out}) + A'_3 (1 + F_{in} F_{out} + F_{in} + F_{out}). \qquad (2\text{-}2)$$

$F_{in}$ and $F_{out}$ have the same functional form with respect to displacement and polarization rotation can be factorized into the product of correlation functions with displacement and polarization rotation separately, so that [109,112]

$$F_{in}F_{out} = F\left(\Delta r_S', \Delta \theta_S\right) F\left(\Delta r_D, \Delta \theta_D\right) = F\left(\Delta r_S'\right) F\left(\Delta \theta_S\right) F\left(\Delta r_D\right) F\left(\Delta \theta_D\right).$$

In the interior of the sample, the field correlation function with displacement $\Delta r$ is given by, $F_E(\Delta r) = \exp(-\Delta r/2\ell)\sin(k\Delta r)/k\Delta r$. [113] The field correlation function on the sample surface, is the Fourier transform of the specific intensity, which is the ensemble average of the intensity per unit solid angle for the distribution of scattered intensity in the far field, $<I(k_{\parallel}/k)>$, where $k_{\parallel}$ is the component of the $k$-vector of transmitted radiation in the plane of the surface [82,114].

The polarization variable has only two independent states for orthogonal polarizations in the input and output planes of the sample. The variation of the correlation function of the field with polarization $\theta$, $E(\theta)$, and field with polarization $(\theta+\Delta\theta)$, $E(\theta+\Delta\theta)$, can be expressed as [109, 112]:

$$\begin{aligned}
F_E\left(\Delta \theta\right) &= <\mathbf{E}\left(\theta\right)\mathbf{E}^*\left(\theta+\Delta\theta\right)> \\
&= <\mathbf{E}\left(\theta\right)\left(\mathbf{E}\left(\theta\right)\cos(\Delta\theta)+\mathbf{E}\left(\theta+\frac{\pi}{2}\right)\sin(\Delta\theta)\right)^* > \\
&= <E\left(\theta\right)E^*\left(\theta\right)\cos(\Delta\theta)> \\
&= \cos(\Delta\theta) \quad .
\end{aligned}$$

in which the field $E(\theta+\Delta\theta)$ is decomposed into two perpendicular linearly polarized components [109]. Thus the field correlation function with polarization rotation has a simple form independent of scattering strength. It is determined exclusively by the cosine of the polarization rotation $\Delta\theta$ of the electric field at the source and detector, $F_E(\Delta\theta) = \cos(\Delta\theta)$ [109,112] which vanishes when the polarization is shifted by 90° This enables an unambiguous separation of the intensity correlation function of a vector wave into short-, long-, and infinite-range components.

By grouping terms with similar dependence upon the source and detector field correlation functions, the intensity correlation function may be expressed as the sum of multiplicative, additive, and constant terms [109],

$$C_I = A_1 F_{in} F_{out} + A_2 (F_{in} + F_{out}) + A_3,$$

where, $A_2 = A'_2 + A'_3$ and $A_3 = A'_3$. For diffusive waves, the multiplicative, additive, and constant terms, which correspond to short-, long-, and infinite-range contributions to $C$ dominate fluctuations of intensity, total transmission and transmittance, respectively. For localized waves, $A'_3$ is much larger than $A'_2$ and 1 so that the infinite-range term dominates correlation and fluctuations. If we fix the source so that $F_{in}$=1, we have [84]

$$C = F_{out} + (A'_2 + 2A'_3)(1 + F_{out}) = F + (A_2 + A_3)(1 + F) = F + \kappa(1 + F). \qquad (2\text{-}3)$$

Here the degree of correlation $\kappa = A_2 + A_3$ equals the value of $C$ at points at which $F$=0. For the autocorrelation function, $F$=1 and $I$=$I'$, $C_I = <(\delta I)^2>/<I>^2 = \text{var}(s_{ab}=I/<I>)$ and $C_I$=1+2$\kappa$. Since $\text{var}(s_{ab})$=2$\text{var}(s_a)$+1, $\kappa = \text{var}(s_a)$ [25,26,82,109]. In the diffusive limit and in the absence of absorption, $\kappa$=2/3$g$, where $g$ is the dimensionless conductance. In this limit, $g$>>1 and $\kappa$ is small. To order 1/$g^2$, the coefficients are $A'_2$=2/3$g$ and $A'_3$=2/15$g^2$, then $A_2$=2/3$g$+2/15$g^2$ and $A_3$=2/15$g^2$.

Since the lengths of the wave trajectories increase with time, the density of such trajectories in space increases so that paths cross upon themselves more frequently and the impact of localization builds up in time. It is therefore natural to study the statistics of localization in the time domain. Because the length of all trajectories reaching the detector at a fixed time delay are equal, the amplitude of all partial waves is reduced equally by absorption and the distribution of paths at a fixed delay is not affected by absorption. The renormalization of scattering parameters is therefore not influenced by absorption in the time domain. Thus the impact of absorption on weak localization can be eliminated in the time domain though these effects are entangled in steady state measurement.

In this chapter, we present steady state and dynamic properties of mesoscopic fluctuations and correlations of localized waves through quasi-1D random samples. We will also describe the dynamics of fluctuations and correlations of localized waves by analyzing the field into its constituent modes.

## 2.2 Samples and experiment setup

Spectra of microwave field transmission coefficients through random waveguides were taken using a vector network analyzer. The sample studied is a copper tube filled with randomly positioned 99.95% alumina spheres of diameter 0.95 cm and refractive index 3.14 embedded in Styrofoam spheres of diameter 1.9 cm and refractive index 1.04 to achieve an alumina filling fraction of 0.068, as shown schematically in Fig. 2-2. The tube was 99.999% copper to minimize losses in reflection. The tube had a diameter of 7.2 cm and supports 30 transverse propagating modes at 10 GHz and 34 modes at 10.24GHz. The quasi-1D sample is locally three-dimensional within the reflecting tube with length much greater than its diameter. The random waveguide was thus similar to a microscopically disordered wire in electronics. The wave incident at different parts of the input surface of this quasi-1D sample was thoroughly mixed on the output surface and intensity statistics were nearly uniform across the output except for fluctuations arising from the finite number of transverse modes. Radiation was strongly scattered within the sample over the frequency range of the experiment which was just above the first Mie resonance of the alumina spheres [115]. The output of the network analyzer was amplified before being directed at the sample.

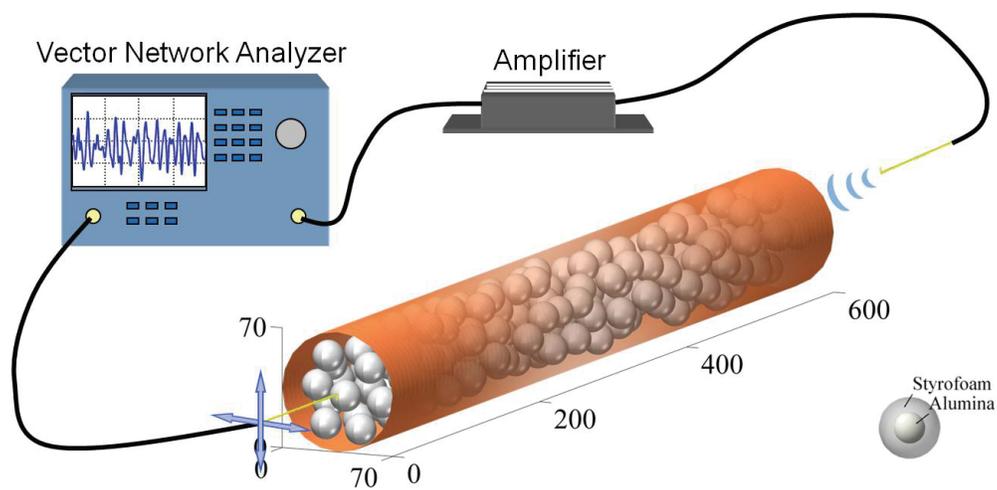

Fig. 2-2 Schematically representation of studied quasi-1D random sample.

Spatial correlation was measured in transmission of a plane wave produced by a horn antenna, placed on the axis of the sample tube and 25 cm in front of the sample, is detected by a 4-mm long wire antenna translated on a 2-mm grid over the output surface of 61-cm-long samples for 200 sample configurations. The antenna was 300 μm in diameter and was connected to the network analyzer through a coaxial cable. The signal detected was the integral over the volume of the antenna of the component of the field directed along the wire axis. The dimensions of the wire were much smaller than the microwave wavelength in air of about 3 cm. In order to measure the field and intensity correlation with shift in polarization, waves are launched and detected using conical horns at the distance of 25 cm from either end of the 50-cm-long sample in more than 4000 configurations. Ensembles of sample realizations were created by rotating the sample tube for several seconds after each measurement.

## 2.3 Result and discussion

### 2.3.1 Steady state correlations of localized waves

We first present the observation of field-field correlation, $<EE^*>$. In order to remove the instrumental response in the measurements of field and intensity, the field is normalized by the square root of the ensemble averaged intensity. The real and imaginary parts of the field correlation function with polarization shift, $F_E(\Delta\theta) = <E(\theta)E^*(\theta+\Delta\theta)>$, and theoretical prediction are displayed by stars and dashed lines, respectively, in Fig. 2-3. Measurements of the real part of field $F_E(\Delta\theta_S)$ and $F_E(\Delta\theta_D)$ are in excellent agreement with $F_E(\Delta\theta)=\cos(\Delta\theta)$ [109,112]. The corresponding imaginary parts are almost zero. This is because

$$F_E\left(\Delta\theta\right) = <E\left(\theta\right)E^*\left(\theta+\Delta\theta\right)> = <E\left(\theta-\Delta\theta\right)E^*\left(\theta\right)> = F_E^*\left(-\Delta\theta\right),$$

and $F_E(\Delta\theta) = F_E(-\Delta\theta)$. So that $F_E\left(-\Delta\theta\right) = F_E^*\left(-\Delta\theta\right)$. Since $F_E$ equals its complex conjugate, $F_E$ must be real. [116]

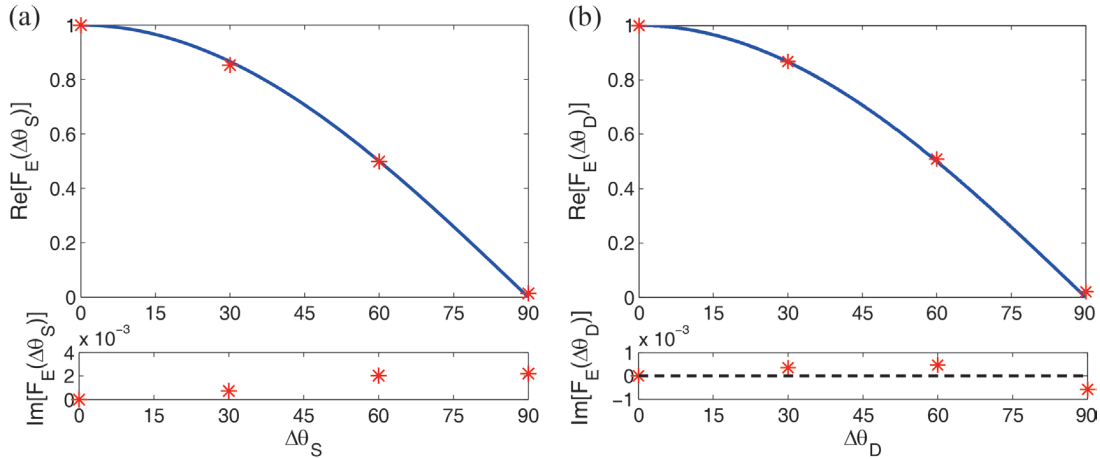

Fig. 2-3 Real and imaginary parts of field correlation function with polarization shift of (a) source ($\Delta\theta_S$) or (b) detector ($\Delta\theta_D$). The curves show the theoretical prediction $F_E(\Delta\theta)=\cos(\Delta\theta)$.

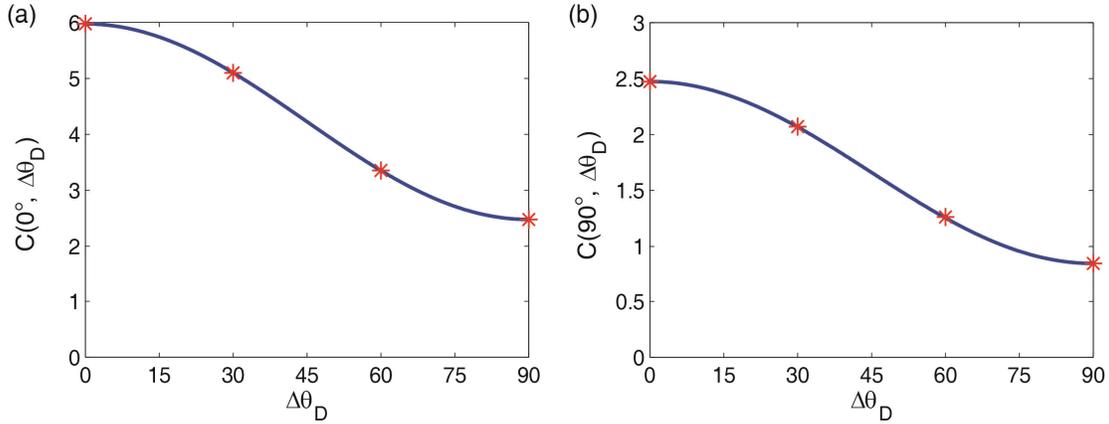

Fig. 2-4 (a) Intensity correlation functions $C$ plotted versus $\Delta\theta_D$ for $\Delta\theta_S = 0°$. The curve is the prediction $C(0, \Delta\theta_D)$. (b) $C$ plotted versus $\Delta\theta_D$ for $\Delta\theta_S = 90°$. The curve is the prediction $C(90°, \Delta\theta_D)$.

The cumulant intensity correlation function is defined as:

$$C(\Delta\theta_S, \Delta\theta_D) = \frac{<\delta I(\theta_S, \theta_D)\delta I(\theta_S + \Delta\theta_S, \theta_D + \Delta\theta_D)>}{<I(\theta_S, \theta_D)><I(\theta_S + \Delta\theta_S, \theta_D + \Delta\theta_D)>}$$
$$= \cos^2 \Delta\theta_S \cos^2 \Delta\theta_D + A_2'(\cos^2 \Delta\theta_S + \cos^2 \Delta\theta_D)$$
$$+ A_3'(1 + \cos^2 \Delta\theta_S \cos^2 \Delta\theta_D + \cos^2 \Delta\theta_S + \cos^2 \Delta\theta_D) \quad . \tag{2-4}$$

where $\delta I$ is the deviation of the intensity from its ensemble average $\delta I = I - <I>$. The measured intensity correlation function for source polarization rotations $\Delta\theta_S = 0$, and $\Delta\theta_S = 90°$, are shown by the crosses together with the fit to the data based on Eq. 2-4 $C(0, \Delta\theta_D) = A_2 + A_3 + (1 + A_2 + A_3)\cos^2(\Delta\theta_D)$ , and $C(90°, \Delta\theta_D) = A_3 + A_2 \cos^2(\Delta\theta_D)$ , respectively, are shown in Fig. 2-4(a) and (b). The solid lines are a fit of the theory to the experiment. Fitting measurements of $C(0, \Delta\theta_D)$ and $C(90°, \Delta\theta_D)$ gives $A'_2 = 0.78 \pm 0.01$ and $A'_3 = 0.85 \pm 0.01$. Since the degree of correlation $\kappa = 2.48$ is greater than the value of 2/3, as shown in Fig. 2-5, the wave is localized. The localization threshold is predicted to occur $\kappa = 2/3$ when the distribution of transmission eigenvalues is bimodal and 1/2 for our sample for which the ratio of the average value of the transmission eigenvalue for adjacent channels is constant.

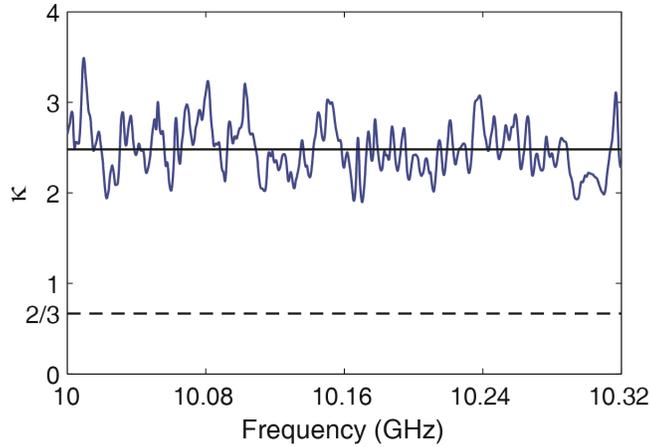

Fig. 2-5 $\kappa$, average of $\kappa$ and localization limit of $\kappa$ plotted with frequency.

Mesoscopic contributions to intensity correlation in this sample exceed those of the field factorization term $C_1$=1. In a longer sample in which the sample is more strongly localized, $A_2$, $A_3$, and $\kappa$ are larger with, $A'_2 = 1.01 \pm 0.05$, $A'_3 = 4.43 \pm 0.03$ with $\kappa = 9.88$ [110]. In the same sample but in a frequency range from 16.95 to 17.05 GHz in which the wave is diffusive, $A'_2 = 0.235$, $A'_3 = 0.029$ and $\kappa$=0.293 [109].

We consider the normalized intensity correlation function, $\Gamma$=1+$C$, for localized waves,

$$
\begin{aligned}
\Gamma(\Delta\theta_S, \Delta\theta_D) &= \frac{< I(\theta_S, \theta_D) I(\theta_S + \Delta\theta_S, \theta_D + \Delta\theta_D) >}{< I(\theta_S, \theta_D) >< I(\theta_S + \Delta\theta_S, \theta_D + \Delta\theta_D) >} \\
&= 1 + \cos^2\Delta\theta_S \cos^2\Delta\theta_D + A'_2(\cos^2\Delta\theta_S + \cos^2\Delta\theta_D) \\
&\quad + A'_3(1 + \cos^2\Delta\theta_S \cos^2\Delta\theta_D + \cos^2\Delta\theta_S + \cos^2\Delta\theta_D) \quad .
\end{aligned}
\tag{2-5}
$$

Thus, for $\Delta\theta_S = 0$, $\Gamma = (1 + A'_2 + 2A'_3)(1 + \cos^2\Delta\theta_D)$. This is confirmed in Fig. 2-6. In the strong localization limit, the correlator $\Gamma$ is predicted to be factorized as follows: $\Gamma = (1 + \cos^2\Delta\theta_S) < s^2 > (1 + \cos^2\Delta\theta_D)$ [110]. This relation is consistent with Eq. 2-5 only if $A'_2$ is equal to 1 in the strongly localized media. The correlator can then be written as, $\Gamma = (1 + \cos^2\Delta\theta_S)(1 + A'_3)(1 + \cos^2\Delta\theta_D)$. We therefore expect $A'_2$ to saturate at 1, while $\kappa$ and $A_3$ increase as the wave becomes more strongly localized within the sample.

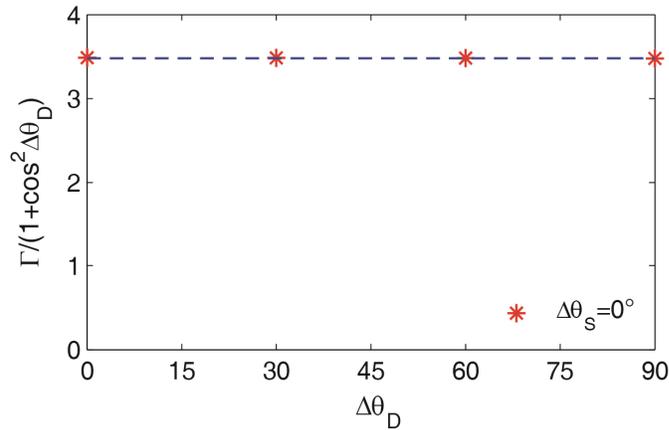

Fig. 2-6 Intensity correlation function with polarization angular shift.

Figure 2-7 shows our results for measurements of the cumulant intensity correlation function with polarization rotation of both source and detector. The experimental results (red crosses) are well fit by Eq. 2-4 (colored grid). This demonstrates that the cumulant intensity correlation function with polarization rotation can be expressed as the sum of multiplicative, additive, and constant terms with respect to the square of the field correlation function with shifts of source and detector polarizations for localized waves.

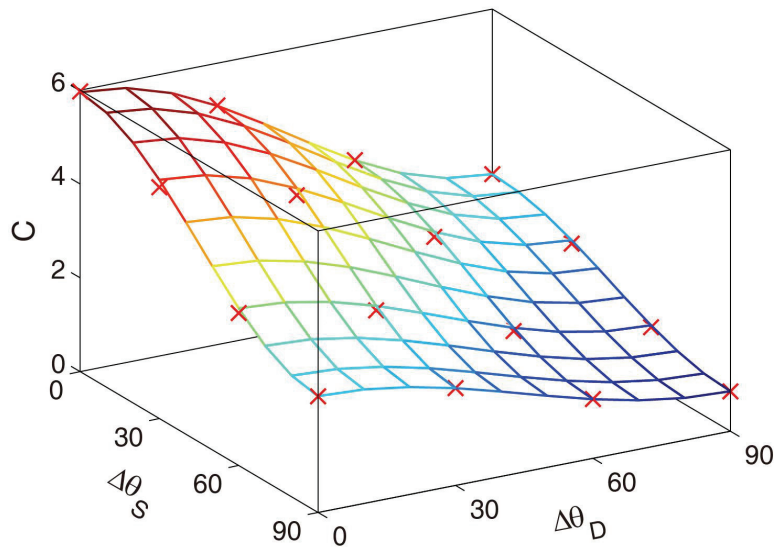

Fig. 2-7 Cumulant intensity correlation function with polarization rotation of both source and detector. Red crosses are the experiments results, colored grid lines is the theoretical fit.

## 2.3.2 Dynamics of fluctuations and correlations of localized waves

In contrast to measurements of the scaling of transmission, which can track the changing impact of weak localization on samples of different size, pulsed measurements may provide the changing contributions of underlying electromagnetic modes with different decay rates in samples of a particular size. A suppression of the decay rate of transmission has been observed in microwave experiments in diffusive quasi-1D samples [117]. A slowing down of the decay rate of pulsed transmission was also observed in optical measurements in a titania powder slab [118]. Measurements of coherent backscattering scattering show that $k\ell$ is larger than unity which indicates the wave is diffusive in the sample [118]. The ensemble average of pulsed transmission was also measured for localized ultrasound just beyond the mobility edge in a slab of sintered aluminum beads [39]. The ultrasound measurements of $<I(t)>$, were well fit by SCLT [54,56,118] with a renormalized diffusion coefficient in space and frequency.

In strongly localized quasi-1D samples, however, the decay rate of microwave transmission fell below predictions for SCLT [58]. The slowing down of transport at long times was explained in terms of a 1D dynamic single parameter scaling (DSPS) model which neglects mode overlap and averages over the distribution of decay rates and associated transmission strengths of localized modes [53,58]. The relative contributions of long-lived localized modes and short-lived "necklace states" is central to understanding dynamics [41,88,90,93,119].

The dynamics of localized waves can be found by Fourier transforming field spectra. The response to a Gaussian incident field pulse, $E_0(t) \sim \exp(-t^2/2\sigma_t^2)\exp(i2\pi\nu_0 t)$ is obtained by taking the Fourier transform of the product of the spectra of the transmitted field and the incident pulse, $E_0(\nu) \sim \exp(-(\nu - \nu_0)^2 / 2\sigma^2)$, where $\sigma = (2\pi\sigma_t^2)^{-1}$. This gives the time-dependent field $E(t)$ and the intensity $|E(t)|^2$.

In the time domain, the cumulant correlation function displays the same dependence upon the field correlation function with the polarization rotation of detector as in the frequency domain, $C_\sigma(t) = F + \kappa_\sigma(t)(1+F)$ [84] with a parameter $\kappa_\sigma(t)$, expressing the degree of correlation which varies with the delay time $t$ following excitation by a Gaussian pulse of spectral width $\sigma$. If the polarization of the source or detector is rotated by 90°, we can obtain the degree of intensity correlation in the time domain directly, $C_\sigma(90^\circ, t) = \kappa_\sigma(t)$. The degree of intensity correlation increases with delay time from a narrow exciting pulse since the density of a given trajectory increases in time since its length increase in time. This enhances the probability that a path may intersect itself. Thus the degree of correlation, $\kappa$, increases continuously. Since the cumulant correlation function displays the same dependence upon the field correlation function with polarization rotation as in the frequency domain, we could express the cumulant correlation functions in the time domain as,

$$\begin{aligned}
C(\Delta\theta_S, \Delta\theta_D; t) = {} & F_{in}(\Delta\theta_S)F_{out}(\Delta\theta_D) + A'_2(t)(F_{in}(\Delta\theta_S) + F_{out}(\Delta\theta_D)) \\
& + A'_3(t)(1 + F_{in}(\Delta\theta_S)F_{out}(\Delta\theta_D) + F_{in}(\Delta\theta_S) + F_{out}(\Delta\theta_D))
\end{aligned} \tag{2-6}$$

Due to reduced signal to noise ratio with increasing delay time, the effect of noise cannot be neglected at long times. Since the signal intensity $S(t)$ and the time independent noise $n(t)$ are uncorrelated, $S(t)$ can be approximately obtained by subtracting $n(t)$ from the square of the Fourier transform of each field spectrum $I(t)$, $S(t)=I(t)-n(t)$. The intensity correlation between two signal $S_1$ and $S_2$, $<S_1S_2> = <(I_1-n_1)(I_2-n_2)> = <I_1I_2> - <I_1><n_2> - <I_2><n_1> + <n_1><n_2>$. By assuming $<n_1> = <n_2> = <n>$, the intensity correlation of two signals can be expressed as the correlation of two intensities with a constant background noise subtracted. The background noise is obtained from the measured intensity at extremely long times at which the true signal has nearly vanished. Thus $<S_1 S_2> = <(I_1 - <n>)(I_2 - <n>)>$. By subtracting a constant noise background, we are able to enhance the dynamic range and to observe the behavior of the coefficients, $A'_2(t)$ and $A'_3(t)$ at relative long times. Figure 2-8 shows that for an incident Gaussian pulse with a narrow frequency linewidth $\sigma = 5$ MHz, $A'_3(t)$ and $\kappa$ increase with delay time $t$ and $A'_2(t)$ increases slowly with time and seems to saturate at unity within the noise of the measurements.

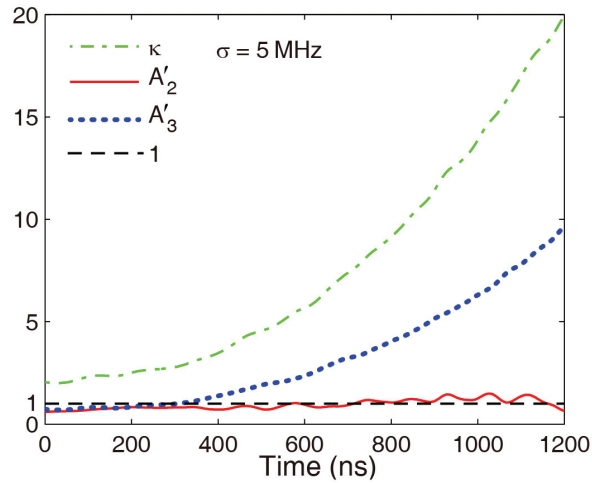

Fig. 2-8 $A'_2(t)$, $A'_3(t)$ and $\kappa$ change with time $t$ after reducing noise.

Cumulant correlation functions with displacement on the output surface, $\Delta r,$ of relative intensity $\hat{I}_a(r,t) = I_a(r,t)/<I_a(r,t)>$, $C_\sigma(\Delta r,t) = <\delta\hat{I}_a(r,t)\delta\hat{I}_a(r+\Delta r,t)>$, for a single incident transverse mode $a$, for different time delays $t$, in quasi-1D samples with $L = 61$ cm are shown in Fig. 2-9. Here $\delta\hat{I}_a(r,t) = \hat{I}_a(r,t) - 1$ is the deviation of relative intensity from the ensemble average value. In the diffusive limit, $\kappa_\sigma \rightarrow 0$ and $C_\sigma \rightarrow F$. Intensity is then correlated over the short range of a speckle spot. The field correlation function with displacement $F_E(\Delta r)$ is the Fourier transform of the normalized specific intensity, which is the angular distribution of transmitted intensity [114,116] and does not change in time since the direction of scattered radiation is independent of time. This is demonstrated in the inset in Fig. 2-9. The cumulant intensity correlation function at different delay times $C_\sigma(\Delta r,t)$ is shown in Fig. 2-9. $C_\sigma(\Delta r,t)$ displays the same dependence upon $F$ but the degree of correlation $\kappa_\sigma(t)$ is time dependent as can be seen in the right vertical axis in Fig. 2-9, which gives the value of $C_\sigma(\Delta r_0,t)$ for displacements $\Delta r_0$ at the first zero of $F$. $C_\sigma(\Delta r,t)$ and $\kappa_\sigma(t)$, as well as $\mathrm{var}(\hat{I}_a(r,t))$ which equals $C_\sigma(\Delta r$=0,$t)$ as seen in Fig. 2-9, drop at early times before increasing.

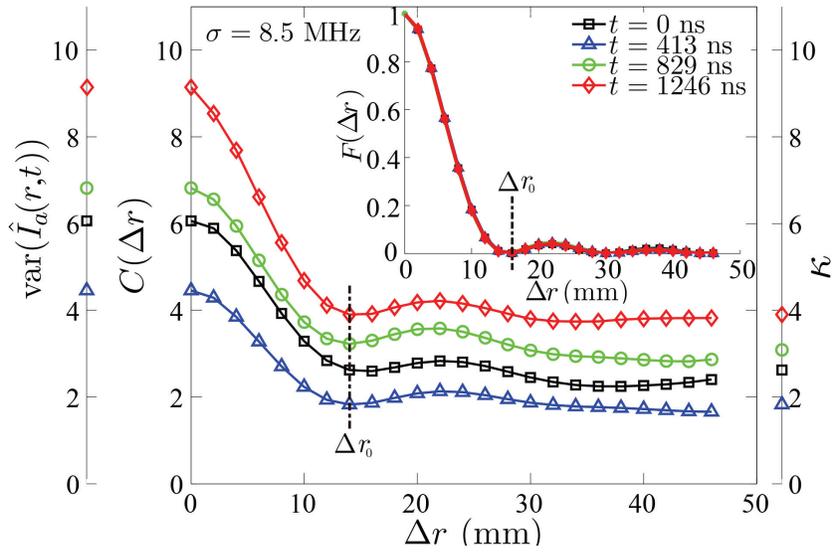

Fig. 2-9 Spatial correlation, for different delay times $t$ from the peak of an incident Gaussian pulse in a sample of $L$=61 cm ~$2\overline{\overline{\xi}}$. $C(\Delta r)$ falls at early times and then rises, while the square of corresponding field correlation functions, $F(\Delta r)$, shown in the inset, are independent of time. The dashed vertical lines at $\Delta r_0$=14 mm indicate the first zero of $F$ at which point, $C_\sigma = \kappa$. $C_\sigma(0) = \mathrm{var}(\hat{I}_a(t))$ and $C_\sigma(\Delta r_0) \equiv \kappa$ are shown on the vertical lines on the right and left sides of the figure, respectively.

It was not possible to study early time dynamics in quasi-1D samples with narrow bandwidth pulses since this broadens the time response. The response of narrow pulses could, however, be found in 1D simulations in samples in which the statistics change little over a wide frequency range [58]. Configurations of random samples are constructed using the model developed in Ref. [58]. Samples of $N=L/a$ layers of equal thickness $a$, are embedded in a vacuum with $\varepsilon = 1$ and wave speed $c$. The dielectric constant in each layer, $\varepsilon_i$, is a uniformly distributed random number between 0.3 and 1.7. The intensity localization length is $\overline{\overline{\xi}} = -L/ < \ln T > = 22a$ and the central frequency is $v_0 = 0.26c/a$ [58].

The temporal variation of relative fluctuations of transmission in 1D following a

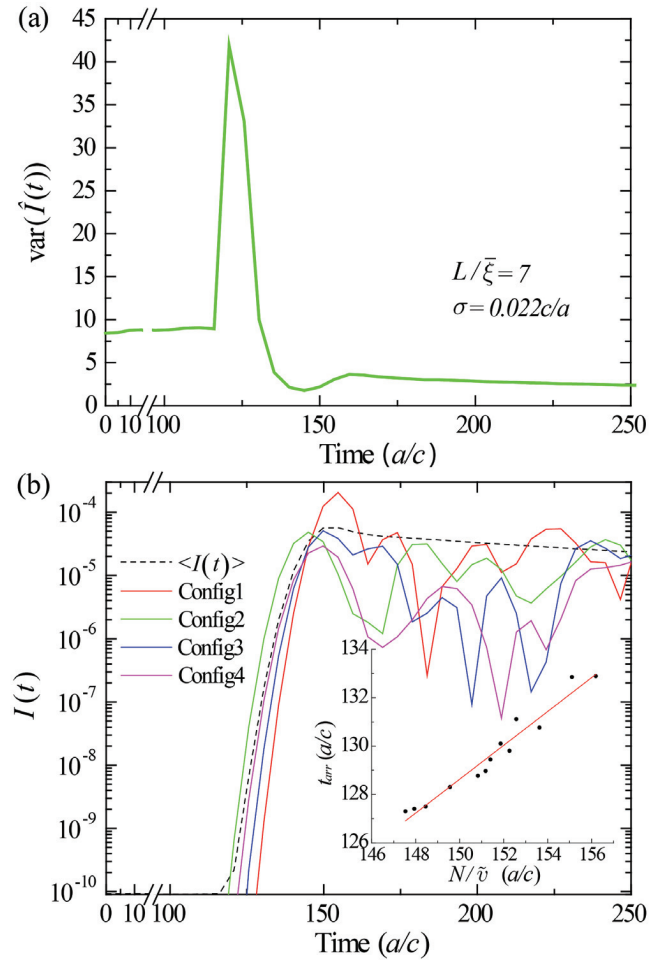

Fig. 2-10 (a) Variance of relative intensity vs. $t$ in 1D. (b) Initial jump in $\mathrm{var}(\hat{I}(t))$ is associated with the different times of arrival $t_{\mathrm{arr}}$ of the rising edge of $I(t)$ in different random configurations. The dashed curve is the configuration average of $I(t)$. The inset shows the arrival time $t_{\mathrm{arr}}$ in different configurations closely matches the calculated ballistic time through the sample with $N$ elements of thickness $a$ and average phase velocity $\tilde{v}$.

Gaussian incident pulse, $\mathrm{var}(\hat{I}(t))$, shown in Fig. 2-10(a), is calculated for 50,000 configurations using spectra of the transmitted field just beyond the output surface. The sharp spike in $\mathrm{var}(\hat{I}(t))$, shown in Fig. 2-10(a), is due to differences in the transit time of the leading edge of the ballistic pulse in different random configurations (Fig. 2-10(b)) which leads to large fluctuations in $\mathrm{var}(\hat{I}(t))$. The time at which the leading edge of the pulse arrives at the sample output, $t_{\mathrm{arr}}$, determined by the time at which $I(t)$ reaches the value $10^{-8}$, is seen to closely track the calculated ballistic arrival time, $\sum_{i}^{N} a\sqrt{\varepsilon_i}\,/\,c \equiv (N\,/\,\tilde{v})(a\,/\,c)$. Once the leading edge of the pulse is transmitted, $I(t)$, remains high for a time equal to the inverse bandwidth of the exciting pulse and relative fluctuations drop to a minimum. We expect that the variance of fluctuations is inversely proportional to the effective number of modes contributing significantly to transmission. At first, only the shortest-lived modes, from which energy leaks rapidly, contribute appreciably to transmission. Because the number of modes contributing substantially is small correlation is relatively high. This leads to a peak in $\mathrm{var}(\hat{I}(t))$ at $\sim 160a/c$. After this time, the contribution of longer-lived modes begins to be felt since energy in the shortest-lived modes has already leaked from the sample. As a result, the effective number of modes contributing to transmission increases and $\mathrm{var}(\hat{I}(t))$ falls.

The dynamics of fluctuations over a broader time scale for measurements in quasi-1D and simulations in 1D are shown in Figs. 2-11, 12 and 13. Measurement in quasi-1D and simulation in 1D show minima in $\mathrm{var}(\hat{I}_a(r,t))$, (Fig. 2-11(a) and 2-12(a)), and $\mathrm{var}(\hat{I}(t))$, (Fig. 2-11(b) and 2-12(b)) at a time $t_m$, which we find is independent of pulse bandwidth as seen in Fig.2-11. This indicates that $t_m$ reflects a property of modes of the medium. The minimum occurs between the times ranges in which either short- or long-lived modes dominate transmission and $t_m$ corresponds to the time within the intermediate time range at which transmission involves the largest number of modes.

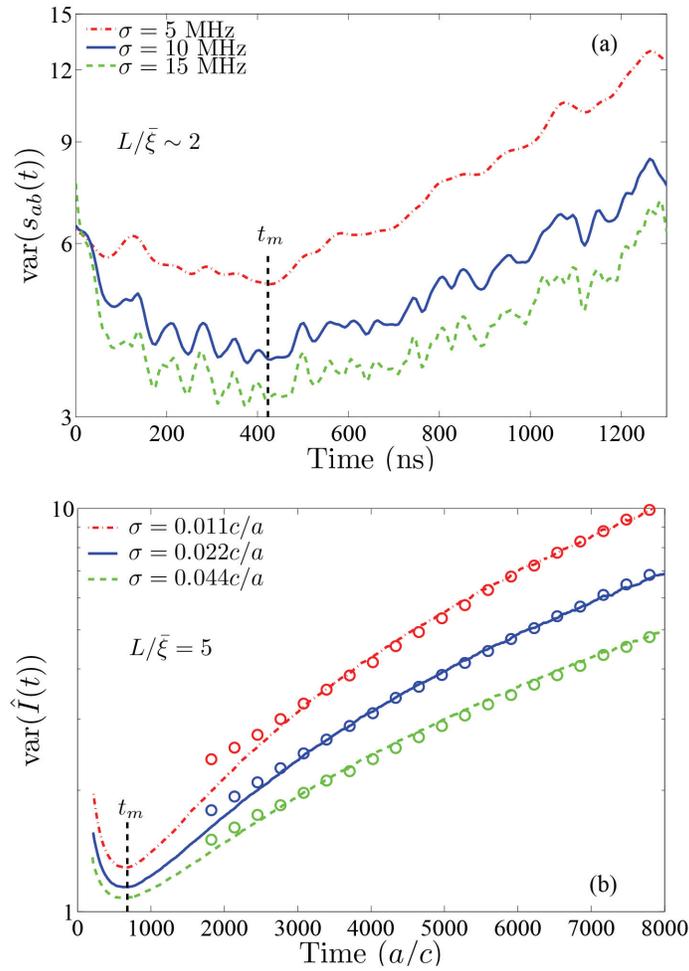

Fig. 2-11 Dynamic variance of relative transmission for three pulse widths for a fixed sample length obtained from (a) measurements in a quasi-1D sample of $L$=61 cm $\sim 2\bar{\bar{\xi}}$, and (b) simulations in 1D sample with $L = 5\bar{\bar{\xi}}$, (curves) and the results of the DSPS model (open circles).

Since the rapid increase in $\text{var}(\hat{I}_a(r,t))$ after $t_m$ arises from the dominance of long-lived modes, its behavior was modeled by the DSPS model proposed in Ref. [58] based on the contribution to transmission of a distribution of modes. We imagine that long-lived modes are not hybridized with resonances in the sample since this would broaden the modes in space and so give increases coupling to the sample boundaries in which case the lifetimes of the modes would decrease. We assume that narrow linewidth bands band modes Following Azbel's treatment of spectrally isolated localized modes, localized modes peaked in space at depth $z$ into the sample have amplitude $A(\gamma, z) = \exp(-\gamma |L - 2z|)$ at the output surface. The Lyapunov exponent, $\gamma = 1/2\xi$, is drawn from a Gaussian distribution, $P(\gamma) = \sqrt{L / 2\pi\overline{\gamma}} \exp[-(\gamma - \overline{\gamma})^2 / (2\overline{\gamma} / L)]$ [53] with a lower cutoff for $\gamma$ limited by the sample length, i.e., $\gamma \geq B/L$, where $B$ is the only adjustable parameter in the model. The time response to a Gaussian incident pulse at long times can be written as,

$$I_{\text{DSPS}}(t) = \left| \sum_{i=1}^{M} E(v_i) A(\gamma_i, z_i) \frac{\Gamma(\gamma_i, z_i)}{2} \exp[-\frac{\Gamma(\gamma_i, z_i)}{2} t - i2\pi v_i t] \right|^2, \qquad (2\text{-}7)$$

where $\Gamma(\gamma_i, z_i)$ is the decay rate of a localized state located at $z_i$ with a localization length and $v_i$ is the frequency of the localized state. The explicit expression for $\Gamma(\gamma_i, z_i)$ is given in Eq. (3) of Ref. [58]. In Monte Carlo simulations of the DSPS model for $I_{\text{DSPS}}(t)$, results are averaged over a window of width $\Delta v = 0.207 c/a$. The number of states $M$ excited inside the window is assumed to follow the Poisson distribution expected when modes are statistically independent, with a mean $\bar{M}$ equals to $\Delta v \rho L$, where the density of states per unit length at $v_0$ is $\rho = 1.32 c^{-1}$, and the frequencies $v_i$ ($i$=1,2, . . . ,$M$) are chosen randomly inside the window. The results of the DSPS model for $t > 2.5 t_m$ are shown as circles in Fig. 2-11(b) and 2-12(b). Excellent agreement between the DSPS model and 1D simulations is found for $t > 3 t_m$. Variances of intensity are seen in Fig. 2-11 and 2-12 to be larger for longer samples at early times as is found in steady state. But at later times, the variance is larger in shorter samples. In this case, the density of trajectories at a given time is larger than in longer samples and the probability of trajectories crossing is greater.

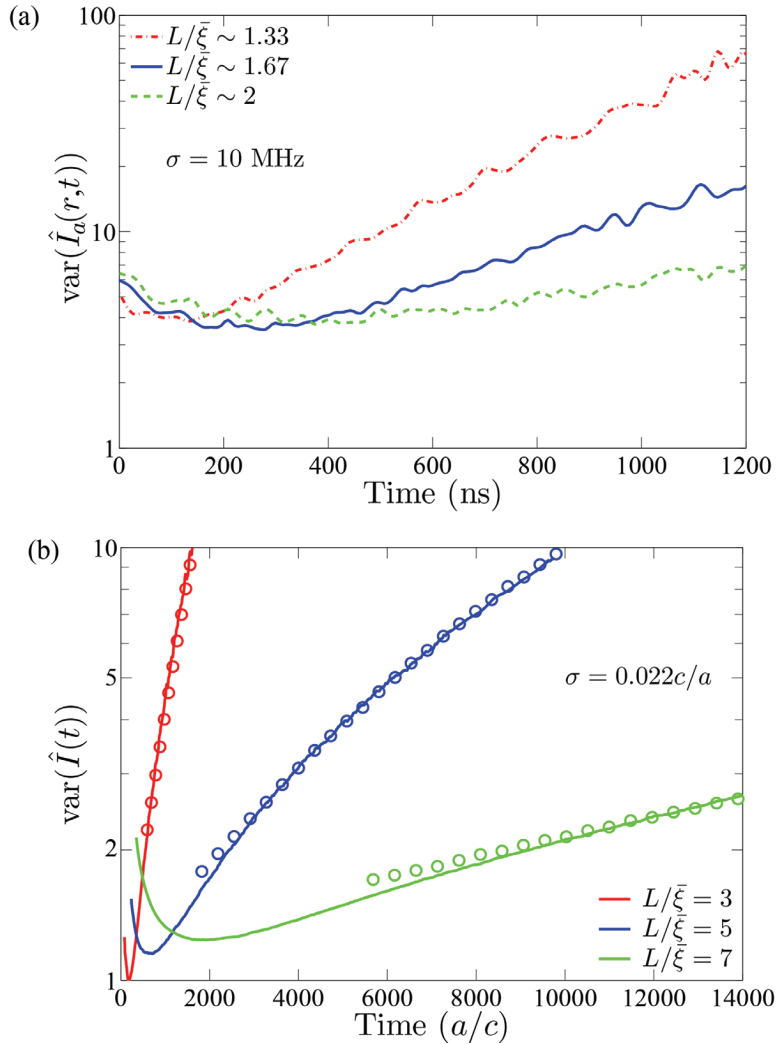

Fig. 2-12 Dynamic variance of normalized transmission for three sample lengths and a fixed pulse width obtained from (a) measurements in quasi-1D for three sample lengths, $L = 40$ cm $\sim 1.33\overline{\xi}$, $L = 50$ cm $\sim 1.67\overline{\xi}$, and $L$=61 cm $\sim 2\overline{\xi}$, and (b) simulations in 1D (curves) and the results of the DSPS model for $t > 2.5t_m$ (open circles).

Though $<I>$ may be compared to either $<I_a(r)>$, $< I_a \equiv \sum_r I_a(r) >$ or $< T \equiv \sum_a T_a >$ since these are the same in 1D, the strength of second-order statistics in quasi-1D depends upon the extent of spatial averaging over the speckle pattern. Because there is no transverse variation of intensity in 1D, the most apt comparison of $\text{var}(\hat{I})$ to second order transmission statistics in quasi-1D is to $\text{var}(s=T/<T>)$, which is equal to the infinite-range correlator, $\kappa_\infty \equiv <\delta \hat{I}_a(r)\delta \hat{I}_{a'}(r')>$ [23,24,81,82,85,109,111], where $a \neq a'$, and $r$ and $r'$ are two position on the output surface at which field correlation vanishes. Short-range correlation of the speckle pattern does not contribute significantly to either $\text{var}(\hat{I})$ or $\kappa_\infty$. On the one hand, $\kappa_\infty$ is independent of the choice of input and output transverse mode or position in quasi-1D, while on the other, there is no transverse intensity variation in 1D. A DSPS calculation of $\text{var}(\hat{I})$ and measurements of $\kappa_\infty$ in quasi-1D for a sample with $L/\overline{\xi} = 1.67$ excited by a pulse of width $\sigma = 5$ MHz are compared in Fig. 2-13. Good agreement is obtained for $t > 700$ ns $\sim 3t_m$. The density of states per unit length in this system is $\rho = 8.67$ ns/cm.

These results demonstrate that measurements in quasi-1D approach Monte Carlo simulations of the 1D DSPS theory in corresponding samples at long times. This convergence of the infinite-range correlator of the quasi-1D medium and the variance of normalized transmitted intensity in the simulation of the 1D DSPS model even in samples for which $L/\overline{\overline{\xi}}$ is not much larger than unity reflects the similar probability distributions for $\hat{I}$ and $s$ for long times. In strongly localized systems, the first transmission eigenvalue for quasi-1D samples is much larger than other eigenvalues, the effective number of transmission channels contributing to the transmission approaches unity $N_{\text{eff}} \rightarrow 1$. Transmission in this eigenchannel has a log-normal distribution as does the distribution for transmission for localized waves in 1D. Here the transmitted intensity in 1D $\hat{I}$ is equivalent to a system with a single transmission channel, while at long times the wave in the quasi-1D system is strongly localized with $N_{\text{eff}} \rightarrow 1$. Thus the relative transmitted intensity $\hat{I}$ through 1D systems and the normalized transmittance $s$ of a strongly localized quasi-1D system will have same statistics. As a result, the variances of these quantities will converge. This explains the convergence of $\text{var}(\hat{I})$ and $\kappa_{\infty}$.

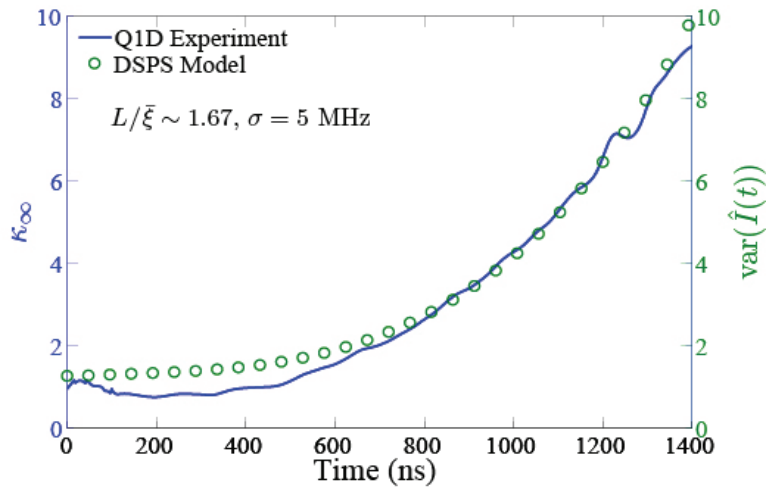

Fig. 2-13 Comparison of measurements of the dynamics of infinite-range correlation in a Q1D sample of $L$=50 cm $\sim 1.67\overline{\xi}$ and calculations of the variance of intensity in the DSPS model for a corresponding 1D sample.

## 2.4 Conclusion

In conclusion, we have measured the statistics of steady state and pulsed transmission of localized waves with polarization rotation of both source and detector and with displacement of detector. The non-monotonic behavior of the dynamics of correlation of localized waves provides a window on the evolving contributions of short- and long-lived electromagnetic modes of the random medium. For $t < t_m$ but somewhat greater than $t_{\text{arr}}$, the transmitted energy is due to modes which release their energy quickly, while for $t > t_m$ a decreasing subset of long-lived modes contribute substantially to transmission leading to an increase in enhanced mesoscopic fluctuations. At $t_m$, both short- and long-lived states contribute to transmission. The relatively large number of modes contributing to transmission leads to reduced fluctuations. These results show that complex mesoscopic transport phenomena for localized waves can be clarified by analyzing the results in terms of modes and transmission channels.

In this chapter, we have qualitatively explained the dynamics of fluctuations and correlations of localized wave in terms of modes. In the next chapter, we will show that strong correlation exists between the speckle patterns of neighboring mode and that this is a key issue in the statistics of random media.

# CHAPTER 3

# TRANSPORT THROUGH MODES IN RANDOM MEDIA

## 3.1 Introduction

Excitations in complex media may be expressed as superpositions of eigenstates that are referred to as 'levels' for quantum systems and 'modes' for classical waves. This is akin to decomposing a musical composition into the tones of the wind, string or percussion instruments each of which is associated with distinct vibrational patterns. The field inside an open random media can be represented as superposition of modes [41,92,120,121].

### 3.1.1 Defect modes in 1D system

The intensity distribution of modes within the interior of a three-dimensional, multiply scattering sample is generally not accessible, but a qualitative understanding of the characteristics of modes in localized random systems can be obtained by considering a related case of few defects embedded in a 1D periodic structure. Simulations of intensity within a periodic structure of binary elements with one or two defects are shown in Fig. 3-1. The sample is composed of periodically distributed 100-nm thick binary elements with refractive indices of 1 and 1.55 with a defect with thickness of one period of 200 nm and refractive index 1. The spatial intensity distributions at the frequency of the transmission peaks for two samples with a single defect placed at the same distance from the input or output end as well as the corresponding sample structure are plotted in Fig. 3-1(a). For a periodic structure with a single defect, the intensity is exponentially peaked at the defect position. The exponential decay length equals to the decay length of the evanescent wave at same frequency in the periodic structure without the defect. In the two structures shown in Fig. 3-1(a), the single defects are placed an equal distance from the input and output end. The transmission spectra for these two samples are equal as can be seen in Fig. 3-1(b). Apart from a modulation on the scale of $\lambda/2$, the intensity first increases exponentially to the defect position, and then falls to the output end with same decay length when the defect is near the input end. When the defect is close to the output end, the intensity first falls exponentially from the incident surface to the position $2d$, where $d$ is the distance of the defect from the middle of the sample, and then increases up to the defect position and then falls again. The transmitted intensity is the same for the two cases, so that the transmission only depends on the distance from the center of the sample $d$ and the exponential decay length $\xi$, with $T=\exp(-2d/\xi)$. The decay length is the same as that for the evanescent wave excited at the same frequency within the band gap in a defect-free structure. When the defect is at the center of the 1D sample, $d=0$, the transmission coefficient is unity. The energy within the sample is then at a maximum relative to samples with the defect positioned away from the center and the leakage rate of energy is at a minimum corresponding to the narrowest spectral linewidth.

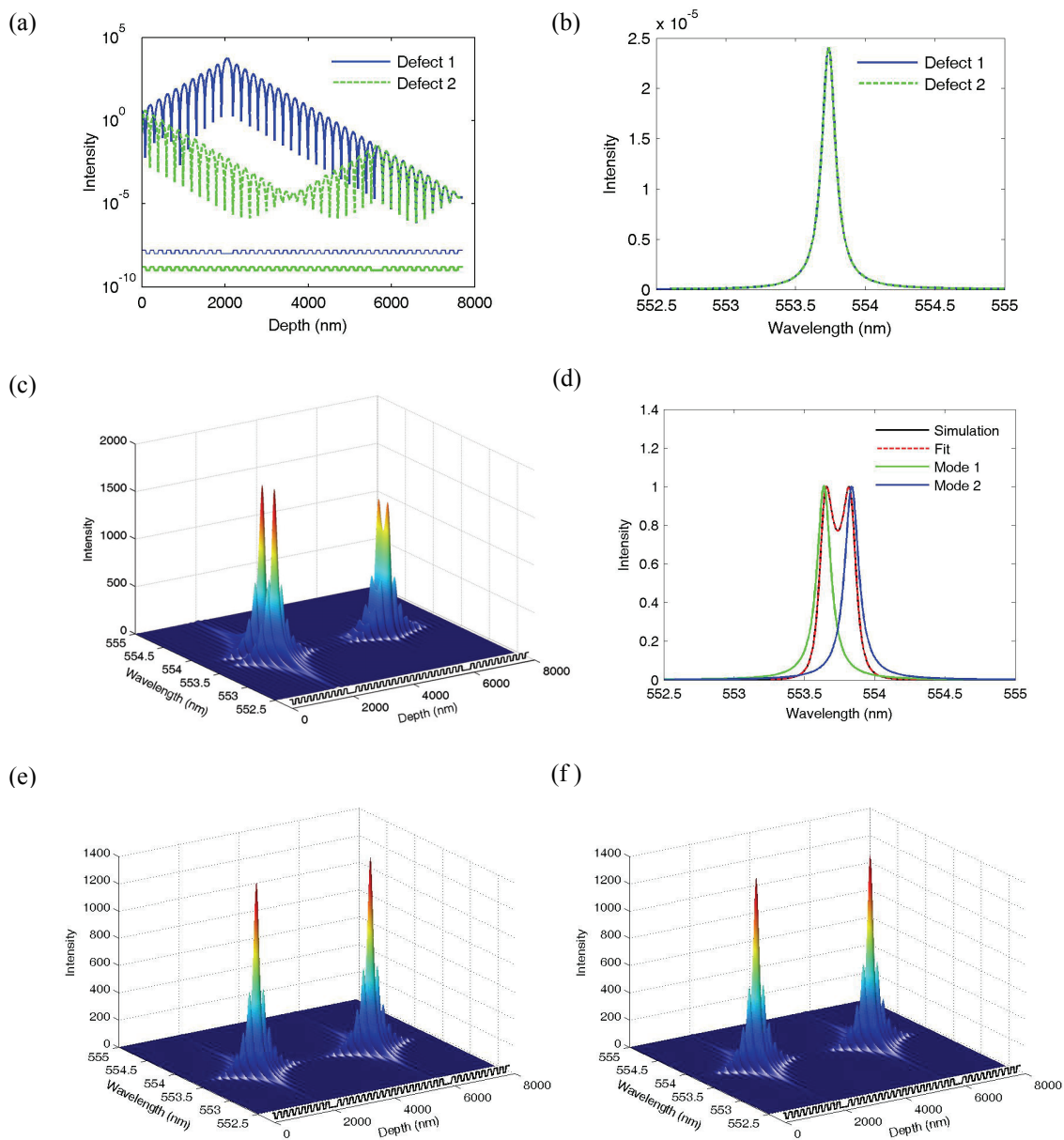

Fig. 3-1 Simulation of wave propagation in 1D sample with defects in a periodic binary structure. The refractive index alternates between 1 and 1.55 in segments of length 100 nm and period 200 nm. (a) Intensity distribution inside and (b) spectra of transmitted intensity for samples with a single defects displaced equally from the input or output ends of the samples. (c) Intensity spectra at different positions inside the sample with two defects symmetrically placed about the center. (d) Spectra of transmission. Black solid line and red dashed line represent the simulation and the fit to the sum of two modes, respectively. The green and blue lines show the transmitted intensity for the two underlying modes individually. (e) and (f) The frequency and spatial profile of intensity distribution inside the sample for each of the two defect modes.

When two defects are placed with equal distance from the center of the otherwise periodic sample, as shown in Fig. 3-1(a), the intensity is again peaked at the second defect after falling exponentially from the first defect (Fig. 3-1(c)) and the transmission spectrum can be decomposed to reveal two modes (Fig. 3-1(d)). The peak of transmission is unity, which is significantly higher than the transmission through the system with a single defect, which is shown in Figs. 3-1(a) and (b). The spatial distribution of intensity in each of the two underlying modes exhibit two identical peaks at the locations of the two defects (Fig. 3-1(e) and (f)). The two modes individually have the same transmission coefficient of unity. The difference in peak intensity and frequency shifts at defects 1 and 2, shown in Fig. 3-1(c), is due to interference between the two hybridized modes. Due to the coupling of resonances at the defect sites, these modes which overlap in frequency and space repel each other and their central frequencies are shifted in opposite directions. In random media, localized modes well separated from others in space and in frequency are exponentially peaked as are the defect modes in a periodic structure. On the other hand, overlapping resonances form a multi-peaked intensity distribution and dominate transmission.

### 3.1.2 Modes of an open medium

In nondissipative open media, modes which are spectrally isolated and exponentially peaked near the center of the open media are long-lived with narrow linewidths. In absorbing samples, the decay rates of such localized modes may be dominated by absorption even for low levels of absorption in a relative long sample since the escape time from the sample is long. In contrast, the leakage rate for energy in multi-peaked and spectrally overlapped modes is high and the mode linewidth and energy are not appreciably changed by the presence of moderate absorption. As a result, the contribution to transmission of long lived modes is more strongly suppressed by absorption than for short-lived modes [122]

The Fourier transform of the mode field spectrum $E_n(\mathbf{r}, \omega) = a_n(\mathbf{r}) \dfrac{\Gamma_n / 2}{\omega - (\omega_n - i\Gamma_n / 2)}$ is a single sided exponential decay with decay rate $\Gamma_n/2$. The linewidth of the $n^{\text{th}}$ mode, $\Gamma_n$, is the sum of the modal energy leakage rate $\Gamma_n^0$ and the absorption rate $1/\tau_a$. The effect of absorption can be effectively eliminated by substituting the linewidth with $\Gamma_n^0 = \Gamma_n - 1/\tau_a$ for $\Gamma_n$ and the complex mode amplitude $a_n^0 = a_n \Gamma_n / \Gamma_n^0$ for $a_n$. An example of the impact of absorption is shown in Fig. 3-2. One-dimensional simulations give spectra of intensity and phase corrected for the impact of absorption which are the same as spectra for nonabsorbing samples as shown in Fig. 3-2(a) and (b).

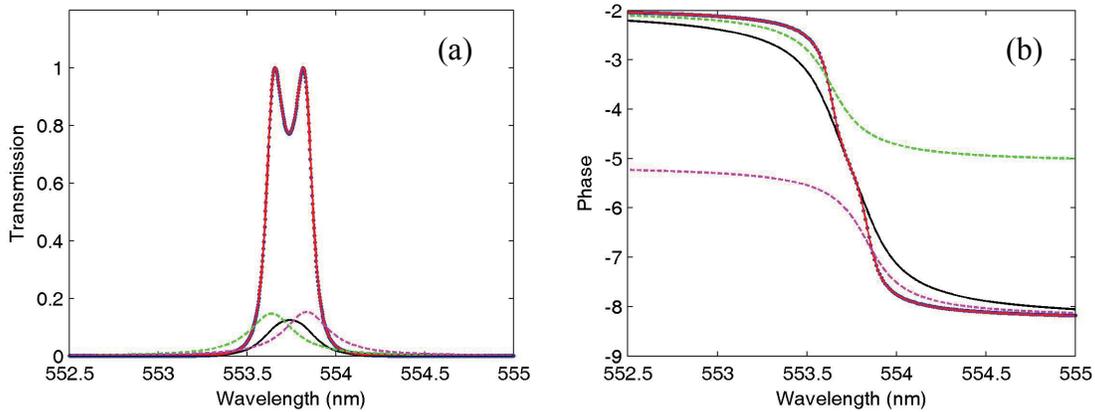

Fig. 3-2 Transmitted field spectra before and after the impact of absorption is removed by substituting $\Gamma_n^0$ and $a_n^0$ for $\Gamma_n$, and $a_n$. The black solid line and blue dotted lines show, respectively, the intensity spectra (a), and phase (b) of transmitted field, through a 1D periodic structure with two defects, with or without absorption. The corrected spectra are shown by red line. The corrected intensity and phase spectra are same as for spectra of samples without absorption. The underlying modes are shown by green and magenta dashed lines.

The affect of absorption can also be approximately removed by compensating the absorption by multiplying the time response of an incident Gaussian pulse by $\exp(t/2\tau_a)$. This process for compensating absorption is shown in Fig. 3-3. Here the sample is the same as the sample with two symmetric defects shown in Fig. 3-1(a) but with absorption. The spectrum of the incident Gaussian pulse, $g(t,\omega_0) \sim \exp(-t^2/2\sigma_t^2)\cos(\omega_0 t)$ is also Gaussian in the frequency domain, $G(\omega,\omega_0) \sim \exp(-(\omega-\omega_0)^2/2\sigma^2)$, where $\omega_0$ is the central frequency and the linewidth is $\sigma = \sigma_t^{-1}$. The time response to the incident pulse is obtained by Fourier transforming the product of the incident Gaussian pulse and the field spectrum into time domain. We then multiply the time response by $\exp(t/2\tau_a)$ for $t > 0$ to compensate for losses due to absorption. The product is then Fourier transformed back to the frequency domain and divide by the spectrum of the incident Gaussian pulse to give the corrected transmission spectrum within the incident Gaussian pulse width. This approximately removes absorption in the transmitted intensity but does not treat the phase. However, this method provides a way of eliminating the effect of absorption upon the intensity without knowledge of the underlying modes. This is especially important for diffusive waves for which mode overlap is strong and a modal analysis is not possible.

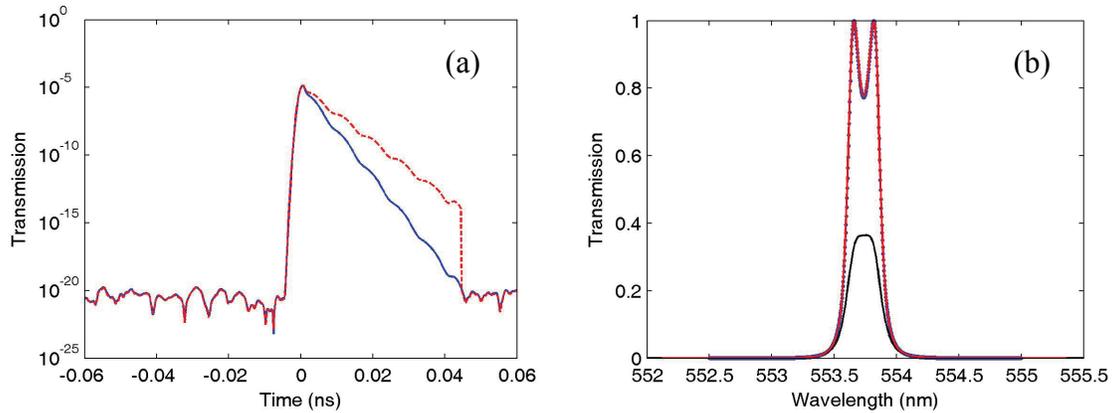

Fig. 3-3 The affect of absorption is compensated for by multiplying the time response of the field to an incident Gaussian pulse by $\exp(t/2\tau_a)$. (a) The time response of an incident Gaussian pulse is shown by the blue line. The response multiplied $\exp(t/2\tau_a)$ is shown by the red dashed line. (b) The black curve shows the spectrum for the absorbing sample, while the red curve shows the spectrum corrected for the impact of absorption. This curve overlaps the transmission spectrum for the sample without absorption shown by the blue dots.

### 3.1.3 Outline

In the remainder of this chapter, we present measurements of the spectral and spatial variation of the microwave field transported through random scatters contained in a copper tube with diameter much less than its length. We show that field speckle patterns of transmitted radiation can be decomposed into a sum of the patterns of the individual modes of the medium and that the central frequency and linewidth of each mode can be found. We develop an algorithm based on the separable nonlinear least square method in order to fit the field spectra measured at different positions on the sample's output surface using a common set of mode central frequencies and linewidths, for each realization of the random sample. We find strong correlation between modal field speckle patterns, which leads to destructive interference between modes. This allows us to explain complexities of steady state and pulsed transmission of localized waves and to harmonize wave and particle descriptions of diffusion.

Section 2 of this chapter provides details of the experimental setup. Observations of the similarities between transmitted field speckle patterns found at peaks of the transmission spectrum are also presented for incident plane waves with different polarizations and for point sources at positions on the input surface.

In section 3, we describe the fitting algorithm in detail. The method of modal decomposition of field spectra is based on the separable nonlinear least square method in which the data are modelled as a linear combination of nonlinear functions.

In section 4, we discuss the modal decomposition of field speckle patterns for microwave radiation transmitted through quasi-1D localized samples. We show the different spectra can be well fit by a superposition of modes. The intensity patterns of highly overlapped modes can be nearly identical and the phase difference between points in the patterns can be nearly constant. This allows for interference between modes which is generally destructive interference and suppresses transmission. Steady-state spectra and dynamics are explained in terms of destructive interference between modes.

Our conclusions are summarized in section 5.

## 3.2 Experimental observations

The decomposition of the transmitted field into the electromagnetic modes of an open medium is carried out on spectra of field speckle patterns on the output surface of the samples described in Chapter 2. The experimental setup and an example of the intensity variation in the speckle pattern are shown in Fig. 3-4(a). That spectral peaks appear in spectra at different positions on the output surface (shown in Fig. 3-4(b)) and for two perpendicular polarized incident plane waves (Fig. 3-5) suggests that there are a single set of modes for each sample configuration and that the transmitted field can be expressed as the summation of the contributions of each of these modes:

$$E(\mathbf{r}, \omega) = \sum_n E_n(\mathbf{r}, \omega) = \sum_n a_{n,j}(\mathbf{r}) \frac{\Gamma_n / 2}{\Gamma_n / 2 + i(\omega - \omega_n)}. \tag{3-1}$$

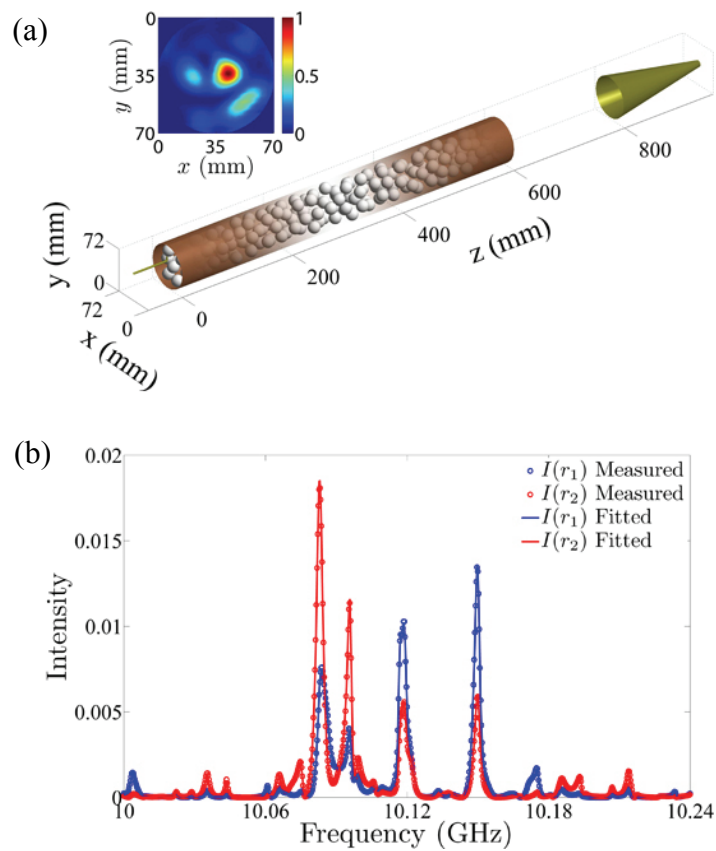

Fig. 3-4 Measurements of transmission through random media. (a) Diagram of the experimental setup. Microwave radiation is launched from a horn placed 20 cm before the sample. Field spectra of transmission through the quasi-1D randomly sample are measured at points on a 2-mm grid over the output surface. Squaring the field at each point gives the typical intensity speckle pattern such as that show at a frequency of 10.435 GHz. (inset). (b) Intensity spectra at two detector positions and the fit of Eq. 3-1 to the data are shown.

Here $E(\mathbf{r}, \omega)$ is the transmitted field spectra measured at different $\mathbf{r}$ on the output surface, and $E_n(\mathbf{r}, \omega)$ is the field spectra for the $n^{\text{th}}$ mode.

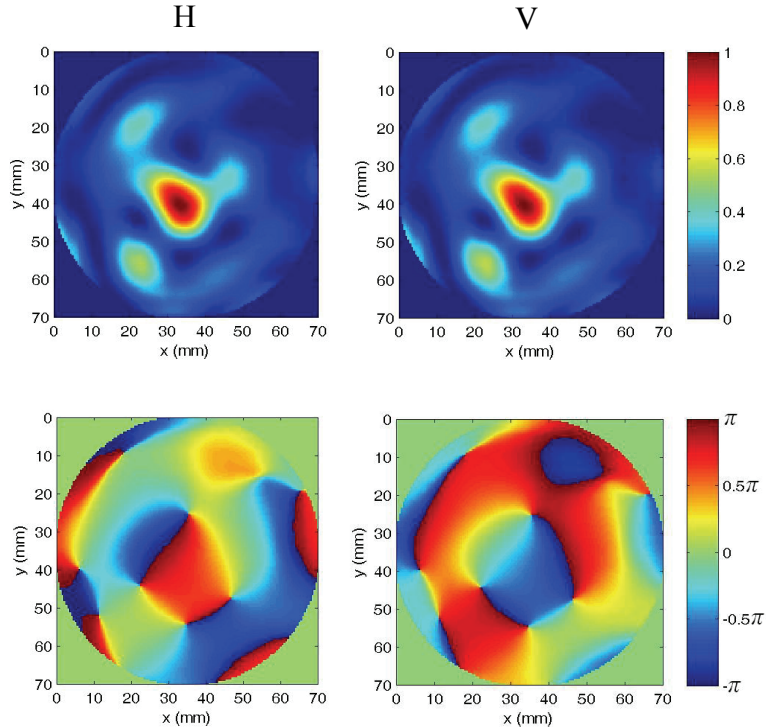

Fig. 3-5 Normalized intensity (upper two panels) and phase (lower two panels) speckle patterns of the transmitted fields of two perpendicular polarized (H and V) plane waves incident to a localized 61 cm long sample. The intensity patterns are normalized to their peak value. The two normalized intensity patterns are almost identical.

## 3.3 Fitting algorithm

To retrieve the set of mode parameters from complex transmitted field spectra through a particular random configuration, we take advantage of the form of Eq. 3-1, in which the field spectra can be written as linear combination of nonlinear functions and apply the nonlinear least square fitting method [123,124] to fit Eq. 3-1 to the measured field. The initially guessed values for the fit are obtained from the time-frequency analysis [125] which will be discussed in detail below. Then transmitted field spectra measured at 45 different output positions were fit at the same time to obtain the mode widths $\Gamma_n$ and central frequencies $\omega_n$. The next step is to solve the complex amplitude $a_n$ for each mode at different positions on the output surface.

### 3.3.1 Time-frequency analysis

To obtain the initial values of the fitting parameters, we use time-frequency analysis [121, 125] to distinguish the contribution of different modes at different delay times. Since different modes contribute differently at a certain time, we can track the evolution of the spectrograms and partially separate the modes as shown in Fig. 3-6. To do this, we window the measured spectra $E(\omega)$ with a Gaussian frequency window $G(\omega, \omega_0) = \exp(-(\omega - \omega_0)^2/2\sigma^2)$, where $\omega_0$ is the carrier frequency of the Gaussian window as shown in Fig. 3-6. We then obtain the inverse Fourier transform of the product to time domain: $E(t, \omega_0) \propto \int_{-\infty}^{+\infty} E(\omega) G(\omega, \omega_0) \exp(it\omega) d\omega$. The spectrogram $E(t, \omega_0)$ is also the temporal response to an incident Gaussian pulse $G(t) = \exp(-t^2/2\sigma_t^2)\cos(\omega_0 t)$, where $\sigma_t = \sigma^{-1}$.

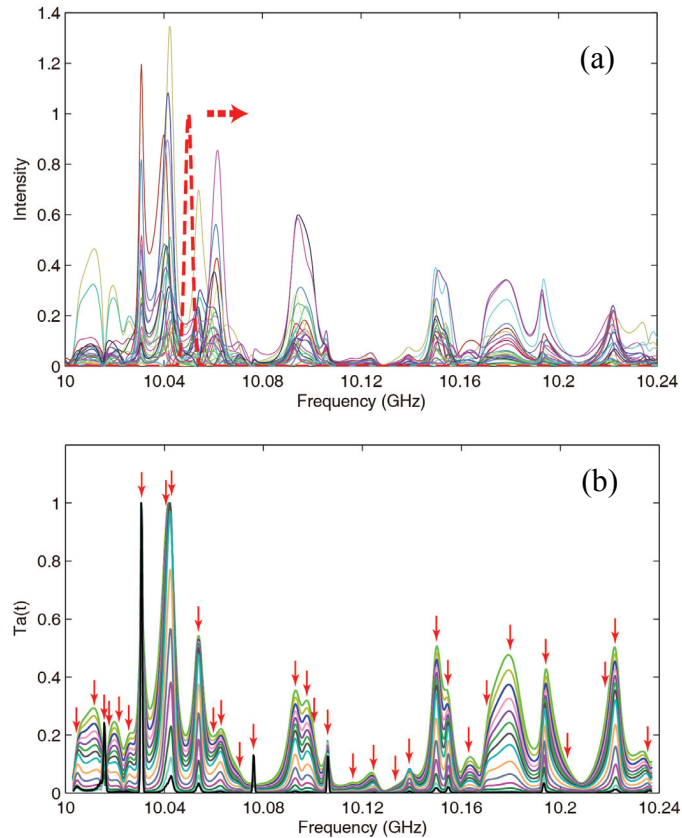

Fig. 3-6 Initial guessing of the modes parameters. (a) Intensity spectra of transmitted field detected at different output positions. Red dash line is the Gaussian pulse used to generate time-frequency spectrogram. (b) Spectrogram of total transmission at different delay time. The arrows indicate the initial guessing positions of underlying modes.

For the function $E_L(\omega) = \dfrac{\Gamma_n/2}{\Gamma_n/2 + i(\omega - \omega_n)}$ , we find the spectrogram

$E_L(t, \omega_0) \propto \int_{-\infty}^{+\infty} E_L(\omega) G(\omega) \exp(i\omega t) d\omega$ , which peaks at $\omega_n$ and decays exponentially with delay time in a decay time $2/\Gamma_n$. This is twice the decay of time for the transmitted intensity following the incident pulse for the isolated mode. The variation with time of the spectrogram of $E_L(t, \omega_0)$ is found to narrow and take on the Gaussian lineshape of the incident pulse as shown in Fig. 3-7. This feature can be exploited to separate the peaks in $|E_L(t, \omega_0)|^2$ to sharpen the determination of the central frequencies of long-lived modes. These values of the central frequency and decay rate

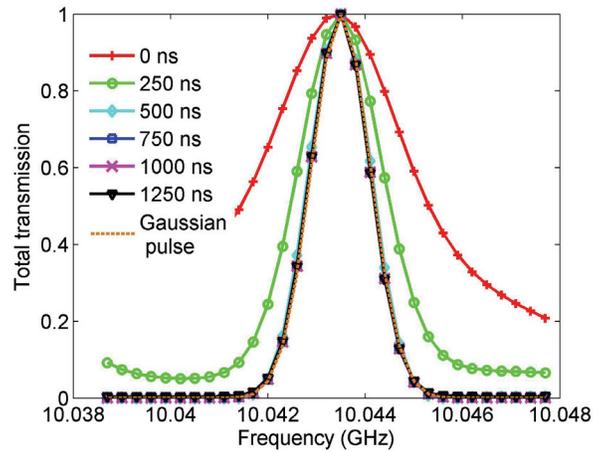

Fig. 3-7 Time evolution of spectra of total transmission with peak normalized to 1.

for the mode are used as initial values for the fit of the superposition of modes to the measured spectra.

### 3.3.2 Separable nonlinear least square fit

Since the transmitted field $E(r,\omega)$ can be viewed as a superposition of modes as in Eq. 3-1, it can be expressed as a linear combination of nonlinear functions, $E(r,\omega) = \sum_n a_n(r)\varphi_n(\omega)$, where the nonlinear function, $\varphi_n(\omega) = \dfrac{\Gamma_n/2}{\Gamma_n/2 + i(\omega - \omega_n)}$, contains the mode central frequencies and linewidths, which do not depend on the position $r$ on output surface. The mode decomposition then becomes a separable nonlinear least square [123,124] fitting problem which minimizes the sum of the square of the difference between measured values and the fit.

The separable nonlinear least square problem has been widely applied for nearly 40 years. For a set of observations $\{x_i, y_i\}$, which can be modeled as a linear combination of nonlinear functions $\Sigma_j a_j \phi_j(x_i; p_k)$, the SNLS problems minimizes the sum of squared residual, the chi-square:

$$\chi^2 = \sum_i (y_i - \sum_j a_j \varphi_j(x_i; p_k))^2, \tag{3-2}$$

where $a_j$ are linear coefficients, and $p$ are $k$ different parameters contained in the nonlinear function $\phi$. When $\chi^2$ is written in matrix format, the SNLS problem become,

$$\min \chi^2 = \min \left\| y - \Phi(x)a \right\|_2^2, \tag{3-3}$$

where the vectors $y$ and $a$ represent observations $\{y_i\}$ and linear coefficients $\{a_j\}$, respectively. Each column of the matrix $\Phi$ is the nonlinear function $\phi_j(x)$ and $\|\cdots\|_2$ represents the vector norm, $\|\alpha\|_2 = (\Sigma_i \alpha_i^2)^{1/2}$ for any vector $\alpha$. Because the linear coefficients $a$ can be expressed in terms of parameters $p$ contained in the nonlinear functions $\phi$, $a = \Phi^\dagger y$, where $\Phi^\dagger$ is the Moore–Penrose pseudoinverse of $\Phi$. For the matrix $\Phi$, which may not have an inverse, since even if one has $\Phi^\dagger \Phi = I$, $\Phi \Phi^\dagger$ may not equal the unit matrix $I$. If $\Phi$ has its inverse then $\Phi^\dagger = \Phi^{-1}$. So the linear parameters in Eq. 3-3 can be eliminated and Eq. 3-3 can be rewritten as,

$$\min \chi^2 = \min \left\| (I - \Phi \Phi^\dagger)y \right\|_2^2. \tag{3-4}$$

To find the minimum of $\chi^2$, an iterative minimization algorithm based on the method of steepest decent, such as the Gauss-Newton and Levenberg-Marquardt methods, [126] can be applied. The gradient of $\chi^2$ with respect to the parameter $p$ is given by,

$$\frac{\partial \chi^2}{\partial p_k} = 2 \sum_{i=1}^{N} f(x_i; p) \frac{\partial f(x_i; p)}{\partial p_k},$$

where $f = y - \sum_j a_j \varphi_j = (I - \Phi \Phi^\dagger)y$, will be zero at the minimum of $\chi^2$. The Jacobian of $f$ with respect $p_k$ is

$$\nabla_k f = -(\nabla_k \Phi)\Phi^\dagger y - \Phi(\nabla_k \Phi^\dagger) y \ .$$

This Jacobian can be simplified and approximately expressed as

$$\nabla_k f \approx -(I - \Phi\Phi^\dagger)(\nabla_k \Phi)\Phi^\dagger y$$

$$= -(I - \Phi\Phi^\dagger) \begin{pmatrix} \dfrac{\partial \Phi_{1k}}{\partial p_k} \\ \vdots \\ \dfrac{\partial \Phi_{mk}}{\partial p_k} \end{pmatrix} a \quad ,$$

The second partial derivative is

$$\frac{\partial^2 \chi^2}{\partial p_k \partial p_\ell} = 2\sum_{i=1}^{N} \left( \frac{\partial f(x_i;p)}{\partial p_k} \frac{\partial f(x_i;p)}{\partial p_\ell} + f(x_i;p) \frac{\partial^2 f(x_i;p)}{\partial p_k \partial p_\ell} \right)$$

When a good fit is achieved, the difference between the measurements and the fit, $f(x,p)$, is small and could be of either sign, so its sum over all $i$ tends to cancel. Including the second derivative of $f(x,p)$ can in fact destabilize the fit if the model is not a good fit to the data [126]. In addition, minor modifications in $\dfrac{\partial^2 \chi^2}{\partial p_k \partial p_\ell}$ has no effect, in practice, on the final value of fitting parameters. These only affect the iterative route that is taken in achieving a good fit since the condition that the $\chi^2$ is minimum is $\dfrac{\partial \chi^2}{\partial p_k} = 0$ for all k, which is independent of the modification of $\dfrac{\partial^2 \chi^2}{\partial p_k \partial p_\ell}$. So we ignore the second derivative of $f(x,p)$ in practice.

There are a number of characteristics of the spectra that are exploited in the separable nonlinear least square fitting procedures:

a)      The data to be fit is the measured field spectra, which is naturally complex. We need therefore to deal with both real and imaginary parts in achieving the best fit.

b)   There are constraints acting on the nonlinear variables, the central frequencies, $\omega_n$ and modes widths $\Gamma_n$. They are required to be positive real numbers. And the modes central frequencies fall in or very close to a small frequency rang 10-10.24 GHz. Most of the modes are only few MHz wide; three orders smaller than the ordinary central frequencies.

c)   The spectra measured at different output positions in a given configuration share a common set of $\omega_n$ and $\Gamma_n$.

To satisfy these conditions, we first center and rescale the frequency to $\tilde{\nu} = (\nu - \overline{\nu})/\sigma_\nu$, where $\overline{\nu}$ = 10.12 GHz is the middle of the frequency range and $\sigma_\nu$ is the standard deviation of $\nu - \overline{\nu}$. Since the mode widths must be positive, we replace $\Gamma_n/2$ by a squared parameter, $w_n^2$ to release the nonnegative constraint of the fitting parameters. By doing this, we sacrifice the efficiency of the fit and the speed of convergence, but we keep the algorithm as simple as possible. This makes it easier to modify the standard separable nonlinear least square algorithm so that it is suitable for our case in which we fit tens or hundreds spectra at the same time for a single sample configuration. Equation 3-1 is then rewritten as,

$$E(\mathbf{r},\mu) = \sum_n c_n(\mathbf{r}) \frac{1}{w_n^2 + i(\tilde{\nu} - \mu_n)} = \sum_n c_n(\mathbf{r}) \psi_n(\tilde{\nu}),$$

where $\mu_n$ and $w_n^2$ are scaled central frequencies and widths, respectively. The actual central frequencies and widths can be retrieved using $\omega_n = 2\pi\nu_n = 2\pi(\mu_n\sigma_\nu + \overline{\nu})$ and $\Gamma_n = 2\pi\gamma_n = 4\pi w_n^2\sigma_\nu$. Now the constraints of the mode central frequencies and widths are removed.

Each sample configuration has one common mode set, $\omega_n$ and $\Gamma_n$, the nonlinear parameters. The difference between spectra taken at different positions is only due to the complex amplitude of each mode $a_n(\mathbf{r})$, the linear parameters, so the spectra $E(\mathbf{r},\nu_n)$ at different positions **r** have same nonlinear function. Instead of writing spectra to be fitted in a vector form for each spectrum, the spectra to be fitted can be written as a matrix, each column of which corresponds to a single spectrum at different position $r_p$. The complex amplitudes are also written as a matrix rather than a vector. The separable nonlinear least square model for $P$ different spectra is then,

$$
\begin{pmatrix} E(r_1,\tilde{\nu}_1) & \cdots & E(r_P,\tilde{\nu}_1) \\ \vdots & \ddots & \vdots \\ E(r_1,\tilde{\nu}_M) & \cdots & E(r_P,\tilde{\nu}_M) \end{pmatrix} = \begin{pmatrix} \psi_1(\tilde{\nu}_1) & \cdots & \psi_N(\tilde{\nu}_1) \\ \vdots & \ddots & \vdots \\ \psi_1(\tilde{\nu}_M) & \cdots & \psi_N(\tilde{\nu}_M) \end{pmatrix} \cdot \begin{pmatrix} c_1(r_1) & \cdots & c_1(r_P) \\ \vdots & \ddots & \vdots \\ c_N(r_1) & \cdots & c_N(r_P) \end{pmatrix}.
$$

Because the nonlinear variables $w_n$ and $\mu_n$ are real number, we need to transform all variable to real numbers and minimize the residual with respect to real parameters. The problem then becomes,

$$
\begin{aligned}
\min \chi^2 &= \min \sum_{\mathbf{r}} \sum_{\tilde{\nu}} \left\{ \left( \mathrm{Re}(E(\mathbf{r},\tilde{\nu})) - \mathrm{Re}(\sum_n c_n(\mathbf{r})\psi_n(\tilde{\nu})) \right)^2 \right. \\
&\quad + \left. \left( \mathrm{Im}(E(\mathbf{r},\tilde{\nu})) - \mathrm{Im}(\sum_n c_n(\mathbf{r})\psi_n(\tilde{\nu})) \right)^2 \right\} \\
&= \min \sum_{\mathbf{r}} \sum_{\tilde{\nu}} \left\{ \left( \mathrm{Re}(E(\mathbf{r},\tilde{\nu})) - \sum_n \big( \mathrm{Re}(c_n(\mathbf{r})) \mathrm{Re}(\psi_n(\tilde{\nu})) - \mathrm{Im}(c_n(\mathbf{r})) \mathrm{Im}(\psi_n(\tilde{\nu})) \big) \right)^2 \right. \\
&\quad + \left. \left( \mathrm{Im}(E(\mathbf{r},\tilde{\nu})) - \sum_n \big( \mathrm{Re}(c_n(\mathbf{r})) \mathrm{Im}(\psi_n(\tilde{\nu})) + \mathrm{Im}(c_n(\mathbf{r})) \mathrm{Re}(\psi_n(\tilde{\nu})) \big) \right)^2 \right\} \quad .
\end{aligned}
$$

We could rewrite the above equation into matrix form,

$$
\min \chi^2 = \min \| E - \Psi C \|_2^2
$$

where $\| \cdots \|_2$ represents the Frobenius norm of a matrix $B$, $\| B \|_2 = \sqrt{\sum_{i,j} |b_{ij}|^2} = \sqrt{\mathrm{tr}(B^*B)}$ . Each column of the $2M \times P$ matrix $E$ represents one spectrum at the output position $r_p$, and instead of using complex numbers to represent the field, field spectra are expressed via real numbers by

writing the real and imaginary parts separately. The $P$ spectra are can then be expressed in a single real matrix $E$ as follows:

$$E = \begin{pmatrix} \mathrm{Re}(E(r_1,\tilde{\nu}_1)) & \cdots & \mathrm{Re}(E(r_P,\tilde{\nu}_1)) \\ \mathrm{Im}(E(r_1,\tilde{\nu}_1)) & & \mathrm{Im}(E(r_P,\tilde{\nu}_1)) \\ \vdots & \ddots & \vdots \\ \mathrm{Re}(E(r_1,\tilde{\nu}_M)) & \cdots & \mathrm{Re}(E(r_P,\tilde{\nu}_M)) \\ \mathrm{Im}(E(r_1,\tilde{\nu}_M)) & & \mathrm{Im}(E(r_P,\tilde{\nu}_M)) \end{pmatrix}.$$

The $2M \times 2N$ matrix $\Psi$ has elements of the form,

$$\Psi_{2m-1,2n-1} = \Psi_{2m,2n} = \mathrm{Re}(\psi_n(\tilde{\nu}_M)) = \mathrm{Re}\left(\frac{1}{w_n^2 + i(\tilde{\nu}_m - \mu_n)}\right),$$

$$\Psi_{2m,2n-1} = -\Psi_{2m-1,2n} = \mathrm{Im}(\psi_n(\tilde{\nu}_M)) = \mathrm{Im}\left(\frac{1}{w_n^2 + i(\tilde{\nu}_m - \mu_n)}\right).$$

The matrix $C$ is accordingly written,

$$C = \begin{pmatrix} \mathrm{Re}(c_1(r_1)) & \cdots & \mathrm{Re}(c_1(r_P)) \\ \mathrm{Im}(c_1(r_1)) & & \mathrm{Im}(c_1(r_P)) \\ \vdots & \ddots & \vdots \\ \mathrm{Re}(c_N(r_1)) & \cdots & \mathrm{Re}(c_N(r_P)) \\ \mathrm{Im}(c_N(r_1)) & & \mathrm{Im}(c_N(r_P)) \end{pmatrix}.$$

Since the optimized linear coefficient $C$ can be approximately expressed as $C = \Psi E$, the separable nonlinear least square problem specific to our system is, $\min \chi^2 = \min \left\| (I - \Psi\Psi^\dagger)E \right\|_2^2$. The approximate Jacobian of the residual of the fit of the spectrum at position $r_p$ on the output surface $R_p = (I - \Psi\Psi^\dagger)E_p$ with respect to the $k^{\text{th}}$ nonlinear coefficient is,

$$\begin{aligned} \nabla_k R_p &\approx -(I - \Psi\Psi^\dagger)(\nabla_k \Psi)\Psi^\dagger E(r_p) \\ &= -(I - \Psi\Psi^\dagger)(\nabla_k \Psi)C_p \\ &= -(I - \Psi\Psi^\dagger)\tilde{\nabla}_k \Psi_p \quad, \end{aligned}$$

where $C_p$ is the $p^{\text{th}}$ column of linear coefficient matrix $C$, $\tilde{\nabla}\Psi_p$ specific for our data model is,

$$\tilde{\nabla}\psi_p = \begin{pmatrix} \cdot & \operatorname{Re}c_{n,p}\operatorname{Re}\dfrac{\partial\psi_n(\tilde{v}_1)}{\partial\mu_n} - \operatorname{Im}c_{n,p}\operatorname{Im}\dfrac{\partial\psi_n(\tilde{v}_1)}{\partial\mu_n} & \cdot & \bigg| & \operatorname{Re}c_{n,p}\operatorname{Re}\dfrac{\partial\psi_n(\tilde{v}_1)}{\partial w_n} - \operatorname{Im}c_{n,p}\operatorname{Im}\dfrac{\partial\psi_n(\tilde{v}_1)}{\partial w_n} & \cdot \\[2mm] \cdot & \operatorname{Re}c_{n,p}\operatorname{Im}\dfrac{\partial\psi_n(\tilde{v}_1)}{\partial\mu_n} + \operatorname{Im}c_{n,p}\operatorname{Re}\dfrac{\partial\psi_n(\tilde{v}_1)}{\partial\mu_n} & \cdot & \bigg| & \operatorname{Re}c_{n,p}\operatorname{Im}\dfrac{\partial\psi_n(\tilde{v}_1)}{\partial w_n} + \operatorname{Im}c_{n,p}\operatorname{Re}\dfrac{\partial\psi_n(\tilde{v}_1)}{\partial w_n} & \cdot \\ \vdots & \vdots & \vdots & \bigg| & \vdots & \vdots \\ \cdot & \operatorname{Re}c_{n,p}\operatorname{Re}\dfrac{\partial\psi_n(\tilde{v}_M)}{\partial\mu_n} - \operatorname{Im}c_{n,p}\operatorname{Im}\dfrac{\partial\psi_n(\tilde{v}_M)}{\partial\mu_n} & \cdot & \bigg| & \operatorname{Re}c_{n,p}\operatorname{Re}\dfrac{\partial\psi_n(\tilde{v}_M)}{\partial w_n} - \operatorname{Im}c_{n,p}\operatorname{Im}\dfrac{\partial\psi_n(\tilde{v}_M)}{\partial w_n} & \cdot \\[2mm] \cdot & \operatorname{Re}c_{n,p}\operatorname{Im}\dfrac{\partial\psi_n(\tilde{v}_M)}{\partial\mu_n} + \operatorname{Im}c_{n,p}\operatorname{Re}\dfrac{\partial\psi_n(\tilde{v}_M)}{\partial\mu_n} & \cdot & \bigg| & \operatorname{Re}c_{n,p}\operatorname{Im}\dfrac{\partial\psi_n(\tilde{v}_M)}{\partial w_n} + \operatorname{Im}c_{n,p}\operatorname{Re}\dfrac{\partial\psi_n(\tilde{v}_M)}{\partial w_n} & \cdot \end{pmatrix}.$$

Here $c_{n,p}$ is the linear coefficient for the $n^{\text{th}}$ mode in the spectrum at position $r_p$. By defining,

$$\alpha = \nabla^2\chi^2 = \sum_p\nabla^2 R^2 = \sum_p J_p^T J_p \text{ and } \beta = -\nabla\chi^2 = -\sum_p\nabla R_p^2 = -J_p(E_p - \Psi C_p),$$

the steepest descent problem becomes,

$$\alpha\delta\mathbf{C}^{nl} = \beta,$$

where $\delta\mathbf{C}^{nl}$ is the vector of increment of the nonlinear coefficients along the downhill direction. In the Levenberg-Marquardt method, a new matrix $\alpha'$ is defined as,

$$\alpha'_{jj} = \alpha_{jj}(1+\lambda),$$
$$\alpha'_{jk} = \alpha_{jk}.$$

The standard Levenberg-Marquardt algorithm can be applied here to obtain the nonlinear coefficients, $w_n$ and $\mu_n$. This can be solved for the linear coefficients $c_{n,p}$. In summary, the fitting procedures shown in the Fig. 3-8 are as follows:

a. Compute the time-frequency spectrogram of total transmission: From the spectrogram of total transmission, guess the number of underlying modes and the mode central frequencies and widths based on the frequencies of the peaks in spectrogram and the decay rate of the peak intensity.

b. Process the data so that it is in suitable for the fit: Center and rescale the spectra, transform all data into real numbers.

c.           Calculate the squared residual $\chi^2$.

d.           Assign a small value to $\lambda$, say $\lambda=0.001$.

e.           Solve for the increment of the nonlinear coefficients $\delta C^{nl}$ and calculate $\chi^2(\mathbf{C}^{nl}+\delta\mathbf{C}^{nl})$.

f.           If $\chi^2(\mathbf{C}^{nl}+\delta\mathbf{C}^{nl}) > \chi^2(\mathbf{C}^{nl})$, increase $\lambda$ by a factor, say 10, and go back to step e.

g.           If $\chi^2(\mathbf{C}^{nl}+\delta\mathbf{C}^{nl}) < \chi^2(\mathbf{C}^{nl})$, decrease $\lambda$ by a factor, say 10, and update the fitting parameters with a new set $\mathbf{C}^{nl}\leftarrow\mathbf{C}^{nl}+\delta\mathbf{C}^{nl}$, and go to step e.

h.           Stop the above loop until the change of $\chi^2$ is small. The change does not have to be the machine accuracy. This saves time and does not affect the accuracy of our fit.

i.           Check the frequency variation of the residuals, which is the sum of the square of absolute value of the difference between measured field and the fit over the whole output, $R^2$; if there is an obvious peak, add a mode with central frequency equal to the peak position and width equal the average of mode width; and go to step c, redo the fit again, until there is no obvious peak in the spectrum of $R^2$. The reason for this step will be explained below.

Spectra of total transmission can be well fit with the number of modes close to the actual number of modes determined from the spectra of the square of the residuals in the fit of spectra between 10 and 10.24 GHz, . This can be seen in the spectrum of total transmission normalized by its ensemble average for a wave launched from an antenna at a point for 40 and 80 cm-long samples and from a horn antenna for the 61 cm-long sample. The source is indicated by $a$, and the normalized total transmission by $s_a=T_a/<T_a>$. Spectra of total transmission are shown in Fig. 3-9(a). Nonetheless, the number of modes can be determined unambiguously from spectra of the squared residual of the fit (Fig. 3-9(b)-(f)). We take the actual number of modes $M$ to be the number of modes $M'$ for which no obvious peaks are seen in the residual, following upon a reduction by unity in the number of peaks as $M'$ is increased. For the spectrum of total transmission through 61cm-long-sample shown in Fig. 3-9, $M$=51. When an additional mode is added in the fit passed this point, the width of the added mode is unphysically broad with a central frequency that falls well beyond the frequency range of the spectrum. The width of the added mode is 60 MHz which is extremely large.

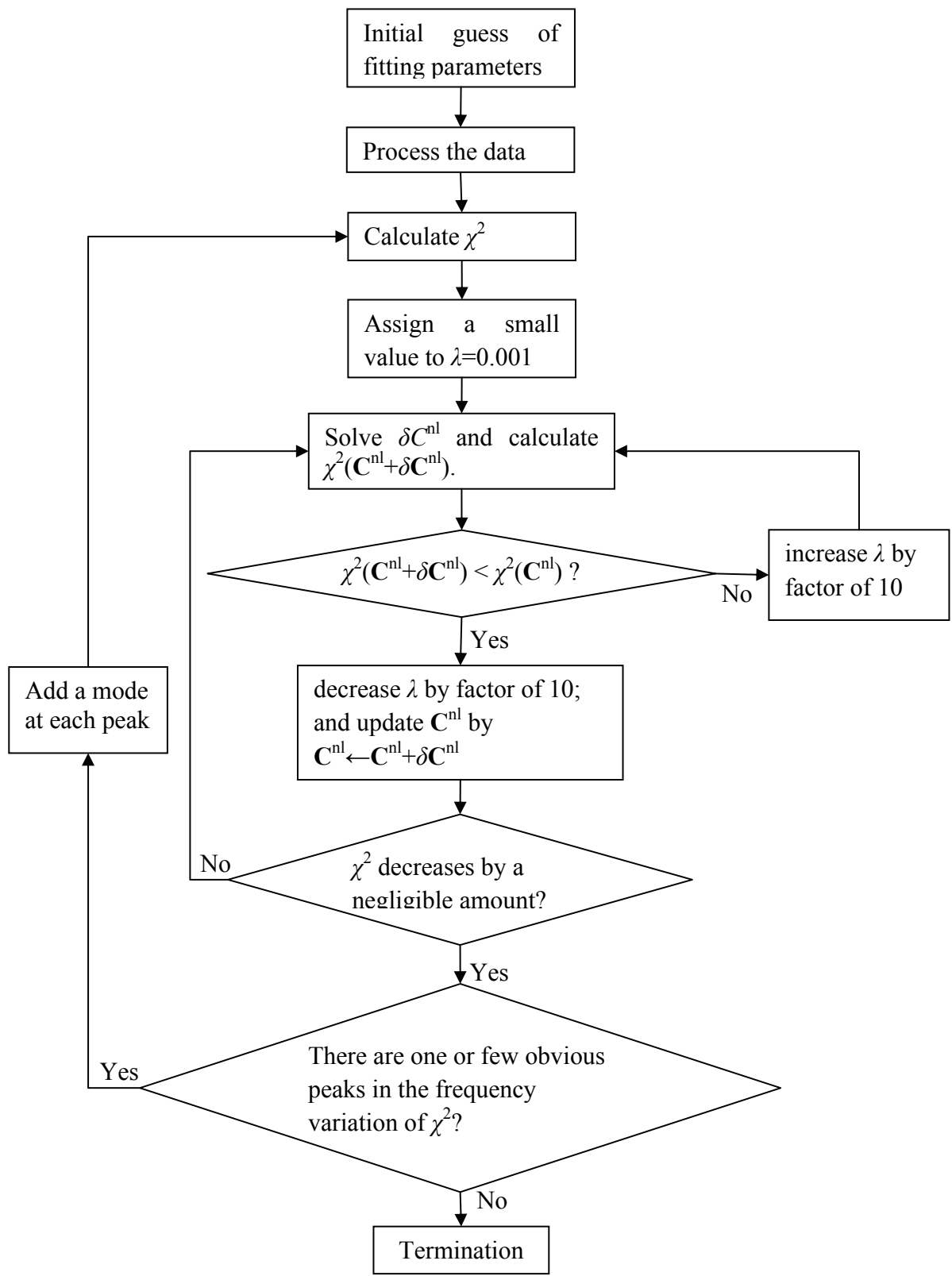

Fig. 3-8 The flowchart of fitting procedures.

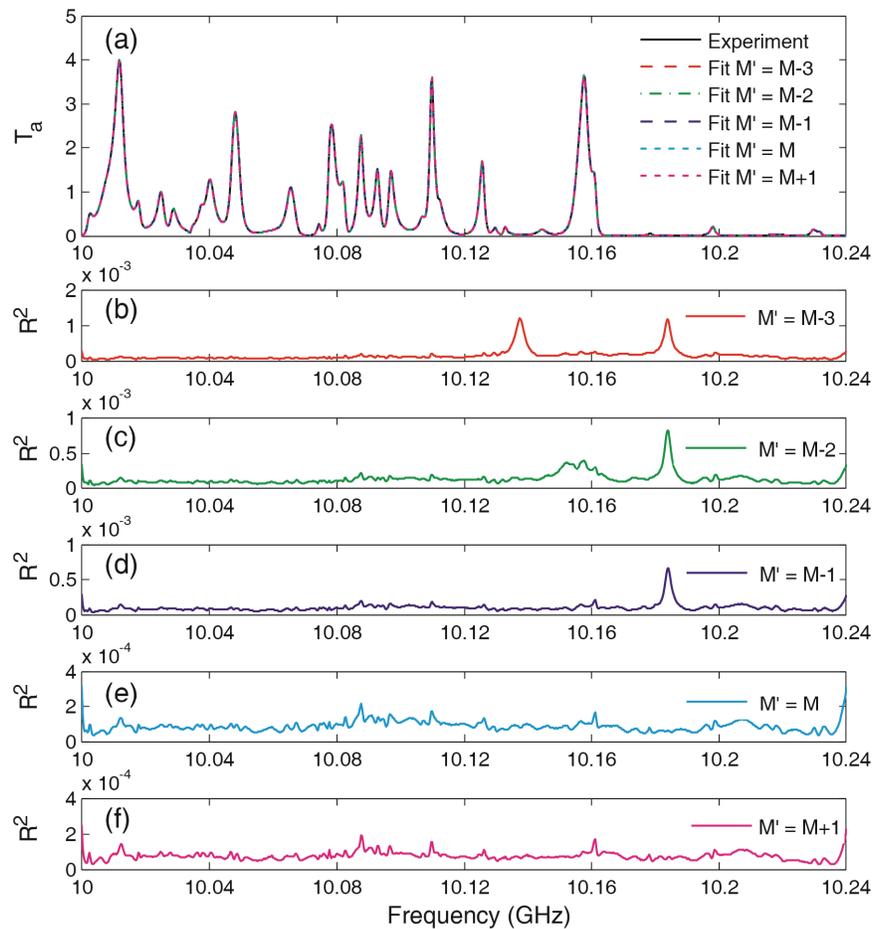

Fig. 3-9 Separable nonlinear least square fit. (a) Total transmission spectrum and the summation of the fit of all spectra with different numbers of modes. (b)-(f) The total squared residual between the measured field spectra and fits with different number of modes.

## 3.4. Result and discussion

To illustrate the way modes contribute to total transmission, we consider a narrow frequency range around the strong peak at 10.15 GHz for the same configuration for which intensity spectra are shown in Fig. 3-4(b). The asymmetrical shape for the line in both intensity and total transmission indicates that more than a single mode contributes to the peak. The modal analysis of the field spectra shows that three modes contribute substantially near this strong peak[121]. Spectra of total transmission for the three modes closest to 10.15 GHz are plotted in Fig. 3-10(a). The integrated transmission for two of these modes corresponding to the 28th and 29th mode in the spectrum starting at 10 GHz are greater than for the measured transmission peak, indicating that these modes interfere destructively. The intensity and phase patterns for each of these modes are shown in Figs. 3-10(b)-(e). Though the average transmission is different for the two modes, the intensity speckle patterns for these modes are nearly identical. At the same time, the distributions of phase shift at 10.15 GHz for the two modes are similar up to a nearly constant phase difference with average value $\overline{\Delta\varphi} = 1.02\pi$ and the standard deviation of the phase difference $0.016\pi$. The surprising similarity between the speckle patterns for these overlapping modes suggests that these spectrally overlapping modes are formed from coupled resonances which overlap spatially within the sample [41,88,90,92]. The similarity in the intensity speckle patterns of adjacent modes and the uniformity of the phase shift across the patterns of these modes makes it possible for the wave to interfere destructively across the entire modal speckle pattern.

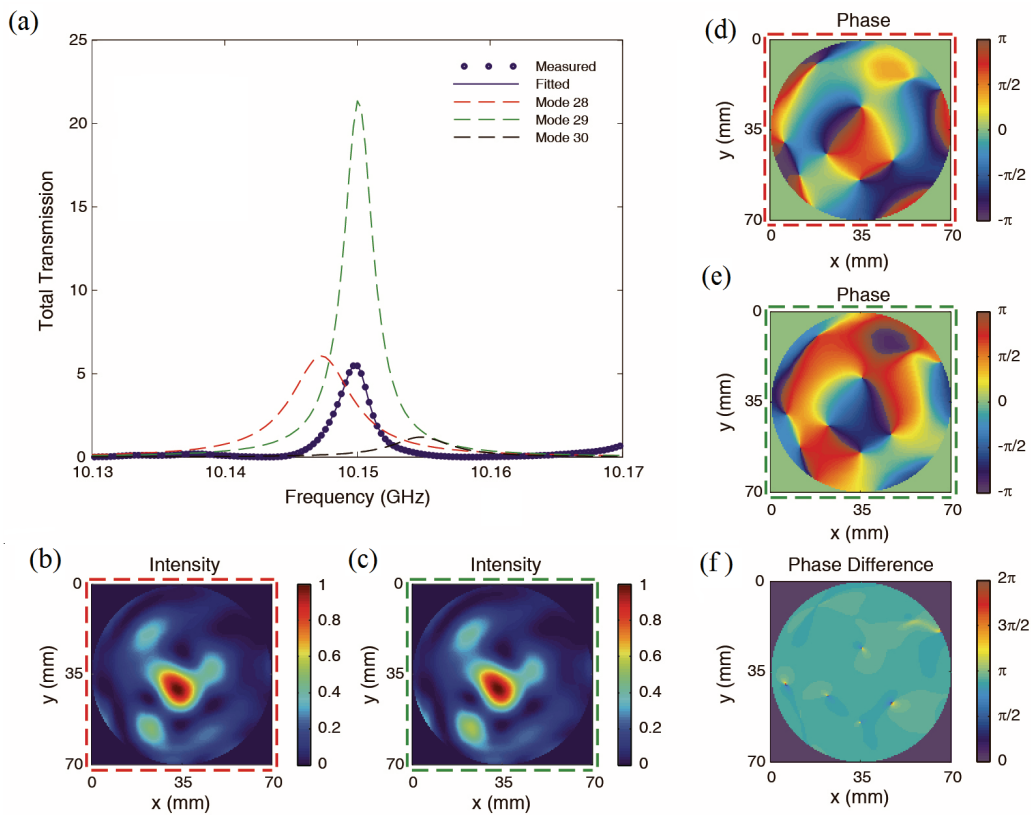

Fig. 3-10 Total transmission and mode speckle patterns. (a) Three modes contribute to the asymmetric peak in the total transmission spectrum (blue) for the same configuration as in Fig.2-9, mode28 (red), mode29 (green) and mode 30 (black). Dashed red (mode 28) and green (mode 29) lines surround their intensity speckle patterns ((b) and (c)) and phase patterns ((d) and (e)). (f) The phase in mode 29 is shifted by nearly a constant of $1.02\pi$ rad relative to mode 28.

The decomposition of field spectra into modes accounts for the characteristics of dynamic as well as of steady-state transmission. We compute the time-frequency spectrogram [125] of the total transmission, $T_a(t,v_0)$, corresponding to the variation with carrier frequency $v_0$ of the sum of intensity over all points on the output surface on which measurements of field are made at delay time $t$ from the peak of an incident Gaussian pulse, $T_a(t,v_0) = \Sigma_{\mathbf{r}}|E(\mathbf{r},t,v_0)|^2$. $T_a(t,v_0)$ is indicated by the color scale in the $x$-$y$ plane in Fig. 3-11. The evolution of the spectrum of total transmission is further indicated by plotting $T_a(t,v_0)$ normalized to the average over each spectrum for four delay times. The increasing impact of long-lived narrow-line modes is manifest in the decreasing number of surviving modes with substantial relative intensity. This results in an increasing variance of normalized transmission with time delay. The variation with time of the spectrogram of total transmission normalized by the peak transmission for the mode with center frequency $\nu_n$ = 10.04352 GHz is seen in Fig. 3-11(b) to narrow and take on the Gaussian lineshape of the incident pulse. The decay of this peak is shown in Fig. 3-11(c). The decay rate found of 1.18 μs$^{-1}$ is nearly identical with the linewidth $\Gamma_n$ = 1.29 MHz. The small difference may be due to the noise background which could decrease the value of the decay rate.

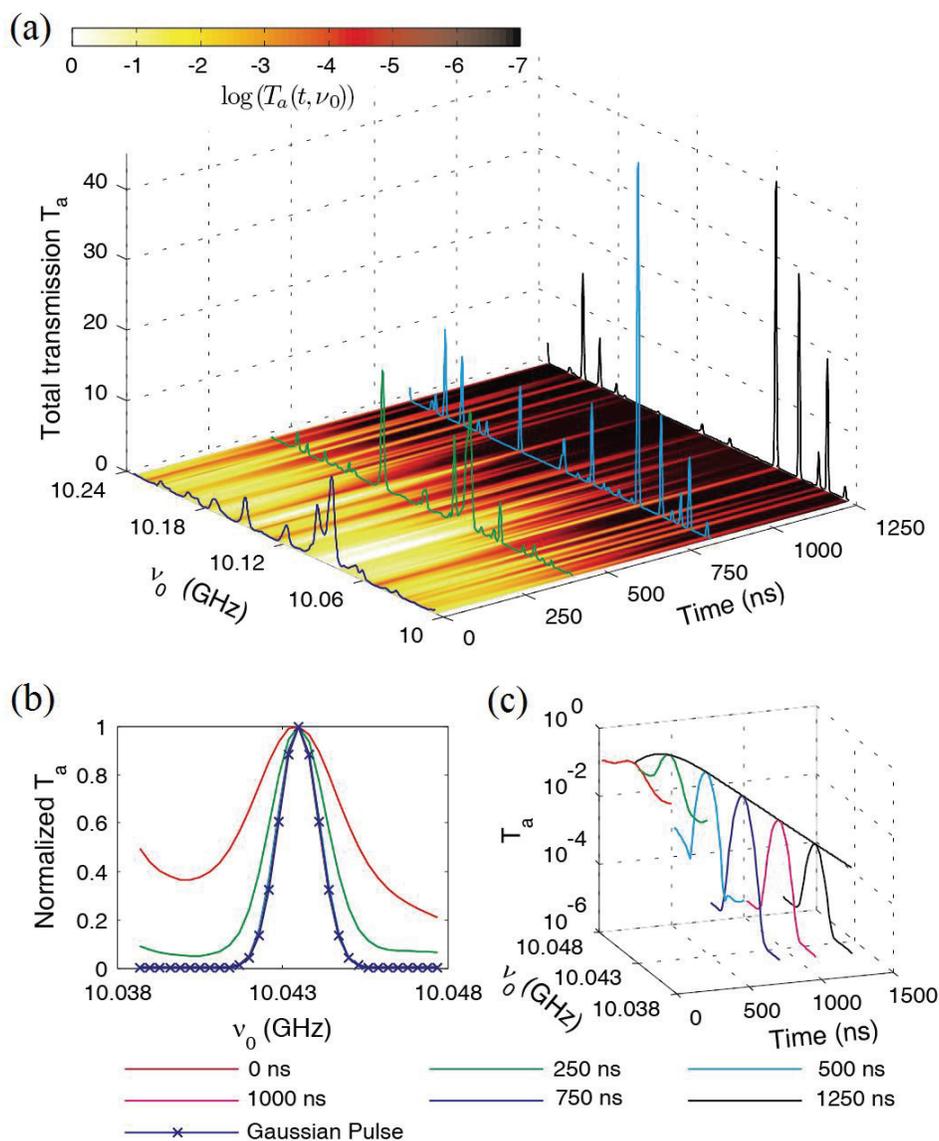

Fig. 3-11 (a) Time–frequency analysis (a) Logarithm of time–frequency spectrogram of total transmission plotted (see color scale) in the *x–y* plane. The central frequency $v_0$ of the incident Gaussian pulse of linewidth $\sigma = 0.85$ MHz is scanned. Each of the four spectra of total transmission at different delay times are normalized to its average. (b) Spectrograms of total transmission for the mode with centre frequency $v_n$=10.04352 GHz are normalized by the peak transmission for different delay times. The color of the curves in (b) and (c) indicate different delay times for the spectrograms in (b) and (c) (see legend at the bottom). All the curves overlap for delays of 500 ns and longer. (c) The decay rate of the peak in (b) of 1.18 $\mu$s$^{-1}$ is essentially equal to the linewidth $\Gamma_n$=1.29 MHz.

The average temporal variation of total transmission, $\langle T_a(t) \rangle$, is obtained by integrating the time-frequency spectrogram over frequency in each configuration at different times and averaging over all configurations. The progressive suppression of transmission in time by absorption may be removed by multiplying $\langle T_a(t) \rangle$ by $\exp(t/\tau_a)$ to give, $\langle T_a^{\,0}(t) \rangle = \langle T_a(t) \rangle \exp(t/\tau_a)$, which gives the decay of transmission due only to leakage from the sample. The same result is obtained by transforming into the time domain using spectra computed from the modal decomposition of the field in which the impact of absorption on each mode is eliminated by substituting $\Gamma_n^0 = \Gamma_n - 1/\tau_a$ for the linewidth $\Gamma_n$ and the complex mode amplitude $a_n^0 = a_n \Gamma_n / \Gamma_n^0$ for $a_n$. The decay of $\langle T_a^{\,0}(t) \rangle$, shown as the solid curve in Fig. 3-12, is seen to slow considerably with time delay [39,58]. This reflects a broad range of modal decay rates. This slowdown in the rate of decay of transmission with delay time from an exciting pulse was first observed in microwave experiments in diffusive quasi-1D samples [117]. A similar slowing down of the decay rate was also observed in optical [118] measurements in diffusive samples and in ultrasound measurements [39] below and just beyond the mobility edge. These measurements are well fit by SCLT. However, measurements of pulsed microwave transmission of more deeply localized waves transmitted which are described here could not be described by SCLT [58,127]. For times up to twice the peak rise time of the transmitted pulse, diffusion theory gives a good description of measurements of <$I(t)$> in the sample with $L$=61 cm, in which the wave is strongly localized [58]. For later times, however, pulsed transmission is higher than predicted by diffusion theory and by SCLT as shown in Ref. [58]. In the following, we show that modal analysis is able to explain the dynamics of deeply localized waves.

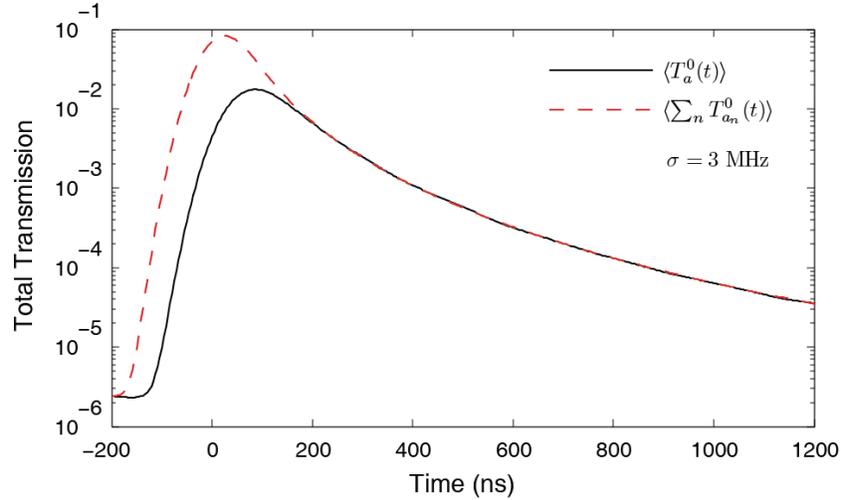

Fig. 3-12 Dynamics of localized waves. Semi logarithmic plot of the ensemble average of pulsed transmission and the incoherent sum of transmission due to all modes in the random ensemble. The impact of absorption is eliminated as described in the text.

The measured decay of pulsed transmission, shown as the solid curve in Fig. 3-12, is compared to the incoherent sum of transmission for all modes in the random ensemble, $\langle \sum_n T_{a_n}^0(t) \rangle$, shown as the dashed curve in Fig. 3-12. $\langle \sum_n T_{a_n}^0(t) \rangle$ is substantially larger than $\langle T_a^0(t) \rangle$ at early times but then converges rapidly to $\langle T_a^0(t) \rangle$. Though transmission associated with individual modes rises with the incident pulse, transmission at early times is strongly suppressed by destructive interference between modes with strongly correlated field speckle patterns such as those shown in Fig. 3-10. The delayed rise of pulsed transmission seen in Fig. 3-12 is a remnant of particle diffusion associated with transport in weakly scattering samples and reflects the low probability of particles traversing the sample by a sequence of scattering events all in the forward direction. At later times, the random frequency differences between modes lead to an additional random phase difference between modes, $(\omega_{n+1} - \omega_n)t$ so that the phase difference between modes is random and the decay is just the incoherent sum of individual decaying modes. This leads to a peak in transmission after a delay time comparable to the inverse of the typical mode linewidth. By this time, interference is no longer destructive on average and approaches the incoherent sum of decaying modes.

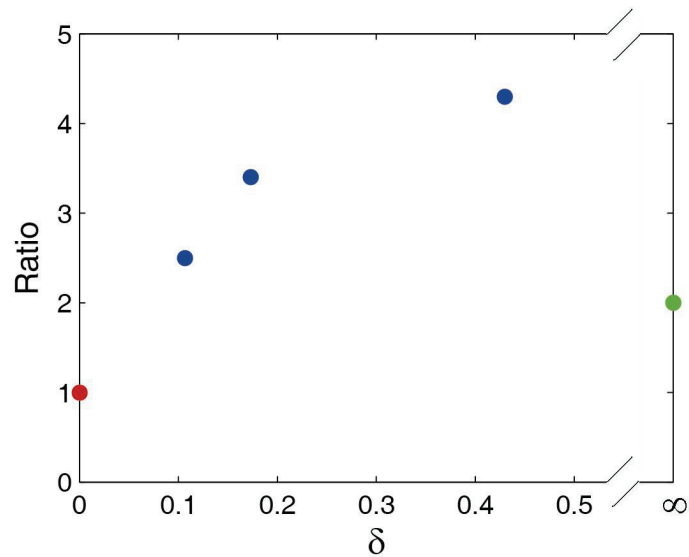

Fig. 3-13 The ratio between the ensemble average of the sum of contributions from individual modes to total transmission and the average of total transmission for three sample length $L$=40, 61 and 80 cm, corresponding to values of the Thouless number of $\delta$=0.43, 0.17 and 0.10, respectively, are shown as the blues circles, while the same ratios in the diffusive and localized limits are shown by the green and red circles, respectively.

The ratios between the time integration of ensemble average of the summations of the total transmission for individual modes and the ensemble average total transmission for three sample lengths $L$=40, 61 and 80 cm, corresponding to values of the Thouless number of $\delta$=0.43, 0.17 and 0.10 , respectively, are shown by the three blue circles in Fig. 3-13. The ratios for three sample lengths are all greater than unity indicating that destructive interference between modes suppresses total transmission. The ratio is found to decrease with sample length for the three lengths for which the modal decomposition was performed. For these lengths, the modes are more and more separated in the samples with increasing length and the overlap and coupling between the modes decrease. In the localized limit, overlap between modes should vanish at any frequency and all modes should be statistically independent. The ratio should therefore equal unity as shown by the red circle in Fig. 3-13. However, overlap between modes should be substantial for diffusive waves. The degree to which interference between modes affects the total transmission for diffusive waves can be calculated by considering transmission excited with a delta pulse, as shown in Fig. 3-14. In the diffusive limit all modes decay at the same rate since the mode is extended throughout the sample and leaks through the sample through a diverging number of uncorrelated speckle spots. The mode decay rate is just the decay rate of transmission at long times, which corresponds to the decay rate of the lowest diffusion mode. The incoherent sum of modal intensities can therefore be obtained by extrapolating the intensity in the long-time tail of the pulsed transmission back to t=0. The ratio for diffusive limit is found to equal 2 as shown by the green circle in Fig.3-13.

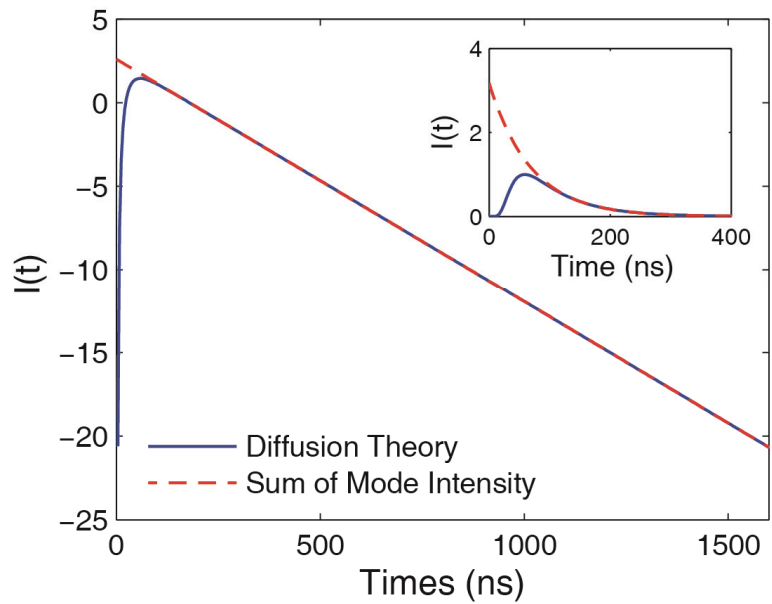

Fig. 3-14 The transmitted intensity and incoherent sum of mode intensities following excitation with delta pulse for diffusive waves are shown by the blue curve and the dashed red curve, respectively. The inset shows linear scale of these two quantities at early time.

## 3.5 Conclusion

In this chapter, we have described the measurement of spectra of fields transmitted through localized quasi-1D random media. We described the algorithm used to decompose the transmitted field speckle patterns into a superposition of the mode patterns and retrieve the central frequency and linewidth of each mode. We find that the intensity patterns of neighboring modes can be highly correlated. The power of the modal analysis of wave transport can be seen in the dynamics of transmission. The modal description explains the suppression of steady-state and pulsed transmission. The correlation between the modes is associated with two important statistical properties of modes: level repulsion, which leads to the low probability of finding mode with central frequencies that are very close, and level rigidity, which gives a small variation of the number of modes in any frequency range in different configurations. The details of the statistics of modes will be discussed in the next chapter.

# CHAPTER 4

# MODE STATISTICS IN RANDOM MEDIA

## 4.1 Introduction

In complex systems it is generally impossible to solve for the eigenvalues of the Hamiltonian or wave equation even when the interactions within the medium are known. However, the nature of wave propagation within different complex systems such as nuclei, classical or quantum chaotic systems and disordered media can be determined by the statistical properties of the spectra of modes or energy levels. One crucial parameter is the Thouless number $\delta$, the dimensionless ratio of the average energy level width to the average spacing between levels in the sample [75,76,121,128]. Since $\delta=g$, $g$ is equal to the average number of levels within the band Thouless energy $E_{Th}= \hbar/\tau_{Th}$, $g = <N(E_{Th})>$ [76,104]. As a result, fluctuations of the dimensionless conductance $g$ are governed by fluctuations in the number of levels in an energy interval of $E_{Th}$, var($g$) = var($N(E_{Th})$) [104]. $\delta$ is also related to many other parameters characterizing different aspects of wave propagation within complex systems such as the return probability $P_{return}$ [14], the variance of total transmission var($s_a=T_a/<T_a>$) [21,23,24,35] and the degree of intensity correlation $\kappa$ [81,82,85].

The Thouless number is the ratio of the average width and spacing of modes in random media. A more detailed characterization of the modes of a random ensemble is the statistics of modal widths and spacings. The statistics of level widths describes the dynamics of levels. The inverse of the level spacings gives the average density of states (DOS). Here we will explore the distribution of the spacings of central frequencies of modes, the level rigidity, and integrated return probability. This exploration suggests that the statistics of the continuous DOS would be of interest and this will be explored in future work. The statistics of discrete levels can be described in terms of statistics of large random matrices [64]. In the study of the distribution of eigenenergies, a quantity of prime interest is the two point correlation function of DOS, which is defined as, $K(\omega) = \overline{\rho(\varepsilon)\rho(\varepsilon-\omega)}\big/\overline{\rho}^2 - 1$ [13], where $\rho(\varepsilon) = \dfrac{1}{V}\sum_n \delta(\varepsilon - \omega_n)$ is the density of states per unit volume. $K(\omega)$ can be simply related to $P(n,s)$ the distribution of spacing between two levels separated by $n$ intervening modes: $K(\omega) = \delta(s) - 1 + \sum_n P(n,s)$, with $\omega = s\Delta\omega$ [13].

Here $s$ is the spacing between two neighboring modes normalized by its average $\Delta\omega$. The Wigner-Dyson distribution of level spacing reflects strong level repulsion. When $s\to 0$, only $P(n=0,s)$ in the sum above contributes to $K(\omega)$, so that $K(\omega)\to\delta(s)-1+P(s)$, where $P(s)$ is the distribution of nearest spacing, $P(s) = P(n=0,s)$. The variance of the number of levels $N(E)$ contained in an energy interval of width $E$, $\Sigma^2(E)$ can also be expressed in terms of $K(\omega)$, $\Sigma^2(E) = \dfrac{2}{(\Delta E)^2}\int_0^E (E-\omega)K(\omega)d\omega$ [13]. $\Sigma^2(E)$ also describes the level rigidity.

In the diffusive limit described in RMT, the Fourier transform of $K(\omega)$, the form factor $\tilde{K}(t)$, can be directly related to the integrated return probability [13,129,130], $\tilde{K}(t) = \dfrac{\Delta^2}{4\pi^2}|t|Z(|t|)$, where, $Z(t) = \int_V P(\mathbf{r},\mathbf{r},t)d\mathbf{r}$, where $V$ is sample volume and $P(\mathbf{r},\mathbf{r}',t) = \dfrac{1}{(4\pi Dt)^{d/2}}\exp(-\dfrac{(\mathbf{r}-\mathbf{r}')^2}{4Dt})$ is the Green function of the diffusion equation. Hence, the spectral statistics of the density of states and the diffusion model are connected through the direct relation between the Fourier transform of the two point correlation function of the DOS and the return probability. In the localized limit, the energy of levels is random so that the two-point correlation function $K(\omega)$ vanishes for $\omega \neq 0$ and $\tilde{K}(t) = 1/2\pi$. Substituting $r = r'$ in $P(r, r', t)$ for diffusive waves gives, $\tilde{K}(t) = \frac{\Delta^2}{2\pi^2\beta} t \frac{\Omega}{(4\pi D t)^{d/2}} \propto t^{1-d/2}$. Thus $\tilde{K}(t)$ is finite for $t \to \infty$ only for $d > 2$ and $d = 2$ is the marginal dimensionality for localization [13].

However, spectral statistics may deviate from the universal properties which are determined only by the symmetries of the system (time reversal and spin) since they also depend upon the character of propagation (the degree of localization) and cannot be explained by RMT. To study how the statistics of levels deviate from predictions for the diffusion limit, different theories and numerical models has been applied. Altshuler and Shklovskii [104] first studied the level fluctuations and correlations in a disordered system using diagrammatic theory. Later the supersymmetry nonlinear $\sigma$ model developed by Efetov [70,71] was used to calculate the level correlation function and various other level statistics [70,71,73,131]. In RMT, the diffusive limit corresponds to ensembles of random symmetric Gaussian matrices, in which the elements of the random matrix are drawn from a Gaussian distribution and obey time reversal and spin. On the other hand the strong localization limit corresponds to the Anderson model, in which off-diagonal elements of the random matrix vanish. It is natural to consider ensembles of random matrices reflecting properties intermediate between the extremes of the diffusive and localized limits and also between the regular and classical chaotic limits of chaotic systems. To this end, Mirlin introduced the power-law random banded matrix in which the random elements decrease from a band around the diagonal in a power law fashion of the distance from the diagonal $\left(\dfrac{1}{|i-j|}\right)^{\alpha}$ [105]. The localization-delocalization transition occurs at the critical value $\alpha = 1$ for which the wavefunction is multifractal [105].

Compared to the substantial theoretical and numerical effort in the study of level statistics in disordered systems, there is little direct experimental measure of the intermediate level statistics. In this chapter, we describe the statistics of level spacings and widths and the rigidity of the mode spectrum for localized waves. Measurements were made in ensembles of samples with lengths of two and three times the localization length for which $\delta$ =0.43 and 0.17, respectively [121]. The level spacing distribution is close to the Wigner surmise expected in the diffusive limit, while the level width distribution is close to a log-normal function for the small linewidth portion of the distribution for deeply localized waves [132,133]. A clear progression towards uncorrelated mode spectra in the longer sample is seen in Dyson's measure of level rigidity [62,134], which specifies the regularity of the modes spacings within a frequency range.

## 4.2 Results and discussions

The experimental setup and random sample studied in this chapter is described in Chapter 2. Samples of length 40 and 61 cm were studied. The mode central frequencies, widths and the complex amplitudes over the output surface are obtained by fitting field spectra at different points to Eq. 3-1. As shown in Fig. 4-1(a) spectra measured at different positions on the output surface for a sample with $L$ = 61 cm are well fitted with a common set of central frequencies and widths of the modes. The spectra of the underlying modes are shown by the Lorentzian lines in Fig. 4-1(b). The Lorentzian lines in the plot are normalized to their integrals over the frequency so that the plot gives the contribution of each of the modes to the density of states.

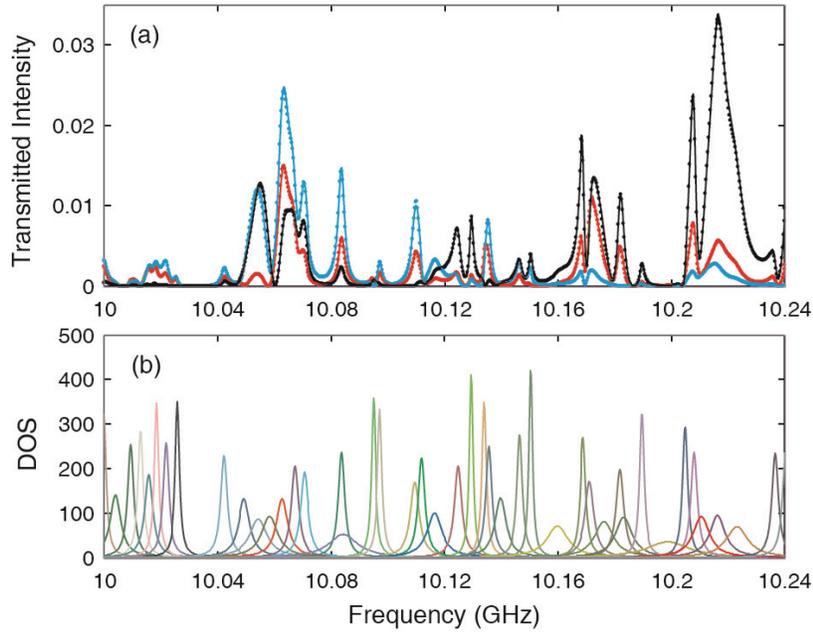

Fig. 4-1 (a) Intensity spectra at three positions in the transmitted speckle pattern for the sample of length $L$=61 cm (dotted lines) and the fit of Eq. 3-1 (solid lines) to the spectra. (b) The solid lines show Lorentzian spectra for each mode normalized to have the same integral, $(2/\pi\Gamma_n)|\varphi_n(\omega)|^2$, representing the contribution of each mode to the density of states.

### 4.2.1 Thouless number

We determined the value of $\delta$ by identifying the average level width, or inverse Thouless time, with the average modal leakage rate $\delta\omega \equiv \tau_{\text{Th}}^{-1} = \left\langle 1/\Gamma_n^0 \right\rangle^{-1}$, and equating the average level spacing $\Delta\omega$ to the average angular frequency difference between neighboring modes. $\delta\omega$ is defined as the inverse of ensemble average of the mode leakage rates rather than the ensemble average of mode widths because it is the dwell time that is indicative of the degree of localization. By defining $\delta\omega$ as the average of the inverse of the Thouless time, the average linewidth is weighted to reflect the dynamics of long-lived modes and hence the degree of localization.

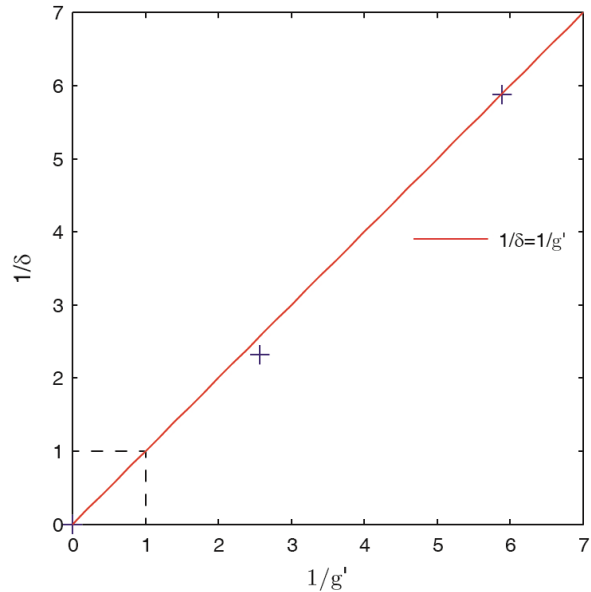

Fig. 4-2 Localization parameters. The degree of localization is tracked by the inverses of the mode overlap parameter, which is equivalent to the Thouless number $\delta$ and the statistical conductance $g'$ which reflects fluctuations in transmission. The dashed lines indicate the localization threshold at $\delta = g' = 1$. Measurements are made for sample lengths of 40 cm and 61cm.

Averaging over 200 sample configurations for the 61-cm-long samples gives $\delta = \delta\omega/\Delta\omega = 0.17$. This equals the statistical conductance $g' = 0.17$, which is determined from the variance of the total transmission in spectra in which the impact of absorption is removed via the relation $g' = 2/3\mathrm{var}(s_a)$. $\delta$ and $g'$ are also found to be close for an ensemble of 40 configurations of samples of length 40 cm with $\delta = 0.43$ and $g' = 0.39$. Figure 4-2 presents the relationship between the inverses of $\delta$ and $g'$, which express the degree of wave localization within the sample. The origin of the plot is the diffusive limit in which the number of modes encompassed within the linewidths diverges and fluctuations of normalized total transmission vanish [21,23, 81,85]. The red line with slope of unity is an extension of the equality of $\delta$ and $g'$ estimated for diffusive waves [21,23] into the localized [35,135] regime. For the systems we studied, the equal spacing found between the logarithms of transmission eigenvalues suggests $g' = 1/2\mathrm{var}(s_a)$ [107]. Thus instead of $\delta = g'$, we have would $\delta = 4g'/3$.

### 4.2.2 Unfolding the spectrum

Because the average mode spacing, and hence the density of states $\rho(\omega)$, in different sections of the spectrum are differ by 10% over the frequency range studied, the statistics of mode spacing in at different frequencies cannot be combined directly. To utilize results over the entire frequency range measured, the spectrum was unfolded [62] by mapping the central frequency of modes to a new frequency scale constructed to give a constant density of states with a value equal to the average value in this interval. We first define a counting function $N(\nu)$ which counts the number of modes in an interval of frequency,

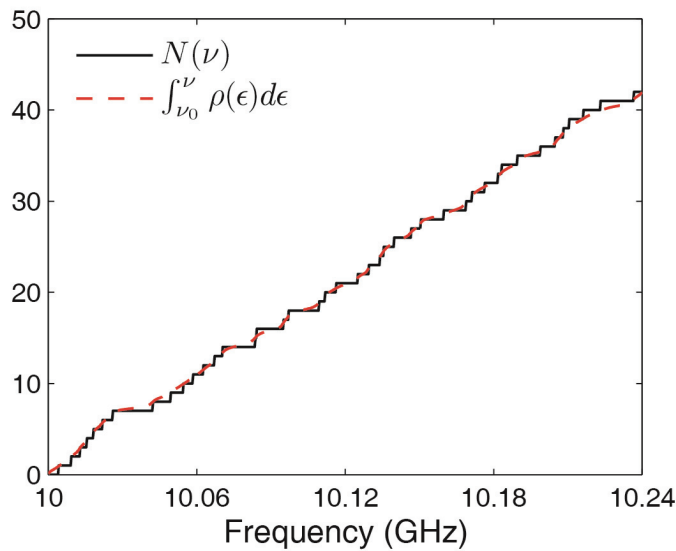

Fig. 4-3 The counting function (black solid line) of the mode spectrum of a configuration of 61-cm-long sample and the integration of the DOS (red dash line).

$$N(\nu) = \sum_n \Theta(\nu - \nu_n) \qquad\qquad (4\text{-}1)$$

where $\Theta$ is the Heaviside function. The counting function increases by unity each time the frequency crosses the central frequency of a mode. The integration of the DOS for the $L$=61 cm sample, obtained by summing up the contributions of modes indicated by the Lorentzian curves in Fig. 4-1(b), is also shown in Fig. 4-3. $N(\nu)$ obtained by counting the number of modes and the cumulative density of states obtained by summing up the contributions of modes with Lorentzian shapes are matched.

The ensemble averages of the counting function for three samples lengths $L$= 40, 61 and 80 cm are shown in Fig. 4-4. When $<N(\nu)>$ is normalized to the maximum value at the end of the frequency range, $<N(\nu)>/<N(\nu=10.24 \text{ GHz})>$ for the three lengths studied overlap. This supports the accuracy of the modal decomposition in cases with different degrees of spectral overlap. We notice that the ensemble average of the counting function deviates from a straight line so that the average of density of states is not uniform across the whole frequency range. To combine the statistics of modes in different frequency segments, we unfold the modal spectra by mapping the sequence of mode central frequencies $\{v_i\}$ to a new sequence $\{x_i\}$ [62]:

$$x_i = <N(v_i)> \tag{4-2}$$

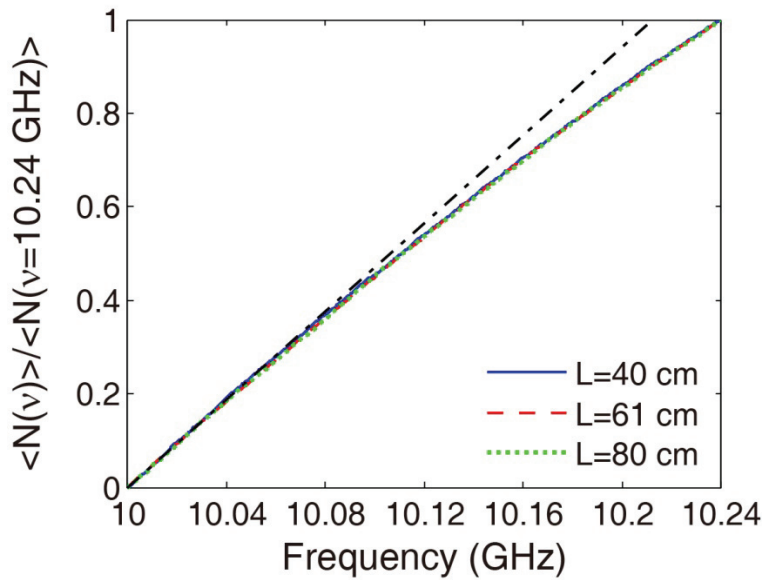

Fig. 4-4 The ensemble averages of the counting functions normalized to their maximum for samples with length $L = 40$ cm (blue solid line), $L = 61$ cm (red dash line) and $L = 80$ cm (green dot line). The black dot-dash line is the best fit to a straight line of the average of counting function at small frequencies.

The new counting function of the sequence $\{x_i\}$, $\hat{N}(x)$ is then forced to be $x$, $\hat{N}(x) = x$, with an average spacing equal to unity over the entire frequency range.

### 4.2.3 Level spacings

The distributions of spacing between the central frequencies of consecutive modes measured for the two sample lengths studied are shown in Fig. 4-5. Distributions for $L=40$ and 61 cm samples are close to the Wigner surmise for GOE systems. The low probability of closely spaced modes demonstrates the presence of strong level repulsion. The reason for this is that even though waves are localized in these samples, modes are still strongly correlated as we can see in the nearly identical intensity patterns and phase patterns with a constant difference seen in Fig. 3-10. This demonstrates that overlapping modes are strongly correlated in these samples. The source of the strong impact of mode overlap can be appreciated by considering the average linewidth $<\Gamma_n^0>$ as opposed to the Thouless linewidth, $\delta\omega$, used to compute $\delta$. $\delta\omega$ is smaller than $<\Gamma_n^0>$ for localized waves since it is weighted to reflect the dynamics of long-lived modes. The ratio of the actual linewidth to the spacing, $<\Gamma_n^0>/\Delta\omega$ equals 0.96 for $L = 40$ cm and 0.42 for $L = 61$cm as compared to the smaller values for $\delta$ of 0.43 and 0.17, respectively. As a result the level spacing statistics are still dominated by correlation between modes as would be the case in the diffusive limit.

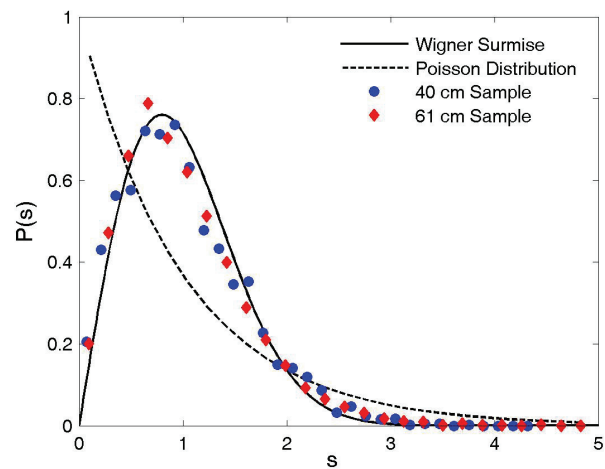

Fig. 4-5 Probability distribution of nearest level spacing in localized samples. The distributions of nearest level spacing are close to the Wigner surmise with $\beta = 1$ (GOE) for both two samples.

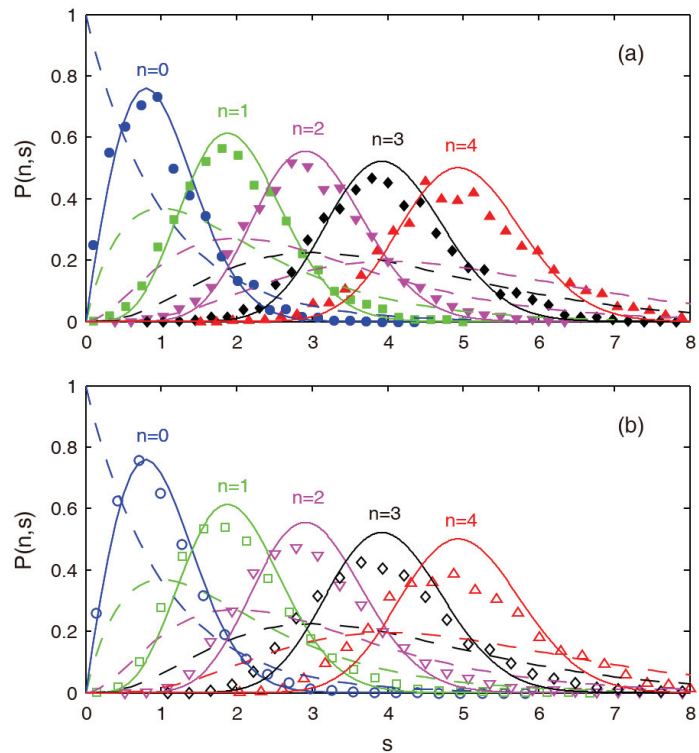

Fig. 4-6 Spacing between two modes separated by $n$ intervening modes, $P(n,s)$, for the two samples studied, $L = 40$ cm (a) and $L = 61$cm (b). The solid and dashed lines show $P(n,s)$ for GOE and Poisson statistics, respectively.

Though the nearest level spacing distributions $P(s)$ are similar for the two localized samples, the differing degrees of level repulsion in these samples can be seen in the level statistics beyond those for nearest neighbors. We first consider the spacing between two modes separated by $n$ intervening modes, $P(n,s)$; for $n = 0$, $P(n,s) = P(s)$. Plots of measurements and of random matrix calculations [64] in the diffusive and localized limits are shown in Fig. 4-6. Measurements of $P(n,s)$ are seen to lie between these limits and to approach the localized limit as $n$ increases. This reflects reduced repulsion between modes compared to the diffusive limit. For more distant neighbors and for more strongly localized waves in longer samples, level spacing statistics depart increasingly from prediction for diffusive samples and approach the prediction for uncorrelated central frequencies, as can be seen in Fig. 4-6(b) .

## 4.2.4 Level rigidity

The rigidity of levels leads to a small variance of the number of modes within a given frequency interval $U$. Normalizing spacings by average spacing gives, $\Sigma^2(U) = <N^2(U)> - <N(U)>^2$ . The variance can be expressed in terms of the 2-point correlation function $K$, $\Sigma^2(x,U) = \int_{x-U/2}^{x+U/2} \int_{x-U/2}^{x+U/2} K(x_1,x_2) dx_1 dx_2$ [13,64]. In the diffusive limit to which RMT applies [20,38], the variance of the number of modes can be expressed as [64]

$$\Sigma^2(U) = \frac{2}{\pi^2}\left[ \ln(2\pi U) + \gamma + 1 + \frac{1}{2}\mathrm{Si}^2(\pi U) - \frac{\pi}{2}\mathrm{Si}(\pi U) - \cos(2\pi U) - \mathrm{Ci}(2\pi U) + \pi^2 U\left(1 - \frac{2}{\pi}\mathrm{Si}(2\pi U)\right) \right]$$

$$= \frac{2}{\pi^2}\left[ \ln(2\pi U) + \gamma + 1 - \frac{\pi^2}{8} \right] + O(U^{-1}), \tag{4-3}$$

where $\gamma \approx 0.5772$ is the Euler constant and the functions $Si(x) = \int_0^x \frac{\sin t}{t} dt$ and $Ci(x) = \gamma + \ln(x) + \int_0^x \frac{\cos t - 1}{t} dt$ are the sine and cosine integrals, respectively. For GOE systems, $\Sigma^2(U)$ increases logarithmically as compared to the linear increase of uncorrelated modes in the localization limit, $\Sigma^2(U) = U$ .

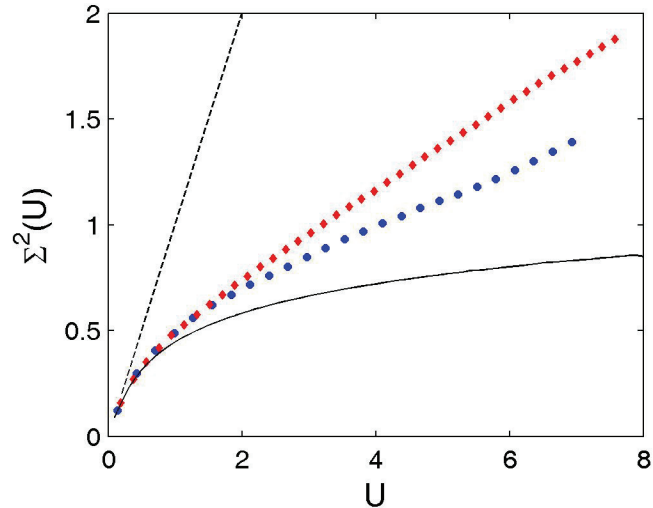

Fig. 4-7 The variance of the number of modes versus normalized frequency interval U for samples of lengths 40 (blue solid circles) and 61cm (red solid diamond), respectively. The curve and dash line represent number variance for the GOE and Poisson ensemble.

The variances of the number of modes in an interval of frequency $U$ normalized by the ensemble average of the nearest neighbor spacing for localized samples for two sample lengths are shown in Fig. 4-7. After a very small frequency interval, these two curves deviate from the curve for GOE systems and increase linearly with the width of the frequency interval [136]. The slope increases with sample length and the crossover from the diffusive limit for a GOE system to the localized limit of Poisson statistics can be seen.

Another measure of the regularity of level spectra is the level rigidity introduced by Dyson and Mehta [134]. It is defined as the least-squares deviation of the staircase function $N(v)$ from the best linear fit over the frequency interval $s$,

$$\Delta(s) = \frac{1}{s} \left\langle \min_{A,B} \int_{x_0}^{x_0+s} (N(x) - Ax - B)^2 \, dx \right\rangle . \qquad (4\text{-}4)$$

as shown in Fig. 4-8(a). The level rigidity is closely related to the number variance and hence to the two-point correlation function $\Sigma^2(U)$, $\Delta(s) = \dfrac{2}{U^4} \int_0^L (s^3 - 2s^2U + U^3)\Sigma^2(U)dU$ [64]. For randomly positioned levels, $\Delta(s) = s/15$. Plots of the spectral rigidity, $\Delta(s)$, for the two sample lengths studied are seen in Fig. 4-8(b) to fall between the limiting curves for GOE and Poisson statistics and to approach the curve for Poisson statistics as the sample length increases. For large $s$, the level rigidity also increases linearly as expected for Poisson statistics, but with a smaller slope.

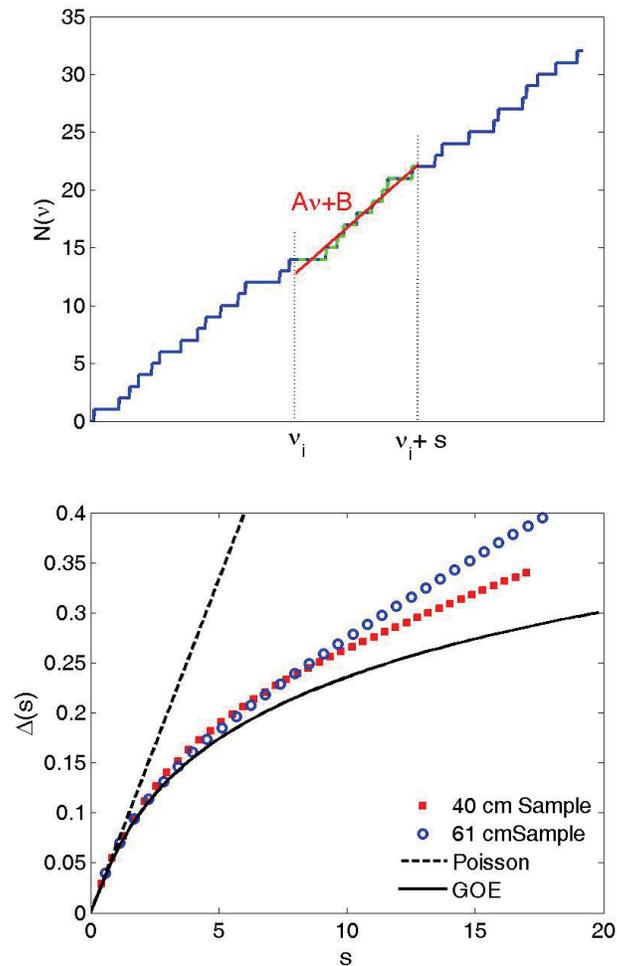

Fig. 4-8 Level rigidity (a) The counting function and its linear fit within an interval *s*. (b) The level rigidity versus frequency interval for samples of lengths 40 and 61cm, respectively.

Thus the spectral rigidity in the localized quasi-1D systems does not disappear but is considerably weaker than for GOE statistics for the diffusive limit. Level rigidity is weakened as the sample length increases and the wave becomes more localized because of level repulsion is reduced as the mode linewidths narrows spatial overlap of modes decreases.

## 4.2.5 Level width

Besides the statistics of mode frequencies, we also studied the statistics of mode widths. For exponentially localized states, the distribution of level widths is predicted to be log-normal based on the two assumptions [132,133], which we also used in the DSPS model [58]: (1) Modes are randomly distributed is space and hence are not correlated in space; (2) modes are exponentially localized with a normal distribution of the inverse localization length. The distribution of level widths normalized by average of level spacing is predicted to be [132,133]

$$P(\tilde{\Gamma}_0) \sim \exp\left(\frac{-\ln^2(\tilde{\Gamma}_0)}{4\dfrac{L}{\xi}}\right) \tag{4-5}$$

where $\tilde{\Gamma}_0$ is the linewidth normalized by average level spacing, $\tilde{\Gamma}_0 = \Gamma_n^0 / \Delta\omega$, $L$ is sample length and $\xi$ is the localization length.

We find that the distribution of normalized linewidths $P(\tilde{\Gamma}_0)$, seen in Fig. 4-9, is close to log-normal, as predicted, for both samples lengths studied. $P(\ln \tilde{\Gamma}_0)$ is seen in Fig. 4-9 to broaden and shift to lower values in the longer sample in which the wave is more strongly localized as predicted in [137]. This agreement with predictions is surprising since the two assumptions are only valid for modes in localized limit. We note that these two assumptions are also made in the DSPS model which is able to model the dynamics of wave transmission at relative long times for lengths which are not much longer than the localization length.

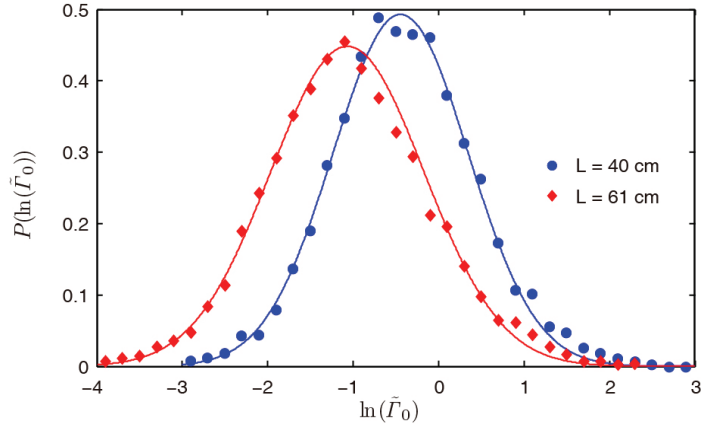

Fig. 4-9 Distribution of mode widths for two localized samples. Curves are the best Gaussian fits to the probability.

The normalized width of fits of a Gaussian to the data are 0.79 and 0.90 for samples with $L$=40 and 61 cm, respectively. The widths of the lognormal distribution are significantly smaller than the predicted value of $(2L_{eff} / \xi)^{1/2}$, 2 and 2.4 for samples with $L$= 40 and 61 cm, which correspond to $L_{eff}$= $L$+$2z_b$ = 52 and 73 cm, where $z_b$ = 6 cm is the extrapolation length [58]. In the diffusive limit, modes tend to have same leakage rate since all the modes overlap strongly with neighboring modes and are well coupled to the boundaries of the sample via a large number of speckle spots. Thus we expect that the width of $P(\ln \tilde{\Gamma}_0)$ to be smaller than given in prediction for deeply localized modes. Thus the small width of the distribution function of linewidths appears to reflect the presence of spatially overlapping states, which were not included in the calculation, but which bring mode linewidth statistics closer to the small width of mode distribution for diffusive waves.

### 4.2.6 Effective number of modes

Aside from parameter $\delta$ obtained by averaging over a random ensemble, the changing degree of mode overlap with frequency shift or with delay time can be described in terms of an effective number of modes, $M_{\text{eff}} = \left( \sum_n I_n \right)^2 \big/ \sum_n I_n^2$. Here $I_n$ is the contribution to the intensity or transmission of the $n^{\text{th}}$ mode at a given frequency or delay time. Fig. 4-10 shows normalized total transmission and the effective number of modes of a configuration of 61-cm-long random sample. Each peak in the spectrum of total transmission corresponds to a dip in the effective number of modes. The average of the effective number of modes is related to the changing variance of total transmission with time as discussed in Chapter 2.

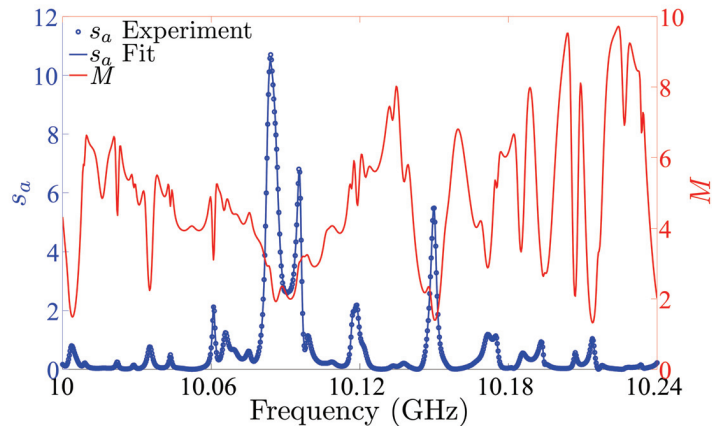

Fig. 4-10 Total transmission and the associated effective number of modes.

### 4.2.7 Correlation between intensity speckle patterns

In chapter 3, we saw that the speckle patterns of two modes which overlap in frequency can be nearly identical. The similarity between the transmitted speckle patterns of neighboring modes and the nearly constant phase difference of the field in these patterns, which can be close to $\pi$, lead to destructive interference between the modes. We believe that the destructive interference between neighboring modes together with the distribution of mode transmission strengths and the spacing and widths of modes can explain the statistics of steady-state and pulsed transmission. To evaluate the correlation between the modes, we define the correlation coefficient between intensity speckle patterns for two modes,

$$C_{i,j}^l(\Delta\nu) = < \frac{\sum_{x,y}(I_i(x,y)-\overline{I}_i)(I_j(x,y)-\overline{I}_j)}{\sqrt{\sum_{x,y}(I_i(x,y)-\overline{I}_i)^2\sum_{x,y}(I_j(x,y)-\overline{I}_j)^2}} > . \qquad (4\text{-}6)$$

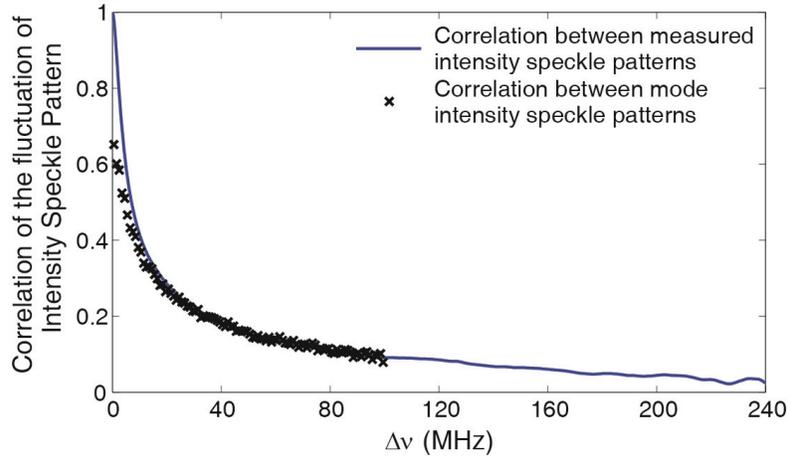

Fig. 4-11 Correlation of intensity speckle pattern with frequency shift and correlation of mode intensity patterns.

Here, $i$ and $j$ are mode indices, $\Delta\nu$ is the frequency difference between modes $i$ and $j$ and $I_j(x,y)$ is the intensity pattern of the $j$th modes. The correlation coefficient varies from unity to zero. Figure 4-11 shows the ensemble average of the variation with frequency shift $\Delta\nu$ of the correlation of transmitted intensity speckle patterns and the correlation of mode intensity patterns with difference in central frequency. For small frequency shift, the correlation of mode intensity patterns is strong but less than the correlation of the transmitted intensity patterns. This can be explained by noting that the intensity pattern is also correlated with other nearby modes. For small frequency shifts, the frequency difference between these modes is greater than difference between the modes shifted only slightly in frequency. One mode can be correlated with modes falling within its width. The mode is therefore not fully correlated with nearby modes. For example, the speckle pattern of mode 21 shown in Fig 4-12 is seen to bear a similarity to the patterns of each of its two nearest neighbors, while the patterns of modes 23 and 24 are not perceptibly correlated. At large frequency shifts, the correlation between mode speckle patterns converges to the correlation function of measured transmission intensity patterns. The residual correlation at large frequency shifts suggests an alternate modal description of long-range correlation.

## 4.2.8 Modes and channels

Wave transport can be described using the scattering matrix or its submatrix, the field transmission matrix $t$. The elements of $t$ relate the transmitted field in propagating channel, $b$, and the channels of the incident field, $a$, $E_b = \sum_{a=1}^{N} t_{ba} E_a$ . $t$ can be expressed as, $t = U \Lambda V^\dagger$, where $U$ and $V$ are unitary matrices and $\Lambda$ is a diagonal matrix with singular values, $\lambda_n$, of $t$ along the diagonal. The dimensionless conductance equals the ensemble average of the sum of all transmission eigenvalues, $g = \sum_n \tau_n$ , where $\tau_n$ is the eigenvalues of the transmission matrix $tt^\dagger$ and $\tau_n = \lambda_n^2$ . The field at the output may be represented as a sum of orthogonal speckle patterns each generated by one of a set of orthogonal incident field patterns corresponding to the singular values $\lambda_n$. Since the field at the output can be expressed as the superposition of the speckle patterns of modes and the complex frequency of each mode, $\omega_i - i\Gamma/2$, is the pole of the scattering matrix, the connection between the modes and eigenchannels is of great interest.

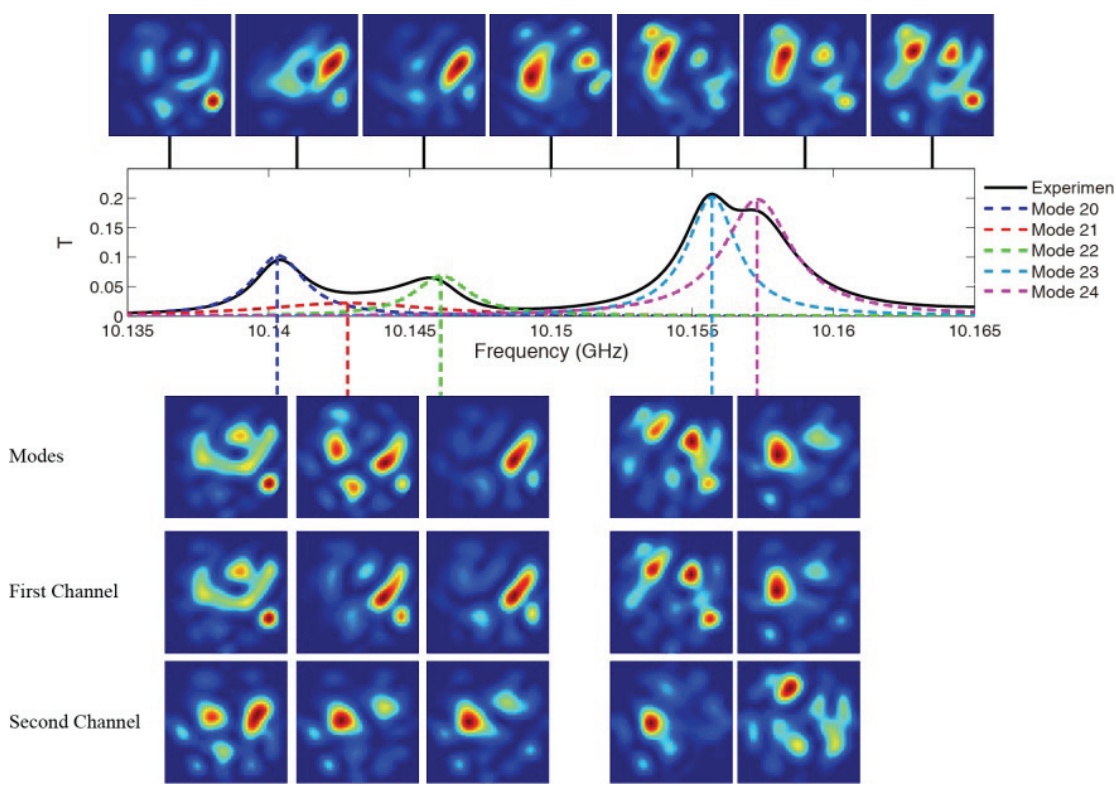

Fig. 4-12 Spectrum of the transmittance and of the transmittance for the most prominent modes together with intensity speckle patterns for transmission generated by a source at the center of the input surface. Patterns for the modes and first two transmission channels are shown below the spectra. The decomposition into modes and channels is described in the text.

Figure 4-12 shows spectra of transmittance over a narrow frequency range in a random 40-cm-long sample, as well as of the transmittance of the most prominent modes found from the global fit of Eq. (3-1) to field spectra. The change in the normalized speckle patterns as the frequency is tuned is shown on the top row of the figure. The mode speckle patterns are shown below the spectra of transmittance. Peaks in total transmission arise when the wave is tuned on or off resonance with modes of the medium. The modes may be highly correlated as seen in the speckle patterns of adjacent modes shown in Fig. 3-10, or more weakly correlated as seen for the two groups of spectrally overlapping modes in Fig. 4-12. The speckle pattern of mode 21 is seen to bear a similarity to the patterns of each of its two nearest neighbors, while the patterns of modes 23 and 24 are not perceptibly correlated. The degree of correlation is linked to the spatial overlap of resonances from which the modes are formed. The speckle patterns of the first two transmission eigenchannels are shown at the bottom of the figure. The transmitted speckle pattern for the first transmission channel in resonance with a mode is seen in Fig. 4-12 to be close to the pattern for the resonant mode. The second channel is necessarily orthogonal to the first and so is made up of speckle patterns of modes that are further off resonance.

## 4.3 Conclusion

In this study, we have compared measurements of the statistics of level width and spacing to long-standing predictions relevant to the diffusive and localized limits and explored the crossover between these limits. The closeness of the distribution of level spacings to the Wigner surmise predicted for diffusive waves reflects significant spatial and spectral overlap of modes, even for localized waves. The small value of the probability distribution at small mode spacings demonstrates that the repulsion between neighboring modes is still strong. Level repulsion is due to the strong coupling and correlation between modes, which is also seen in the strong correlation between the intensity patterns of the modes when $\delta < 1$. Other statistics of modes including the correlation between mode speckle patterns and the effective number of modes can be introduced to provide a fuller description of wave transport. The statistics of level spacing may also open a window on dynamics within the random medium. The correlation of mode density with frequency shift may provide a direct measure of the impact of weak localization by giving the average number of photon returns to typical coherence volumes within the medium. In addition, the connection between modes and transmission eigenchannels may help clarify the connection between statistics of intensity, total transmission and transmittance.

# CHAPTER 5

# CONCLUSION

The ratio between average linewidth and spacing of energy levels, known as the Thouless number, has been understood to govern the transition between diffusion and localization for the last 40 years and has been a touchstone in the study of wave propagation in disordered system. But before this work, the statistical properties of modes in random media have not been determined experimentally other than in one-dimensional studies. This has delayed a comprehensive modal description of wave propagation.

In our initial studies, we have measured the steady state and dynamic characteristics of fluctuations and correlation of localized microwave radiation transmitted through random waveguides. Correlation was observed with polarization rotation of both the source and detector or with displacement of a wire antenna detector. We found that mesoscopic fluctuations and correlation of localized waves did not increase monotonically with delay times. The degree of correlation first decreased and then increased with time. This was explained in terms of the variation with time of the effective number of modes contributing to transmission. The effective number of modes changed because modes decay with time with a wide distribution of decay rates. The non-monotonic behavior of correlation with time was explained in terms of the changing contributions of short- and long-lived electromagnetic modes of the random medium. But this explanation is qualitative and not complete, the distribution of mode linewidths as well as possible correlation between modes was not known.

To retrieve modes from complex field spectra, we developed an algorithm to decompose the transmitted field speckle patterns into a superposition of the mode patterns and found the central frequency and linewidth of each mode. We find strong correlation between modal field speckle patterns, which leads to destructive interference between modes. This allows us to explain complexities of steady state and pulsed transmission of localized waves. The correlation between the modes leads two important statistical properties of modes: level repulsion, which suppresses the probability of central frequencies of two modes being close, and level rigidity, which gives a small variation of the number of modes in any frequency range. A number of other statistical properties of modes were studied including the correlation between modal speckle patterns and the effective number of modes. We have found the connection between speckle patterns of modes and speckle patterns of the first two transmission eigenchannels and have explored the link between two powerful approaches, the modal and channel descriptions of wave transportation.